\documentclass[%
reprint,
superscriptaddress,
amsmath,amssymb,
aps,
prd,
]{revtex4-1}

\usepackage{graphicx}
\usepackage{dcolumn}
\usepackage{bm}
\usepackage{hyperref}
\usepackage{subfigure}
\usepackage{float}

\begin{document}

\preprint{APS/123-QED}

\title{Charged Compact Boson Stars and Shells in the Presence of a Cosmological Constant} 

\author{Sanjeev Kumar}
\email{sanjeev.kumar.ka@gmail.com}
\affiliation{Department of Physics and Astrophysics, University of Delhi, Delhi-110007, India}
\author{Usha Kulshreshtha}%
\email{ushakulsh@gmail.com, ushakuls@iastate.edu}
\affiliation{Department of Physics, Kirori Mal college, University of Delhi, Delhi-110007, India}%
\affiliation{Department of Physics and Astronomy, Iowa State University, Ames, 50010 IA}
\author{Daya Shankar Kulshreshtha}
\email{dskulsh@gmail.com, dayakuls@iastate.edu}
\affiliation{Department of Physics and Astrophysics, University of Delhi, Delhi-110007, India}%
\affiliation{Department of Physics and Astronomy, Iowa State University, Ames, 50010 IA}

\date{\today}

\begin{abstract}
In this work we study the boson stars and boson shells in a theory involving {\it massive} complex scalar fields coupled to the U(1) gauge field and gravity in a conical potential in the {\it presence} of a cosmological constant ${\Lambda}$ which we treat as a free parameter taking positive and negative values and thereby allowing us to study the theory in the de Sitter and Anti de Sitter spaces respectively. Boson stars are found to come in two types, having either ball-like or shell-like charge density.  We have studied the properties of these solutions and have also determined their domains of existence for some specific values of the parameters of the theory. Similar solutions have also been obtained by Kleihaus, Kunz, Laemmerzahl and List in a theory involving {\it massless} complex scalar fields coupled to the U(1) gauge field and gravity in a conical potential  in the {\it absence} of a cosmological constant~${\Lambda}$. 
\end{abstract}
\maketitle

\section{Introduction}
Boson stars and boson shells representing the localized self-gravitating solutions were introduced long ago \cite{Feinblum:1968,Kaup:1968zz,Ruffini:1969qy} and they have been studied vary widely in the literature \cite{Jetzer:1991jr,Lee:1991ax,Mielke:2000mh,Liebling:2012fv,Friedberg:1976me,Coleman:1985ki,Kleihaus:2009kr,Kleihaus:2010ep,Hartmann:2012da,Hartmann:2012wa,Hartmann:2013kna,Kumar:2016oop,Kumar:2014kna,Kumar:2015sia,Astefanesei:2003qy,Radu:2012yx,Prikas:2004yw,Brihaye:2013hx,Pugliese:2013gsa,Brihaye:2014gua,Kichakova:2013sza,Dzhunushaliev:2014bya,Arodz:2008jk,Arodz:2008nm,Arodz:2012zh}. Such theories are being considered in the presence of positive \cite{Hartmann:2013kna,Kumar:2016oop,Kumar:2014kna,Kumar:2015sia} as well as negative \cite{Kumar:2016oop,Kumar:2015sia,Astefanesei:2003qy,Radu:2012yx,Prikas:2004yw,Brihaye:2013hx} values of the cosmological constant $	 {\Lambda}$. The theories with positive values of $	 {\Lambda}$ (corresponding to the de Sitter (dS) space) are relevant from observational point of view as they describe a more realistic description of the compact stars in the universe since all the observations seem to indicate the existence of a positive cosmological constant. Such theories are also being used to model the dark energy of the universe. However, the theories with negative values of $	 {\Lambda}$ (corresponding to the Anti de Sitter (AdS) space) are meaningful in the context of AdS/CFT correspondence \cite{Maldacena:1997re,Witten:1998qj,Brodsky:2011sk}.

In Ref. \cite{Kumar:2014kna}, we studied the boson stars and boson shells in a theory of complex scalar field coupled to  $U(1)$ gauge field $A_{\mu}$ and the gravity in the presence of a {\it positive} cosmological constant $ {\Lambda}$ (i.e. in the dS space) and in Ref. \cite{Kumar:2016oop}, we studied the {\it boson stars} in a theory of complex scalar field coupled to  $U(1)$ gauge field $A_{\mu}$ and the gravity in the presence of a positive as well as negative cosmological constant ${\Lambda}$ allowing us to study the theory in the dS as well as in the AdS space.

In the present work we study {\it not only the boson stars but also the boson shells} in this theory of complex scalar field coupled to  $U(1)$ gauge field $A_{\mu}$ and the gravity and a cosmological constant ${\Lambda}$ which we treat as a free parameter and {\it which takes positive as well as negative values} and thereby allowing us to study the theory in the dS as well as in the AdS space.  As in Ref. \cite{Kumar:2016oop}, for our present investigations also we study the theory in the presence of a potential: $V(|\Phi|) := (m^2 |\Phi|^2 +\lambda |\Phi|) $ (with $m$ and $\lambda$ being constant parameters). We investigate the  properties of the solutions of this theory and determine their domains of existence for some specific values of the parameters of the theory. 

Similar solutions have also been obtained by Kleihaus, Kunz, Laemmerzahl and List in a theory involving {\it massless} complex scalar fields coupled to the U(1) gauge field and gravity in a conical potential  in the {\it absence} of a cosmological constant $	 {\Lambda}$ \cite{Kleihaus:2009kr,Kleihaus:2010ep}. They have obtained explicitly the domain of existence of compact boson stars and boson shells. They have also considered the boson shells, which do  not have an empty inner region $r<r_i$, but instead they harbour a Schwarzschild black hole or a Reissner-Nordstr\"om black hole in the region $r<r_i$ \cite{Kleihaus:2009kr,Kleihaus:2010ep}. Boson stars have also been studied in the presence of polynomial potentials \cite{Pugliese:2013gsa,Brihaye:2014gua,Dzhunushaliev:2014bya,Kichakova:2013sza} 

In the present work, we construct the boson star and boson shell solutions of this theory numerically and we study their properties, where we assume the interior of the shells to be empty space (dS-like or AdS-like).  The action and the equations of motion are given in section II. In section III the equations of motion are re-expressed in terms of the rescaled variables. The boundary conditions and the global charges are considered in section IV. The numerical solutions for boson stars and boson shells are studied in section V and finally the summary and conclusions are given in section VI.

\section{The action, Ans\"atze and equations of motion}

The action of the theory under consideration reads:
\begin{eqnarray}
S&=&\int \left[ \frac{R-2\Lambda}{16\pi G}   +\mathcal L_M \right] \sqrt{-g}\ d^4\,x\label{action}
\,,\nonumber\\ \mathcal L_M& =&	- \frac{1}{4} F^{\mu\nu} F_{\mu\nu}
   -  \left( D_\mu \Phi \right)^* \left( D^\mu \Phi \right)
 - V(|\Phi|)\,,\nonumber\\
 D_\mu \Phi &=& (\partial_\mu \Phi + i e A_\mu \Phi)\,,\ \ F_{\mu\nu} = (\partial_\mu A_\nu - \partial_\nu A_\mu)\,.
\end{eqnarray}
Here $R$ is the Ricci curvature scalar, $G$ is Newton's Gravitational constant and $	 {\Lambda}$ is cosmological constant. Also, $g = det(g_{\mu\nu})$ where $g_{\mu\nu}$ is the metric tensor and the asterisk in the above equation denotes complex conjugation. Using the variational principle, equations of motion are obtained as
\begin{eqnarray}
 G_{\mu\nu}&\equiv& R_{\mu\nu}-\frac{1}{2}g_{\mu\nu}R = 8\pi G T_{\mu\nu}-\Lambda g_{\mu\nu}
\,,\nonumber \\ 
 \partial_\mu \left ( \sqrt{-g} F^{\mu\nu} \right)
&= &  -i\, e \sqrt{-g}\, [\Phi^* (D^\nu \Phi)-\Phi (D^\nu \Phi)^* ] 
\,,\nonumber \\
D_\mu\left(\sqrt{-g}  D^\mu \Phi \right) &=& 2 m^2 \sqrt{-g}\, \Phi +\frac{\lambda }{2}\sqrt{-g}\,\frac{\Phi}{|\Phi|}\,,\nonumber \\
\left[D_\mu\left(\sqrt{-g}  D^\mu \Phi \right)\right]^* &=& 2 m^2 \sqrt{-g}\, \Phi^* +\frac{\lambda }{2}\sqrt{-g}\,\frac{\Phi^*}{|\Phi|}
  \label{vfeqH}
 \end{eqnarray}
The energy-momentum tensor $T_{\mu\nu}$ is given by 
\begin{eqnarray}
T_{\mu\nu} &=& \biggl[ ( F_{\mu\alpha} F_{\nu\beta}\ g^{\alpha\beta} -\frac{1}{4} g_{\mu\nu} F_{\alpha\beta} F^{\alpha\beta})
\nonumber\\ & &+ (D_\mu \Phi)^* (D_\nu \Phi)+ (D_\mu \Phi) (D_\nu \Phi)^*  
\nonumber \\ & & -g_{\mu\nu} \left((D_\alpha \Phi)^* (D_\beta \Phi)    \right) g^{\alpha\beta}
 -  g_{\mu\nu}\; V( |\Phi|) \biggr]\,.  \ \ \   \label{vtmunu}
\end{eqnarray}
To construct spherically symmetric solutions we adopt static spherically symmetric metric with Schwarzschild like coordinates \cite{Kleihaus:2009kr,Kleihaus:2010ep}:
\begin{equation}
ds^2= \biggl[ -A^2 N dt^2 + N^{-1} dr^2 +r^2(d\theta^2 + \sin^2 \theta \;d\phi^2) \biggr].\  \ \
\end{equation}
This leads to the components of Einstein tensor ($G_{\mu\nu}$) 
\begin{eqnarray}
G_t^t &=& \biggl[ \frac{-\left[r\left(1-N\right)\right]'}{r^2} \biggr] ,\, 
G_r^r = \biggl[ \frac{2 r A' N -A\left[r\left(1-N\right)\right]'}{A\ r^2} \biggr],\nonumber \\
G_\theta^\theta &=& \biggl[ \frac{2r\left[rA'\ N\right]' + \left[A\ r^2 N'\right]'}{2 A\ r^2} \biggr]
\  \   = \  G_\varphi^\varphi\,.
\end{eqnarray}
Here the arguments of the functions $A(r)$ and $N(r)$ have been suppressed. For solutions with vanishing magnetic field, the Ans\"atze for the matter fields have the form \cite{Kleihaus:2009kr,Kleihaus:2010ep}:
\begin{eqnarray}
 \Phi(x^\mu)=\phi(r) e^{i\omega t}
\ \ ,\ \ 
A_\mu(x^\mu) dx^\mu = A_t(r) dt.
\end{eqnarray}
With these Ans\"atze, the Einstein equations
\begin{eqnarray}
\;G_t^t = 8 \pi G\ T_t^t -\Lambda\ \ ,\ \  \;G_r^r =  8 \pi G\  T_r^r -\Lambda\ \,,\nonumber \\ \;G_\theta^\theta =  8 \pi G\ T_\theta^\theta-\Lambda\ \ ,\ \  \;G_\varphi^\varphi =  8 \pi G\ T_\varphi^\varphi-\Lambda
\end{eqnarray}
with the arguments of $A(r)$, $N(r)$, $\phi(r)$ and $A_t(r)$ being  suppressed, reduce to:
\begin{widetext}
\begin{eqnarray}
 \frac{-1}{r^2}\left[r\left(1-N\right)\right]' 
& = & \frac{-8\pi G}{2A^2 N e^2} \bigg[ N [(\omega +e A_t )']^2  +(\omega + e A_t )^2 (\sqrt{2} e \phi)^2 \,,\nonumber \\
& & \hspace{1.5cm}+ A^2 N^2 (\sqrt{2} e \phi')^2 \left.+ A^2 N m^2(\sqrt{2}\, e\, \phi)^2 +\frac{2\,e\,\lambda}{\sqrt 2}(\sqrt{2}\,e\,h)  \right] -\Lambda\\
\frac{2 r A' N -A\left[r\left(1-N\right)\right]'}{A r^2}
& = & \frac{8\pi G}{2A^2 N e^2}\bigg[ -N [(\omega +e A_t )']^2 +(\omega + e A_t )^2 (\sqrt{2} e \phi)^2\,,\nonumber \\
& & \hspace{1.5cm}+ A^2 N^2 (\sqrt{2} e \phi')^2  \left.- A^2 N m^2(\sqrt{2}\, e\,\phi)^2 -\frac{2\,e\,\lambda}{\sqrt 2}(\sqrt{2}\,e\,\phi) \right]  -\Lambda\\
\frac{2r\left[rA'N\right]' + \left[A r^2 N'\right]'}{2 A r^2} & = &\frac{8\pi G}{2A^2 N e^2} \bigg[ N [(\omega +e A_t )']^2 +(\omega + e A_t )^2 (\sqrt{2} e \phi)^2 \,,\nonumber \\
& & \hspace{1.5cm}-A^2 N^2 (\sqrt{2} e \phi')^2  \left.- A^2 N m^2(\sqrt{2}\, e\,\phi)^2 -\frac{2\,e\,\lambda}{\sqrt 2}(\sqrt{2}\,e\,\phi)\right]  -\Lambda \label{dtheta}
\end{eqnarray}
\end{widetext}
\noindent Here the prime denotes differentiation with respect to $r$ and the equation $G_\varphi^\varphi =  8 \pi G\ T_\varphi^\varphi-\Lambda$ also leads to an equation identical with Eq. \ref{dtheta}.

\section{The equations of motion in terms of rescaled variables}
We redefine $\phi(r)$ and $A_t(r)$ as

\begin{equation}
 h(r)=\frac{\sqrt{2} \;e\, \phi(r)}{m}\,, \ \ \ \ \ \  b(r)=\frac{\omega+e A_t(r)}{m}\label{hb}.
 \end{equation}
We introduce new dimensionless constant parameters:
\begin{equation}
 \alpha = \frac{4\pi G\,m^2}{e^2}\,, \ \ \ \ \ \  \tilde{\lambda}=\frac{\lambda\,e}{\sqrt{2}\; m^3} \,, \ \ \ \ \ \tilde{\Lambda}=\frac{\Lambda}{m^2}.
\end{equation}
Introducing a dimensionless coordinate $\hat{r}$ defined by $\hat{r}=m\,r$~(implying $\frac{d}{dr}=m\frac{d}{d\hat{r}}$),  Eq. (\ref{hb}) reads as:
\begin{equation}
 h(\hat{r})=\frac{\sqrt{2} \;e\, \phi(\hat{r})}{m} \,, \ \ \ \ \  b(\hat{r})=\frac{\omega+e A_t(\hat{r})}{m}\label{hb1}.
 \end{equation}

 Equations of motion in terms of $h(\hat{r})$ and $b(\hat{r})$ [where the primes denote differentiation with respect to $\hat{r}$] read
 
\begin{eqnarray}
\left( A N \hat{r}^2 h'\right)' &=& \frac{\hat{r}^2}{A N}\left[A^2 N(h+\tilde{\lambda}\, {\rm sign}(h)) - b^2 h\right]\label{vheq},\\
\left[\frac{ \hat{r}^2\  b'}{A} \right]' &=& \left[ \frac{\hat{r}^2 h^2 b }{A\ N} \right]\label{vbeq}.
\end{eqnarray}

where 

$$ {\rm sign}(h) = \left\{ \begin{array}{ll}
\pm1 \ \ & h>0,\ h<0 \\ 0 & h=0 \end{array}\right.$$

With the above Ans\"atze (with the primes denoting the differentiation with respect to $\hat{r}$) we obtain

\begin{widetext}
\begin{subequations}
\begin{eqnarray}
\frac{1}{\hat{r}^2}\left[\hat{r}\left(1-N\right)\right]'&=&\bigg[\tilde{\Lambda} + \frac{\alpha}{A^2 N}\left(b^2 h^2 + N b'^2+A^2 N^2 h'^2+ A^2 N (h^2 +2 \tilde{\lambda} h)\right)\bigg]\label{E_00},
\\
\frac{2 \hat{r} A' N -A\left[\hat{r}\left(1-N\right)\right]'}{A \hat{r}^2} 
&=&\bigg[ -\tilde{\Lambda} +\frac{\alpha}{A^2 N}
\left(b^2 h^2-N b'^2+A^2 N^2 h'^2   - A^2 N (h^2 + 2 \tilde{\lambda} h)\right)\label{E_rr},\\
\frac{2r\left[rA'N\right]' + \left[A r^2 N'\right]'}{2 A r^2}&=&\bigg[ -\tilde{\Lambda} +\frac{\alpha}{A^2 N}
\left(b^2 h^2+ N b'^2-A^2 N^2 h'^2   - A^2 N (h^2 +2\tilde{\lambda} h )\right)\bigg] ,
\\
h''&=& \biggl[\frac{A^2 N (h+\tilde{\lambda} \, {\rm sign}(h)) - b^2 h}{A^2 N^2} -\frac{2\,h'}{\hat{r}}-h'\left(\frac{A'}{A}\frac{N'}{N}\right) \biggr] \ \ \ \label{heq1}, \\
b''&=&\biggl[ \frac{b\,h^2}{N}+\frac{b'A'}{A}-\frac{2\,b'}{\hat{r}} \biggr] \label{beq1}.
\end{eqnarray}
\end{subequations}
\end{widetext}

Simplifying Eqs.(\ref{E_00}) and (\ref{E_rr}) for $A'$ and $N'$ and also using eqs (\ref{heq1}) and (\ref{beq1}) we get

\begin{subequations}
\begin{eqnarray}
h'' & = &\bigg[ \frac{\alpha\, \hat{r} h'}{A^2N} \left(A^2 h^2 +2 A^2 h \tilde{\lambda} + \,b'^2\right)
-\frac{h'\big(1+N-\tilde{\Lambda} \hat{r}^2\big)}{\hat{r}N}\nonumber\\ & &+\frac{A^2 N  h+A^2 N \tilde{\lambda} \, {\rm sign}(h) - b^2 h}{A^2 N^2}\bigg]\label{eq_H},\\
b'' & = & \bigg[\frac{\alpha}{A^2 N^2} \hat{r} b'\left(A^2 N^2 h'^2 + b^2 h^2\right) -\frac{2 b'}{\hat{r}} + \frac{b h^2}{N}\bigg]\label{eq_b},\\
N' & = &\bigg[ \frac{1-N-\tilde{\Lambda} \hat{r}^2}{\hat{r}} -\frac{\alpha \hat{r}}{A^2 N}
\bigg(A^2 N^2 h'^2 + N b'^2 +  b^2 h^2\nonumber\\ & & + A^2 N  h^2 + 2 A^2 N h \tilde{\lambda}\bigg)\bigg] \label{eq_N},\\
A'  &=& \bigg[ \frac{\alpha \hat{r}}{A N^2}\left(A^2 N^2 h'^2 + b^2 h^2\right) \bigg]\ \label{eq_A}.
\end{eqnarray}	
\end{subequations}	

To solve equations (\ref{eq_H}), (\ref{eq_b}) ,(\ref{eq_N}), (\ref{eq_A}) numerically, we introduce a new coordinate $x$ as follows \cite{Kleihaus:2009kr,Kleihaus:2010ep}
\begin{equation}
\hat{r}= \hat{r}_{i} +x (\hat{r}_{o}-\hat{r}_{i})\ , \ \ \ \ 0\leq x \leq 1 \label{vcoord_x}.
\end{equation}
implying that $ \hat{r}=\hat{r}_{i} \ \ {\rm at} \ \ x=0 $ and 
$ \hat{r}=\hat{r}_{o}\  {\rm at} \ x=1$.
Thus the inner and outer boundaries of the shell are always at $x=0$ and $x=1$ respectively, 
while their radii $\hat{r}_{i}$ and $\hat{r}_{o}$ become free parameters. 
\section{The boundary conditions and global charges}
For the metric function $A(r)$ we choose the 
boundary condition
\begin{eqnarray}
A(\hat{r}_{\rm o})=1 \, ,\ \ \label{aro}
\end{eqnarray}
where $r_{\rm o}$ is the outer radius of the shell. For constructing globally regular ball-like boson star solutions, we choose
\begin{eqnarray}
 N(0)&=&1 \,, \ \ \ \ b'(0)=0 \,, \ \ \   h'(0)=0 \,,\nonumber\\  
h(\hat{r}_{o})&=&0\,, \ \ \ \ h'(\hat{r}_{o})=0  \label{bcstar}
\end{eqnarray}

For the boson stars, for the positive $\tilde \Lambda$ we match the exterior region $\hat r >\hat r_o$ , with the Reissner-Nordstr\"om de Sitter solutions and for the negative $\tilde \Lambda$ we match with the Reissner-Nodrstro\"om Anti de Sitter solutions.

For globally regular boson shell solutions with empty space-time in the interior of the shells, $r<r_{i}$, we choose the boundary conditions
\begin{eqnarray}
N(\hat{r}_i)&=&1-\frac{\Lambda}{3} \hat{r}_i^2\,, \ \ \ b'(\hat{r}_{i})=0 \ ,\ \  h(\hat{r}_{i})=0 \ ,\nonumber \\
h'(\hat{r}_{i})&=& 0\,, \ \ \ \ \ \ \ \ \ \ \ \  h(\hat{r}_{o}) = 0\,, \ \ \ h'(\hat{r}_{o})=0\label{bcshell}.
\end{eqnarray}
where $\hat{r}_{i}$ and $\hat{r}_{o}$ are the inner and outer radii of the shell.

For the boson shells, for the positive ${\tilde \Lambda}$ we match the interior region $\hat{r}<\hat{r}_i$ , with the de Sitter vacuum solution and  the exterior region $\hat{r}>\hat{r}_o$ , with the Reissner-Nordstr\"om de Sitter solutions. However, for the negative ${\tilde \Lambda}$ we match  the interior region $\hat{r}<\hat{r}_i$ , with the Anti de Sitter vacuum solution and the exterior region $\hat{r}>\hat{r}_o$ , with the Reissner-Nordstr\"om Anti de Sitter solutions.

\begin{figure*}
\begin{center}
\mbox{\subfigure[][]{\includegraphics[scale=0.65]{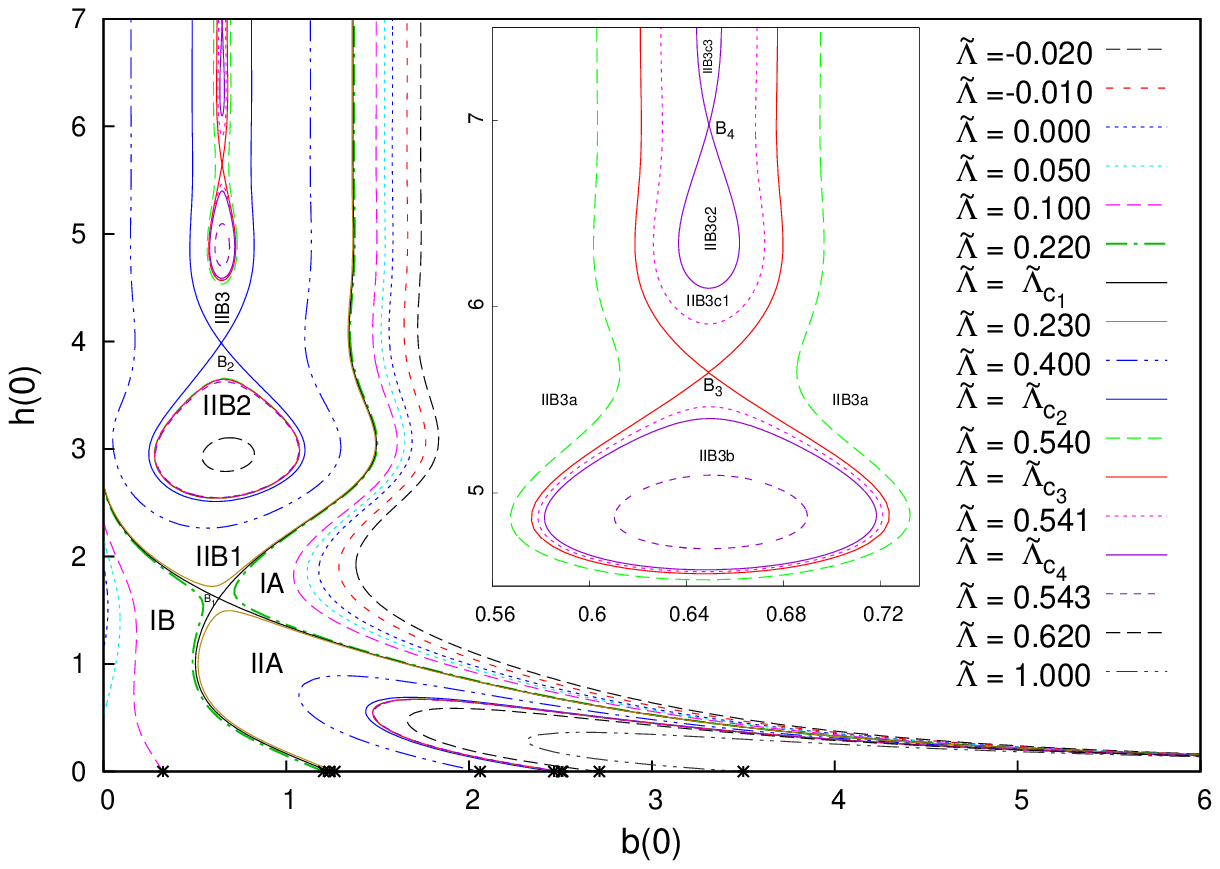}\label{fig:pa}}\hspace{1cm}
\subfigure[][]{\includegraphics[scale=0.65]{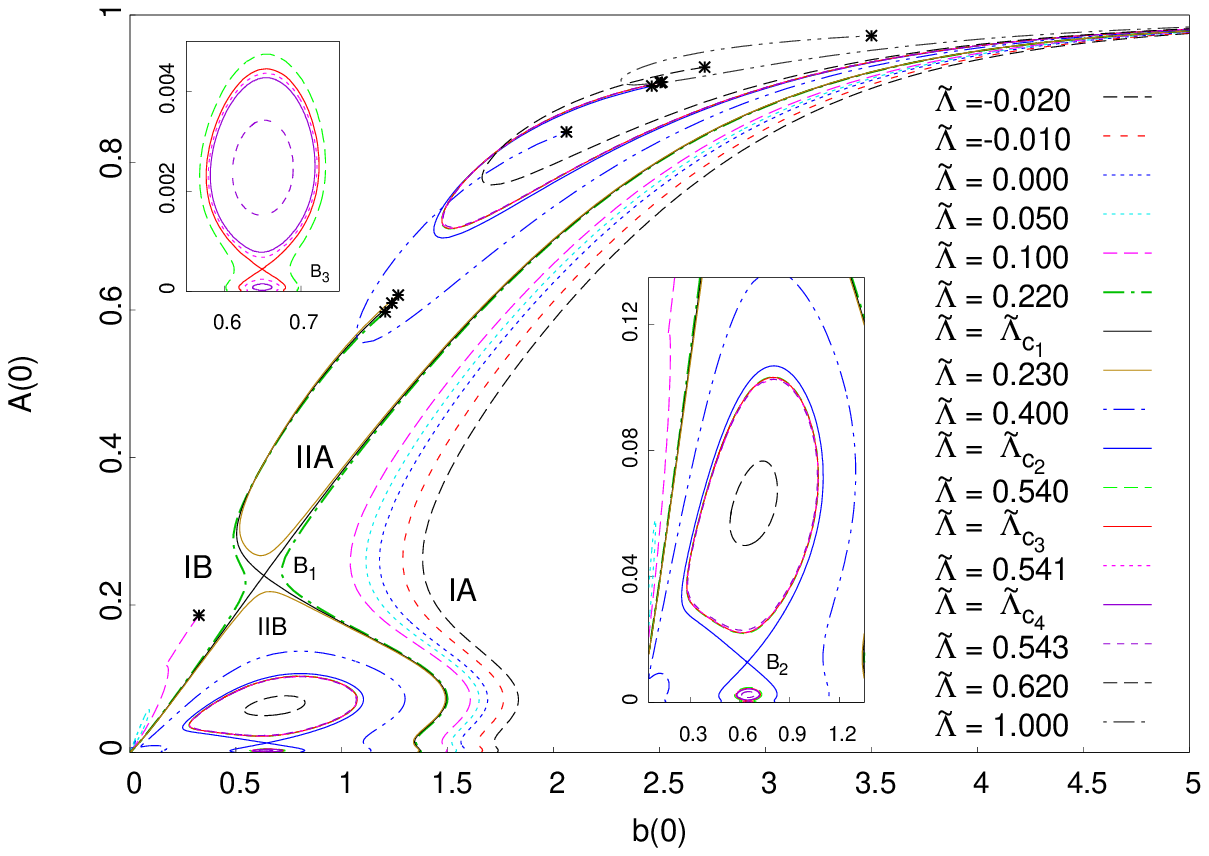}\label{fig:pb}}}
\mbox{\subfigure[][]{\includegraphics[scale=0.65]{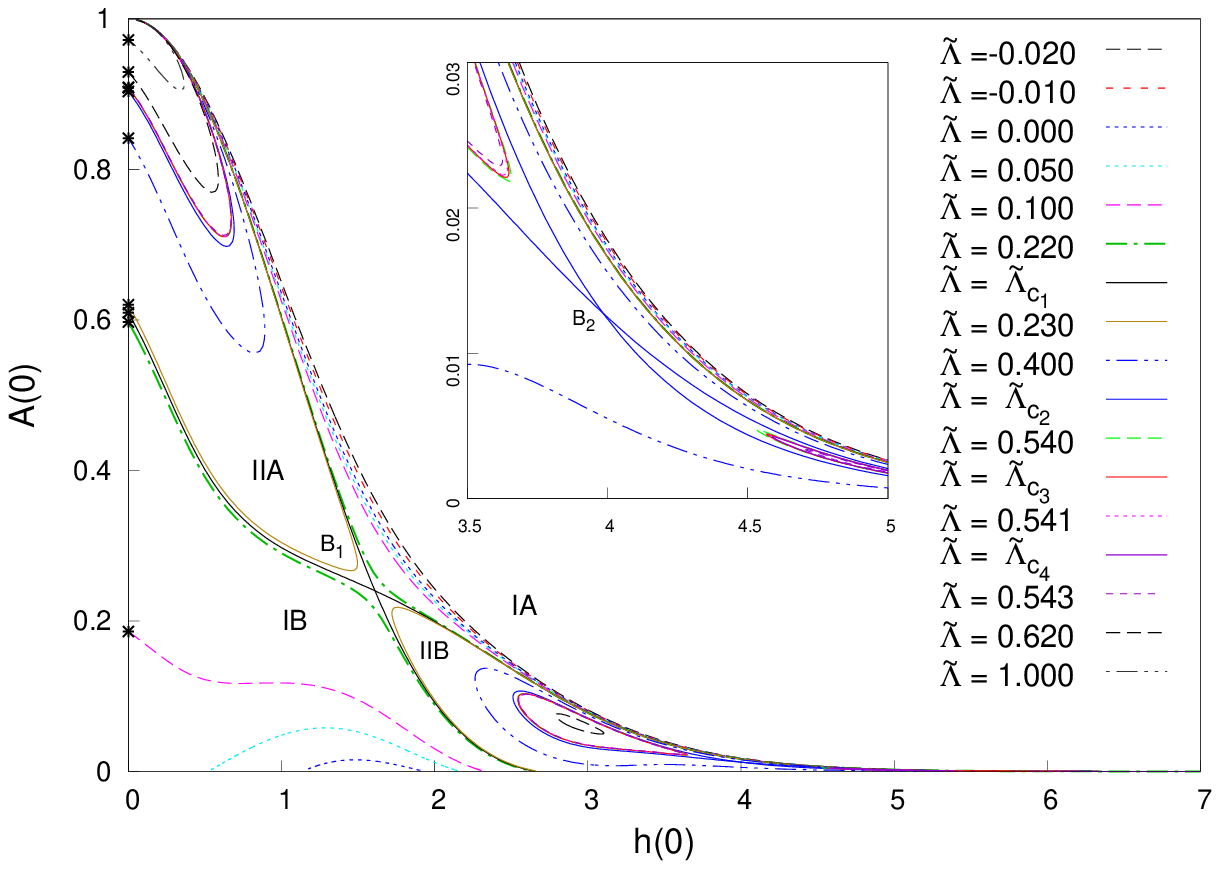}\label{fig:pd}}\hspace{1cm}
\subfigure[][]{\includegraphics[scale=0.65]{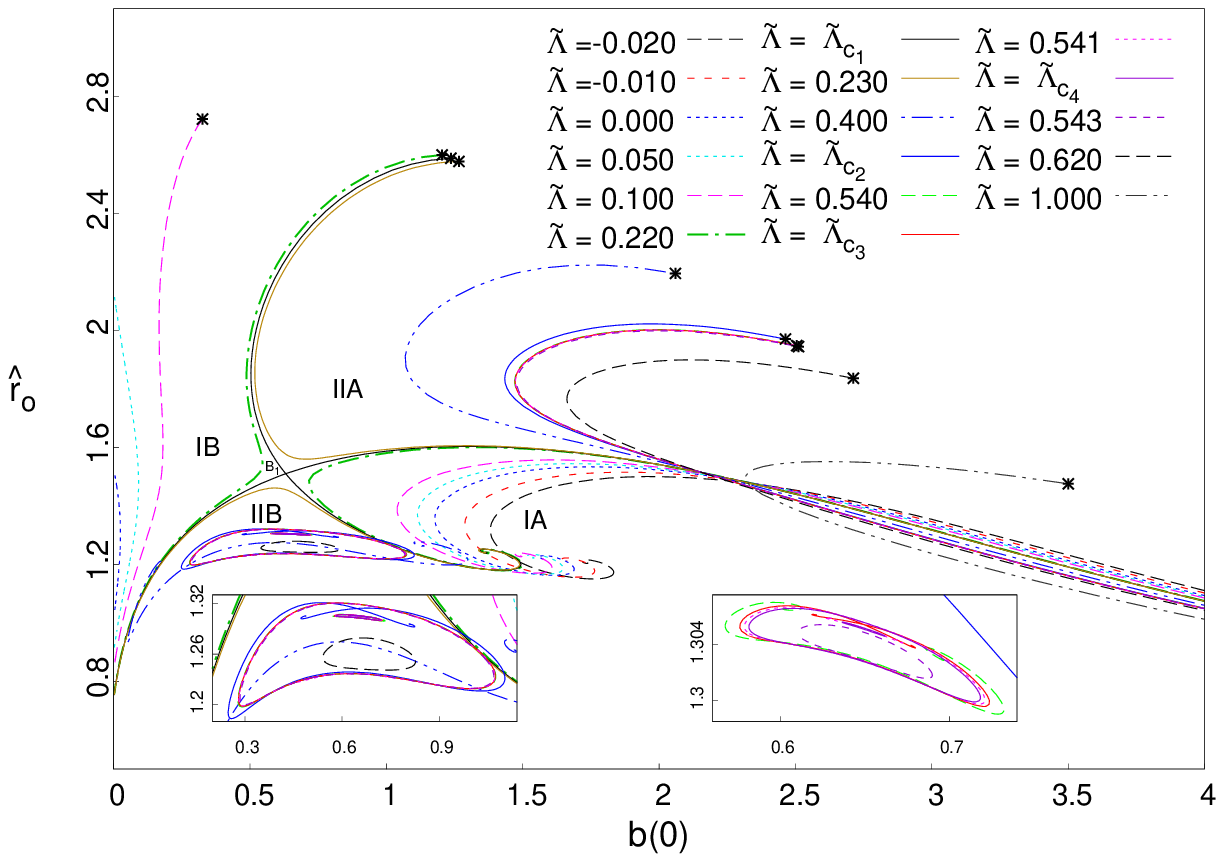}\label{fig:pc}}}
\caption{Figs. (a) to (c) depict the phase diagrams of the theory showing respectively the plots of $h(0)$ versus $b(0)$, $A(0)$ versus $b(0)$ and $A(0)$ versus $h(0)$ (where $h(0)$, $b(0)$ and $A(0)$  denote the values of these fields at the centre of the star) and Fig. (d) depicts a plot of $\hat{r}_o$ versus $b(0)$. Also, all these four figures show the plots for different values of $\tilde \Lambda$ ranging from $\tilde \Lambda =-0.020$ to $\tilde \Lambda =+1.000$ (covering the AdS as well as dS spaces). The figures shown in the insets in Figs. (a)--(d) represent particular sections of these figures with better precision.\label{fig:1f}}
\end{center}
\end{figure*}

\begin{figure*}
\begin{center}
\mbox{\subfigure[][]{\includegraphics[scale=0.65]{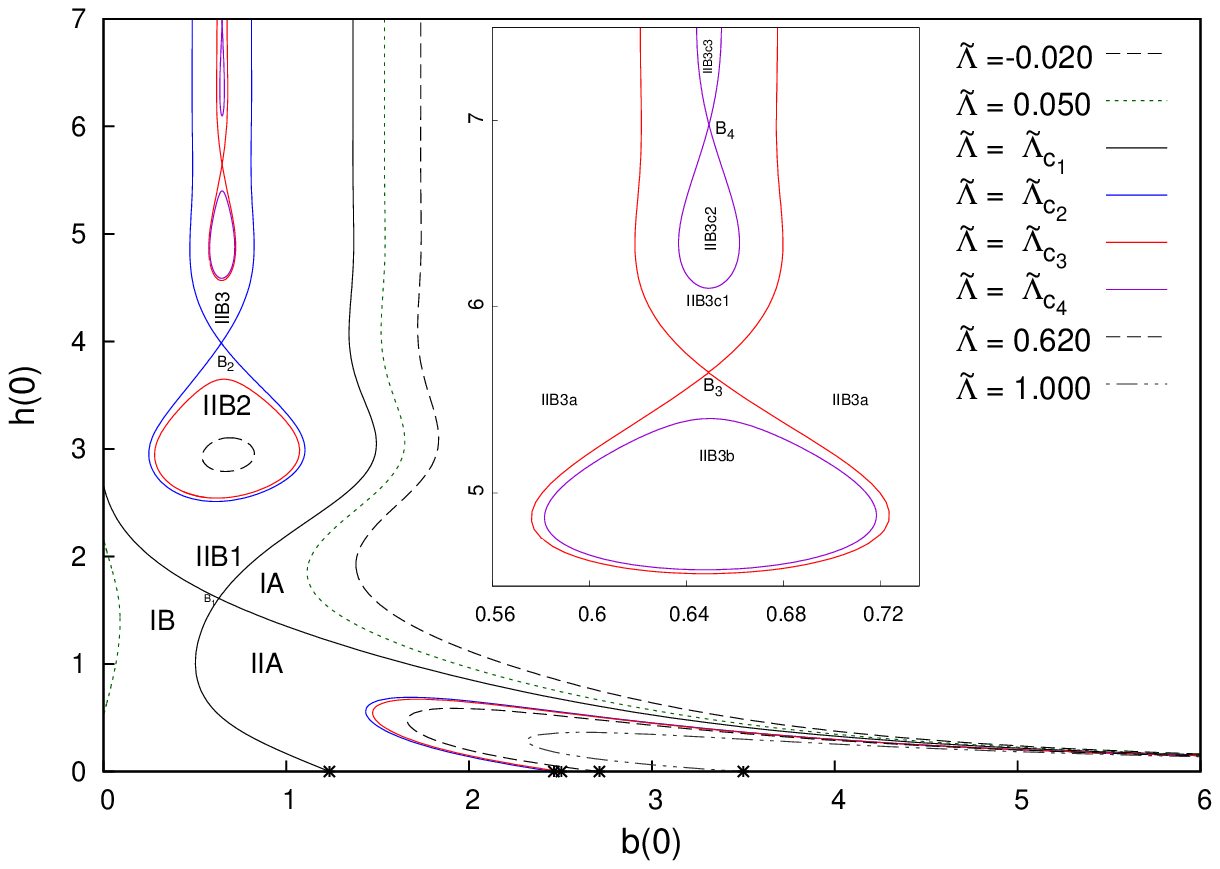}\label{fig:spa}}\hspace{1cm}
\subfigure[][]{\includegraphics[scale=0.65]{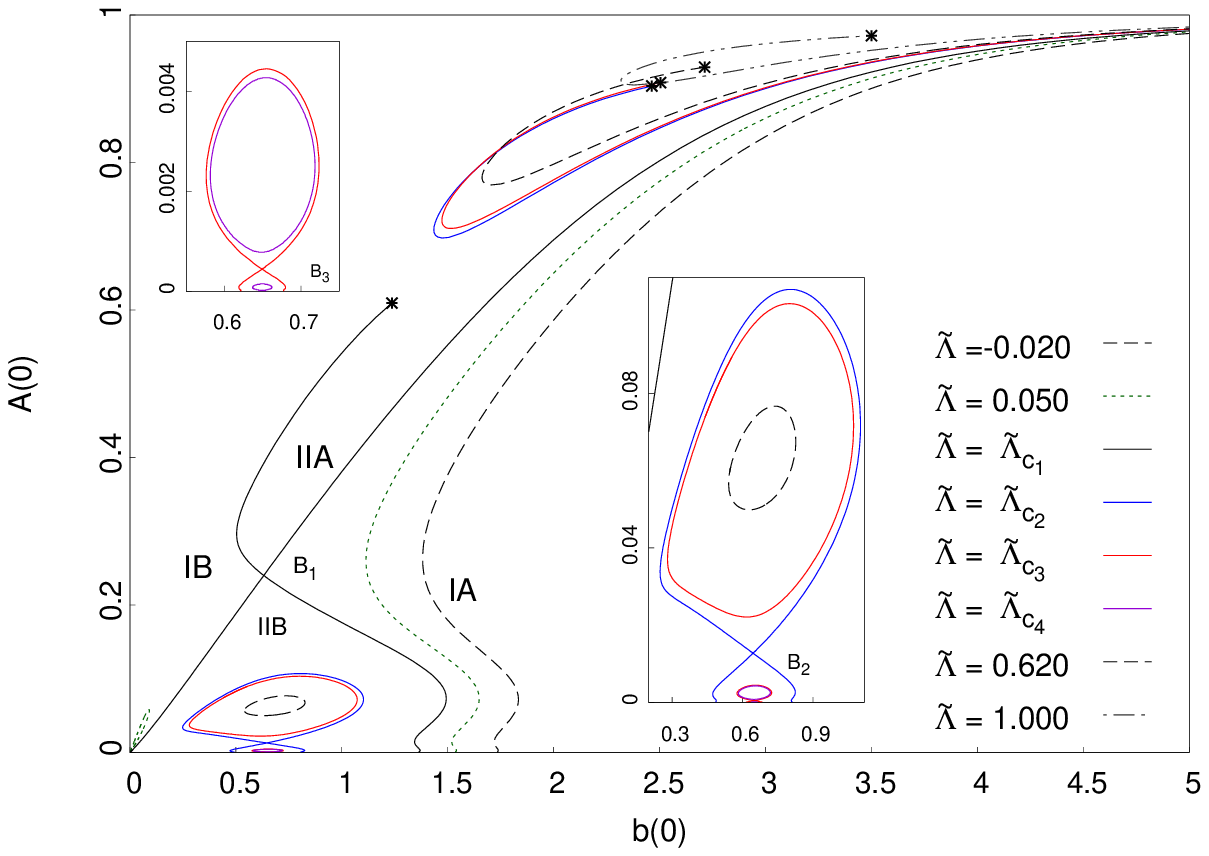}\label{fig:spb}}}
\mbox{\subfigure[][]{\includegraphics[scale=0.65]{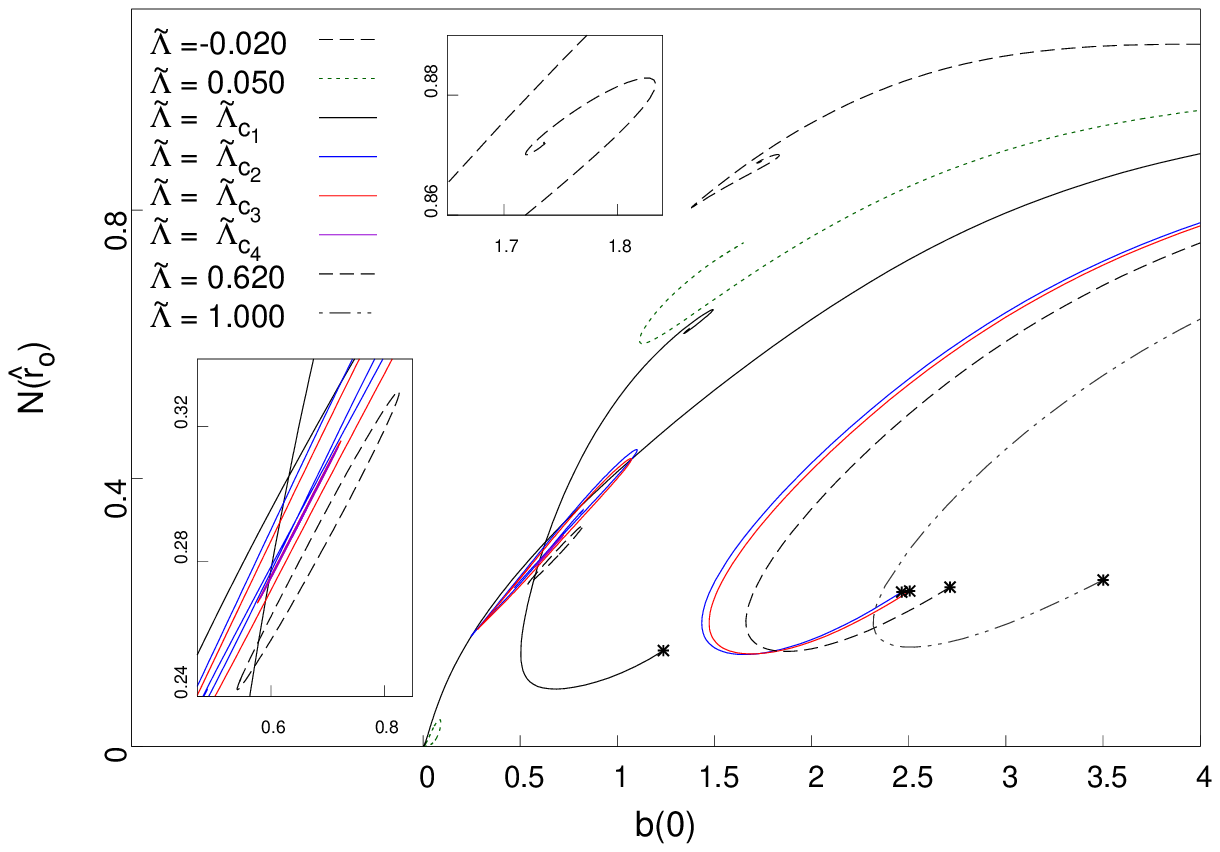}\label{fig:s1f1}}\hspace{1cm}
\subfigure[][]{\includegraphics[scale=0.65]{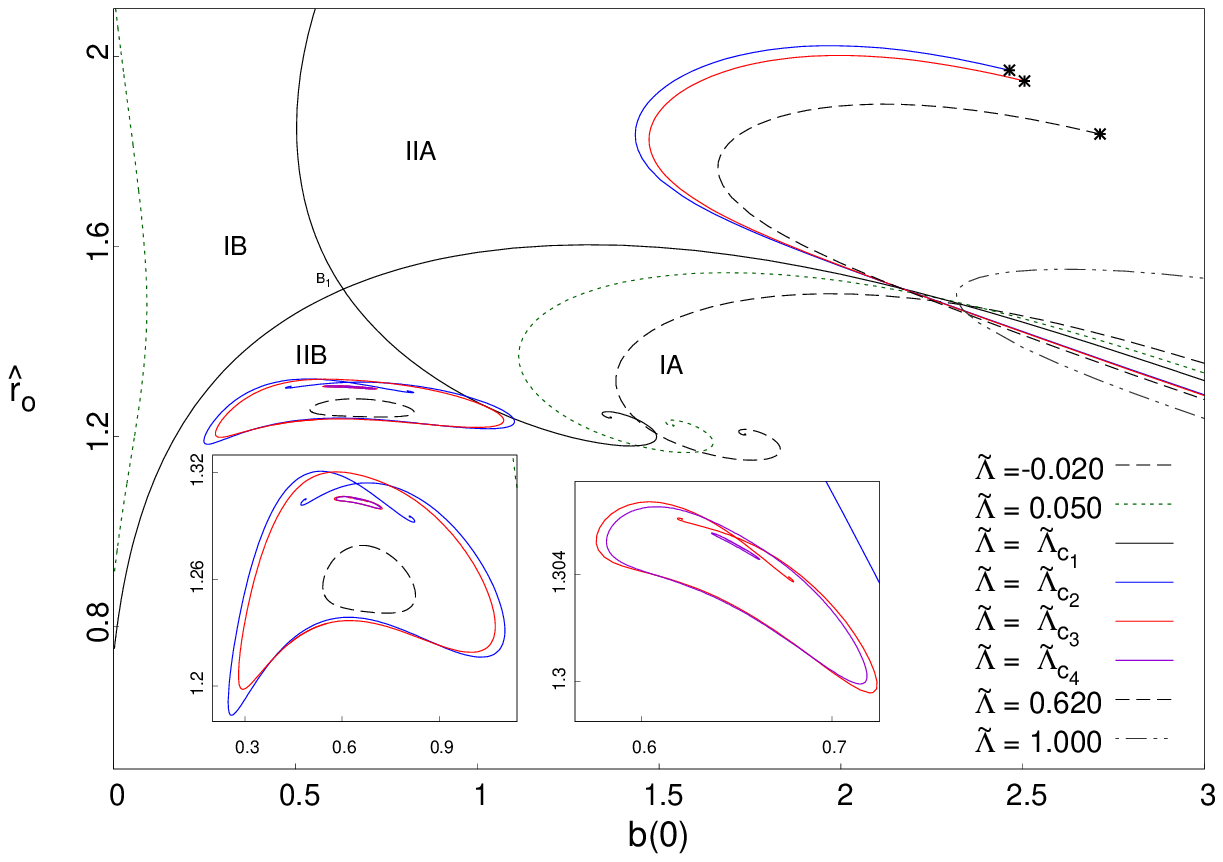}\label{fig:spc}}}
\caption{Figs. (a) and (b) depict the phase diagrams of the theory showing respectively the plots of $h(0)$ versus $b(0)$ and $A(0)$ versus $b(0)$ (where $h(0)$, $b(0)$ and $A(0)$  denote the values of these fields at the centre of the star). Fig. (c) depicts the phase diagram of the theory involving $N(\hat{r}_o)$ and $b(0)$ (where $N(\hat{r}_o)$ denotes the value of the field at the outer radius of the star) and Fig. (d) depicts a plot of $\hat{r}_o$ versus $b(0)$. Also, all these four figures show the plots for different values of $\tilde \Lambda$ ranging from $\tilde \Lambda =-0.020$ to $\tilde \Lambda =+1.000$ (covering the AdS as well as dS spaces). The figures shown in the insets in Figs. (a)--(d) represent particular sections of these figures with better precision.\label{fig:1f1}}
\end{center}
\end{figure*}

\begin{figure*}
\begin{center}
\hspace{-1.0cm}
	 \mbox{\subfigure[][]{\includegraphics[scale=0.64]{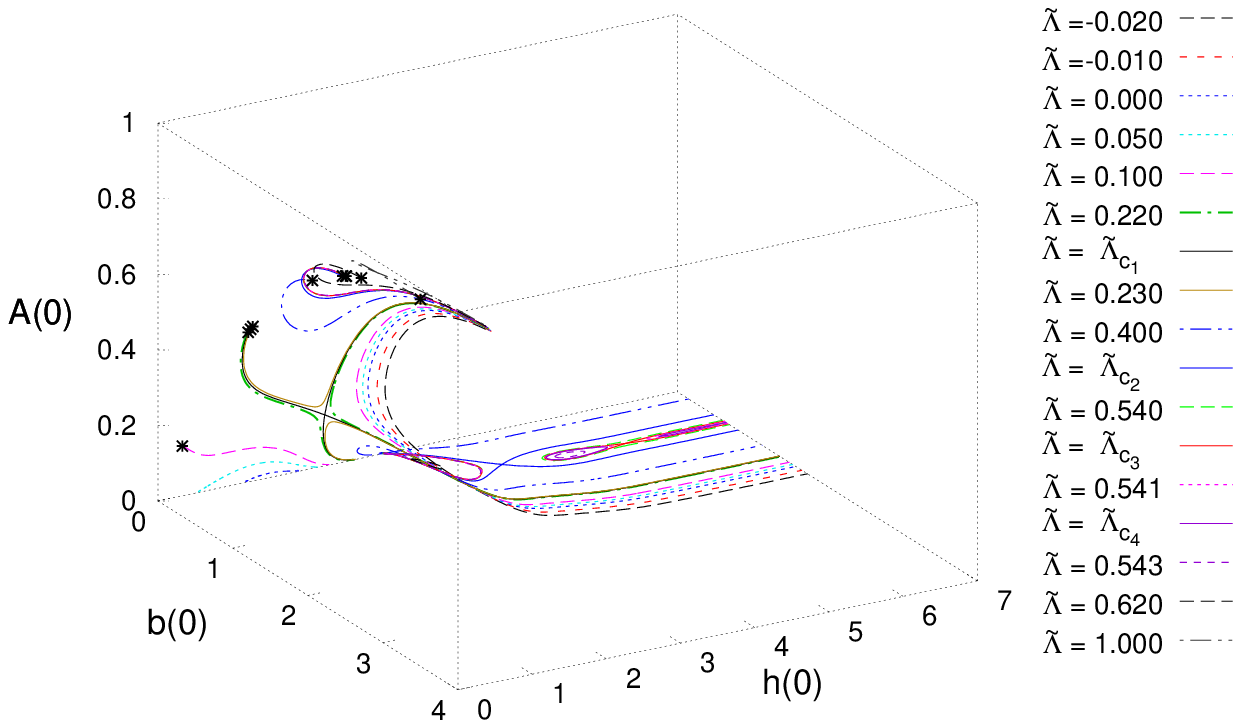}\label{fig:fa}}\hspace{1cm}
	 \subfigure[][]{\includegraphics[scale=0.64]{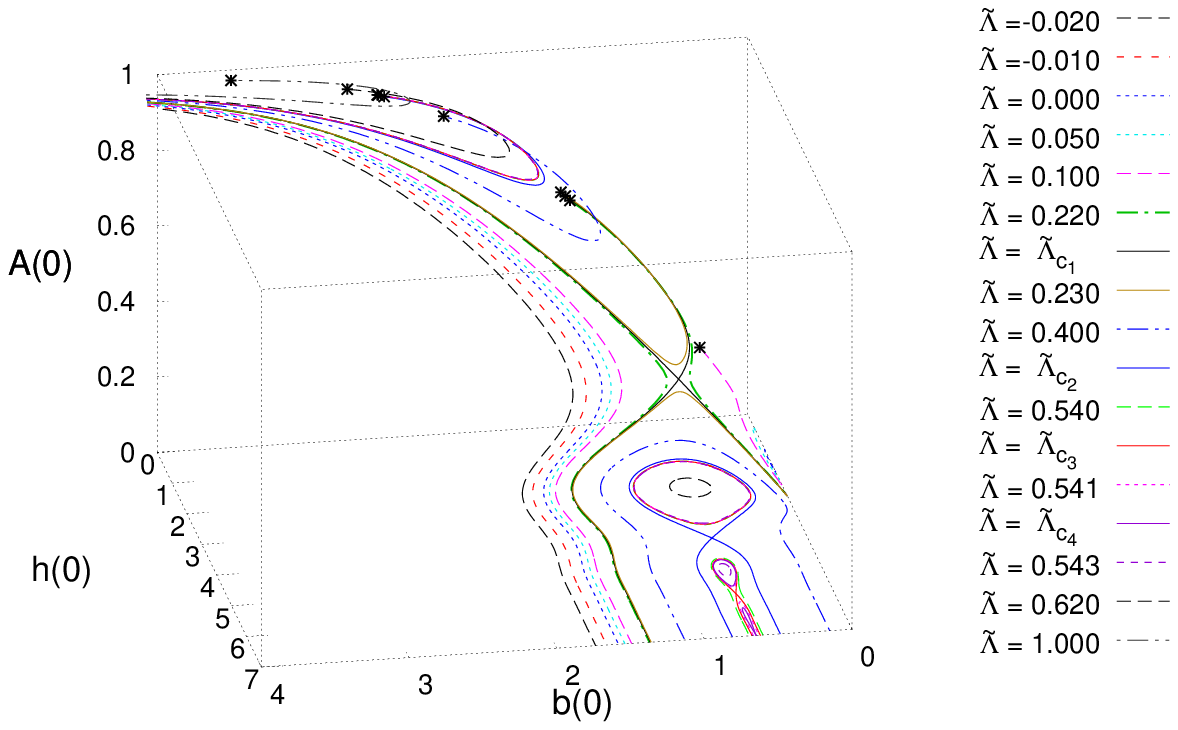}\label{fig:fb}}}
        \mbox{\subfigure[][]{\includegraphics[scale=0.64]{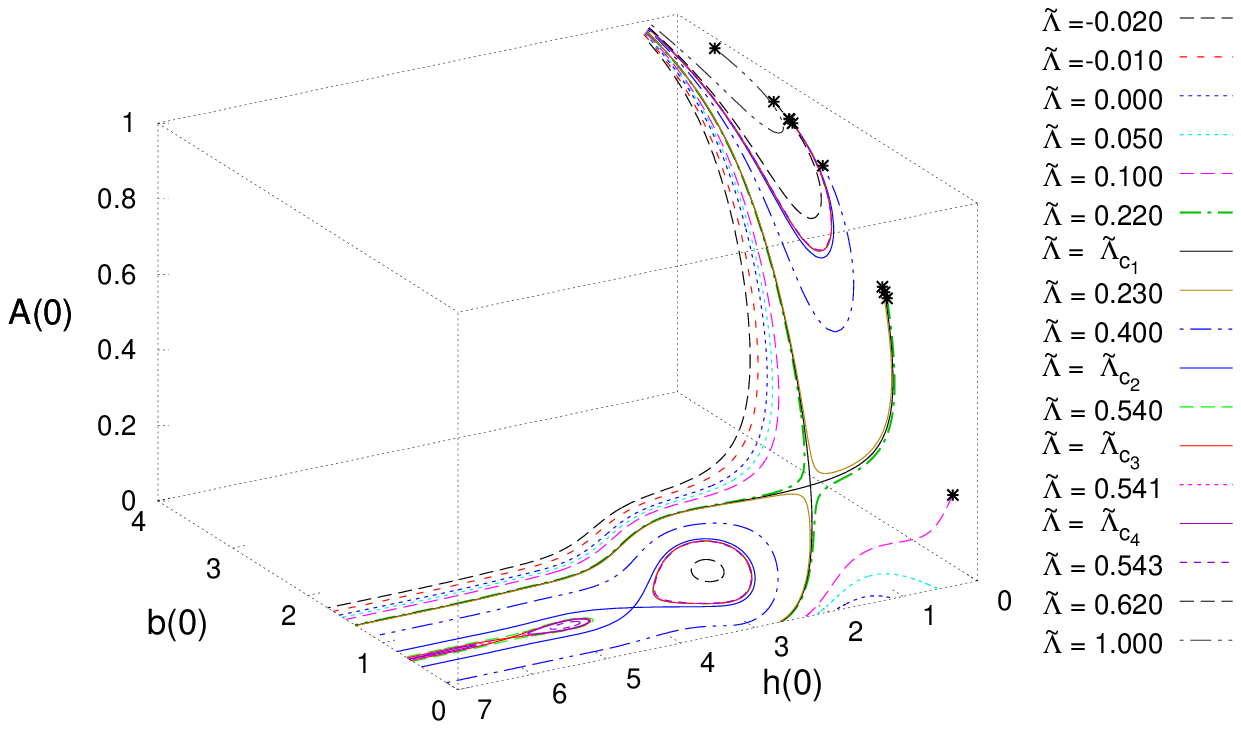}\label{fig:fc}}\hspace{1cm}
        \subfigure[][]{\includegraphics[scale=0.64]{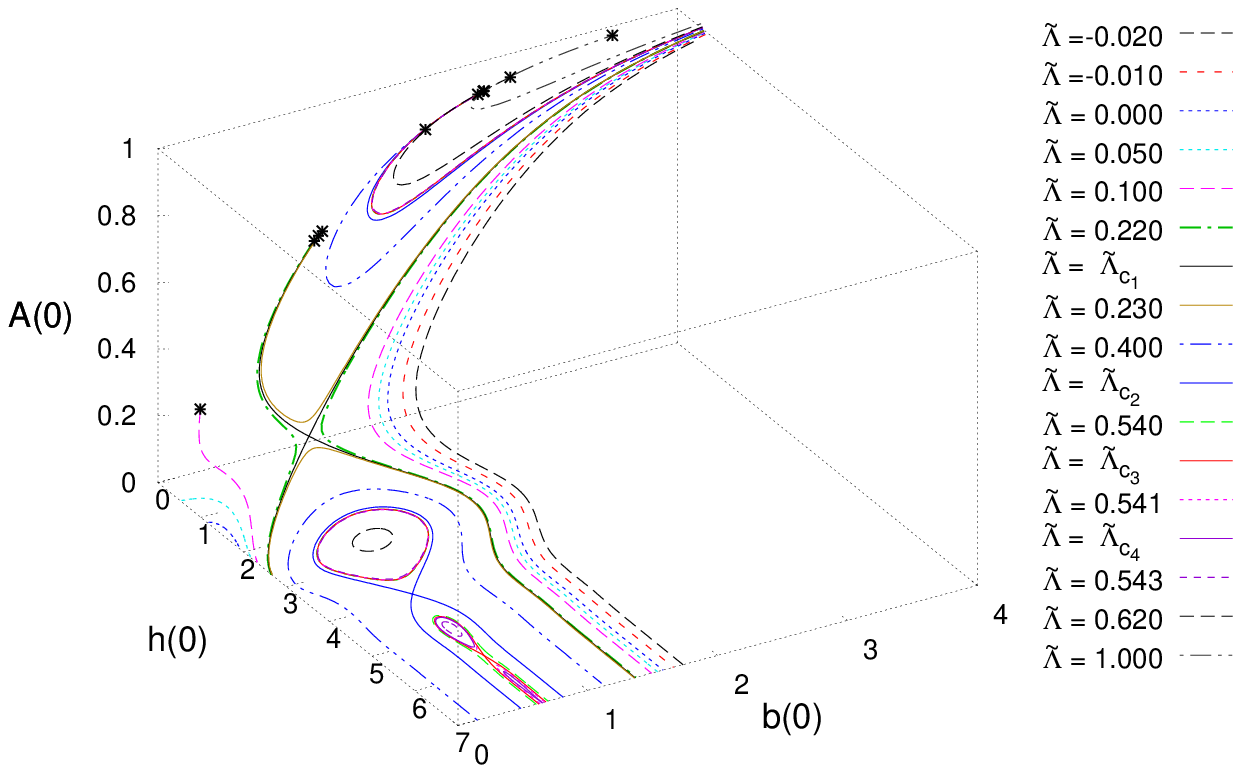}\label{fig:fd}}}
        \mbox{\subfigure[][]{\includegraphics[scale=0.64]{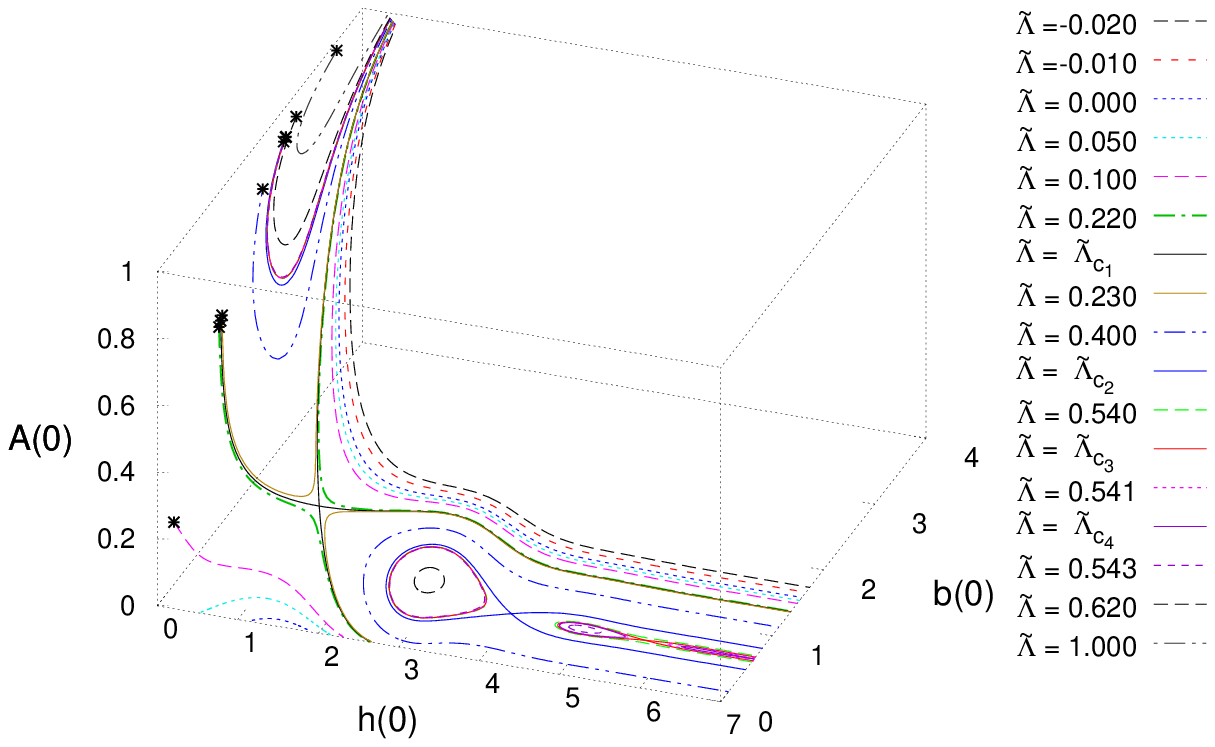}\label{fig:fe}}\hspace{1cm}
        \subfigure[][]{\includegraphics[scale=0.64]{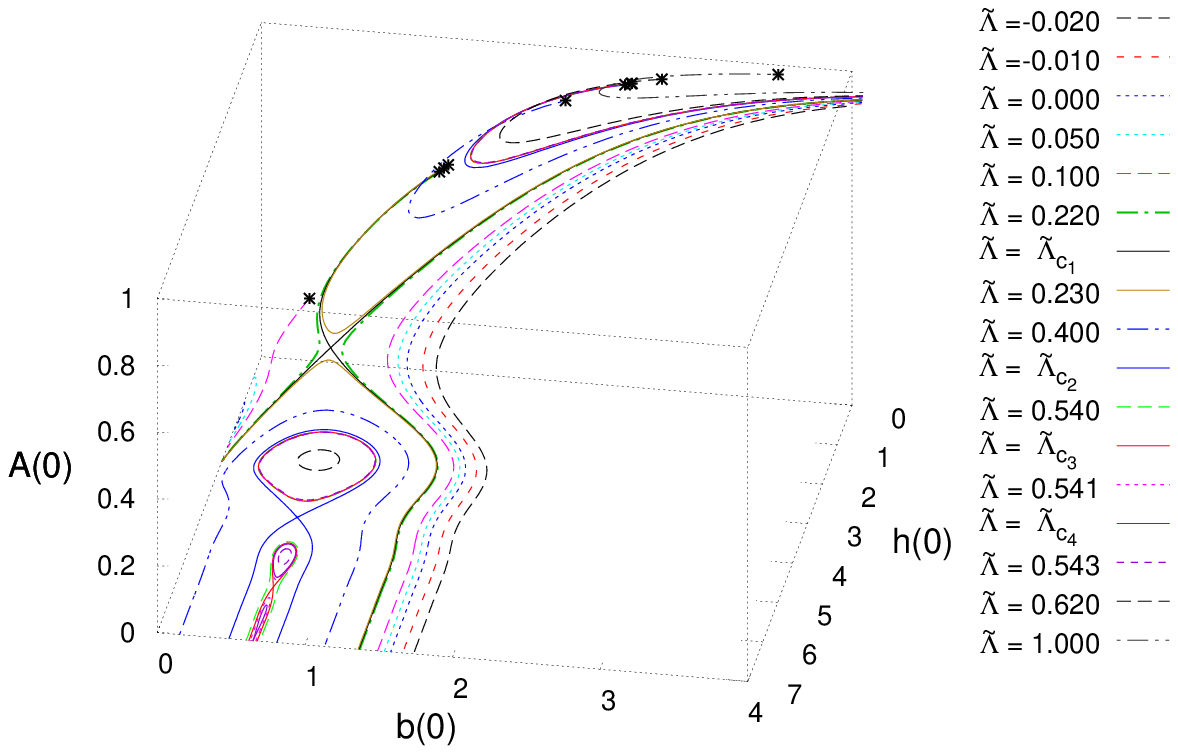}\label{fig:ff}}}
    \end{center}
      \caption{Properties of gravitating boson stars:  Figs. (a), (b), (c), (d), (e) and (f) show the 3D plot (with six different viewing angles) of the scalar field $h$ and the U(1) guage field $b$ versus the  metric field $A$ at the centre of the boson star for several values of the cosmological constant $\tilde\Lambda$ ).  For the viewing angles denoted by $(\theta, \phi)$ we use the convention such that $\theta$ denotes the angle of rotation about the $ox$-axis in the anticlockwise direction and it can take values between 0 to $\pi$ and $\phi$ is the angle of rotation along the $oz^\prime$-axis also in the anticlockwise direction and it can take values between 0 to $2\pi$. Figs. (a)-(f)  correspond respectively to the values: $(\theta,\phi)\equiv(60,60),\ (60,170),\ (60,240),\ (130,30),\ (130,70) \mbox{ and } (130,350)$.\label{fig:1df}}
\end{figure*}

\begin{figure*}
\begin{center}
\hspace{-1.0cm}
	  \mbox{\subfigure[][]{\includegraphics[scale=0.64]{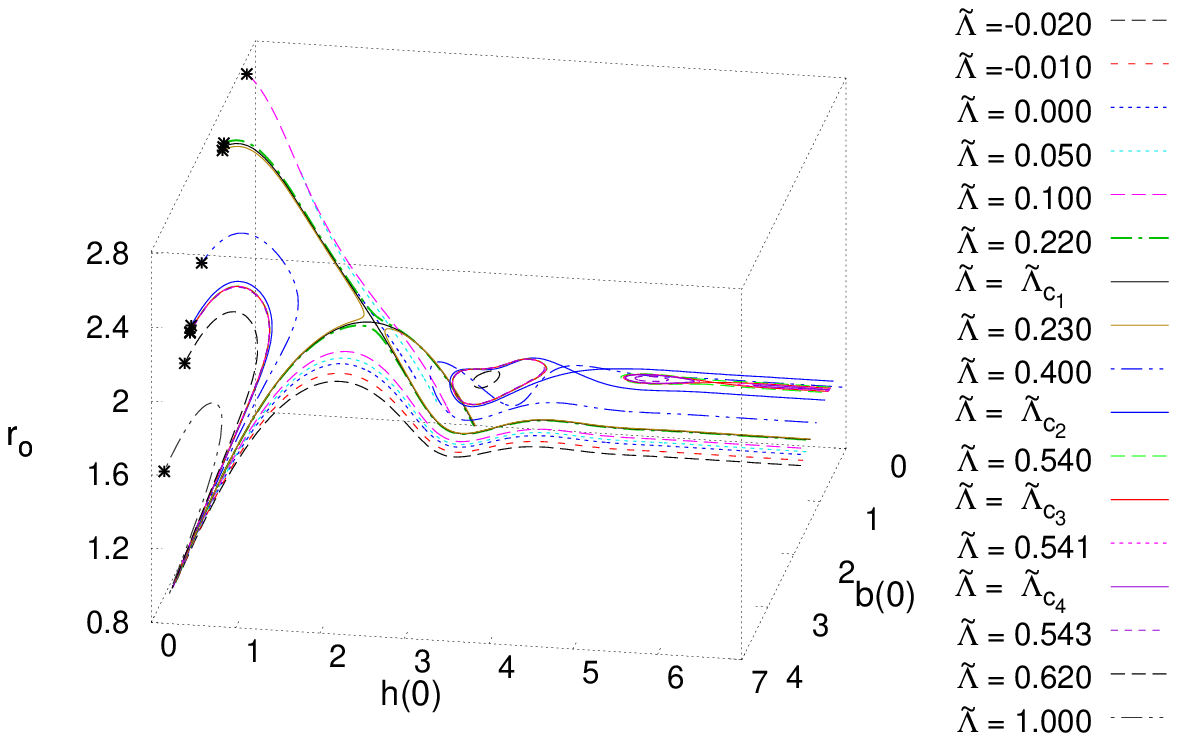}\label{fig:fa2}}\hspace{1cm}
	 \subfigure[][]{\includegraphics[scale=0.64]{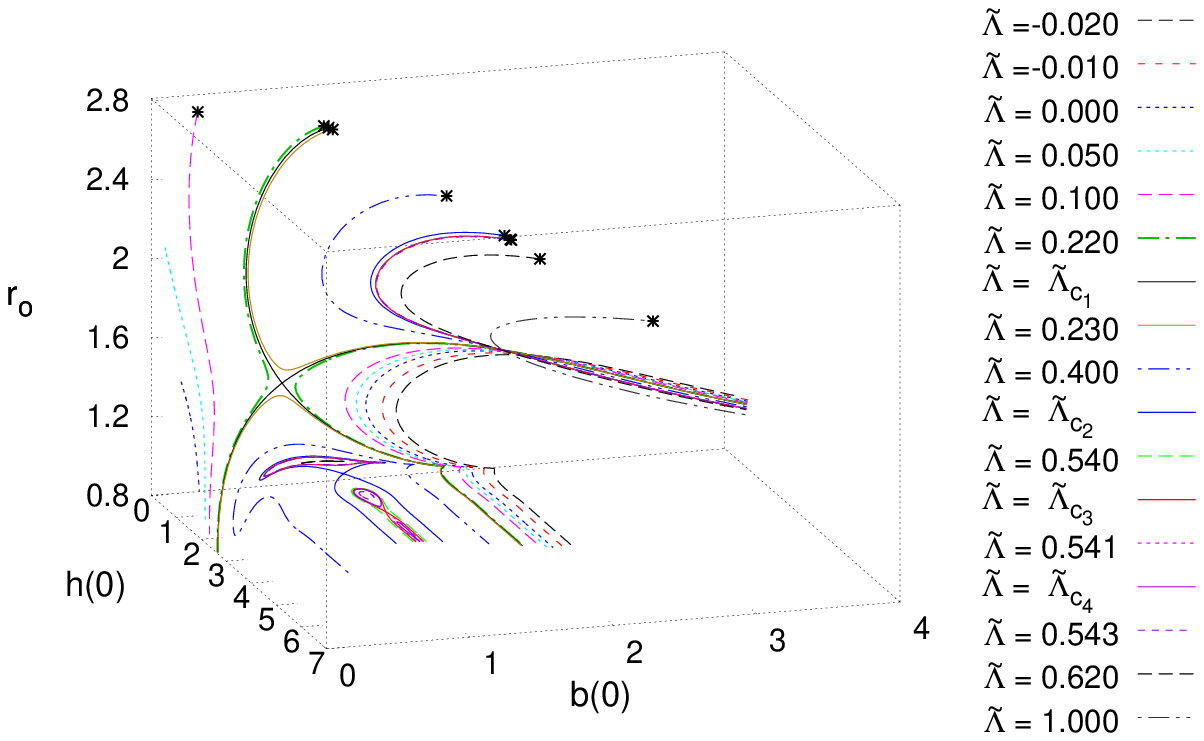}\label{fig:fb2}}}
        \mbox{\subfigure[][]{\includegraphics[scale=0.64]{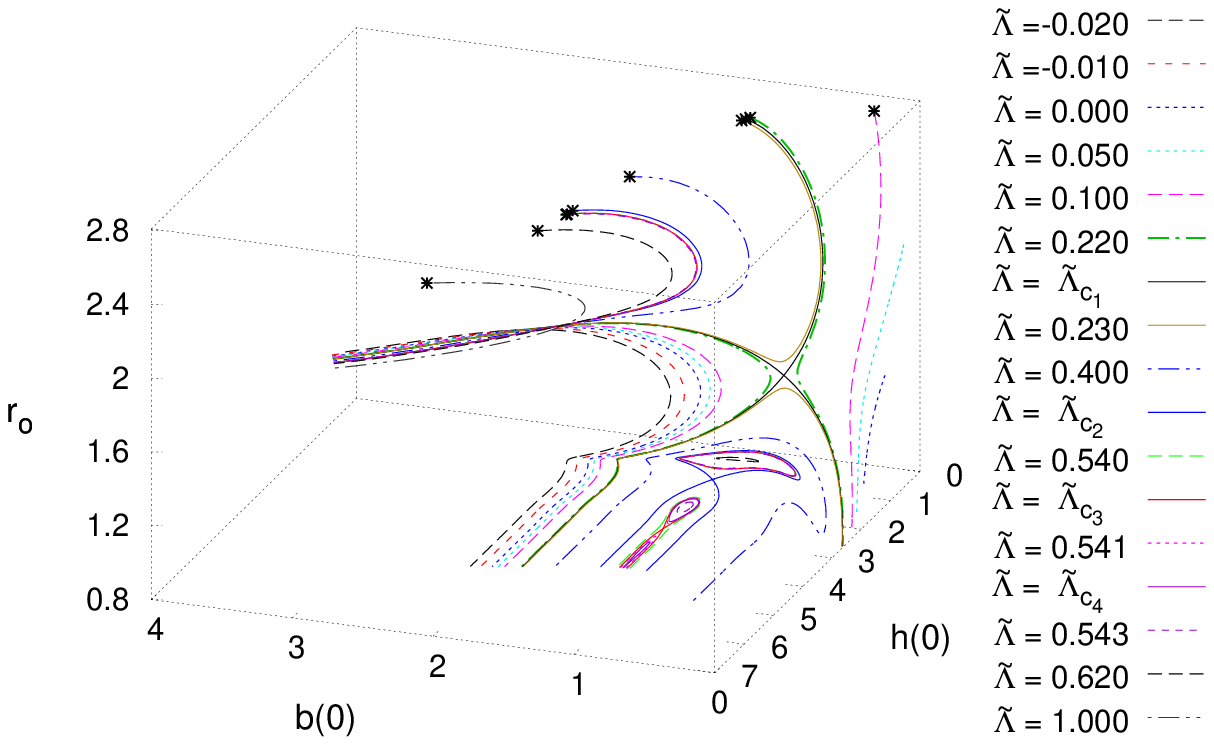}\label{fig:fc2}}\hspace{1cm}
        \subfigure[][]{\includegraphics[scale=0.64]{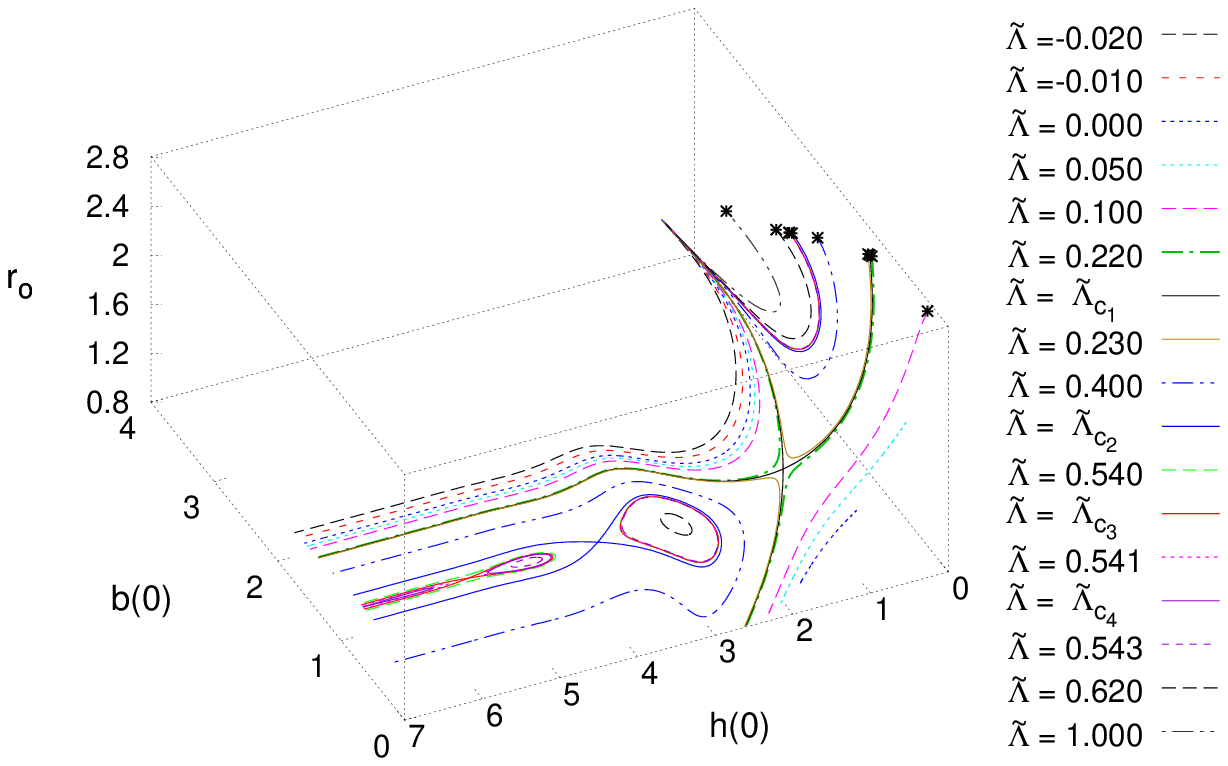}\label{fig:fd2}}}
        \mbox{\subfigure[][]{\includegraphics[scale=0.64]{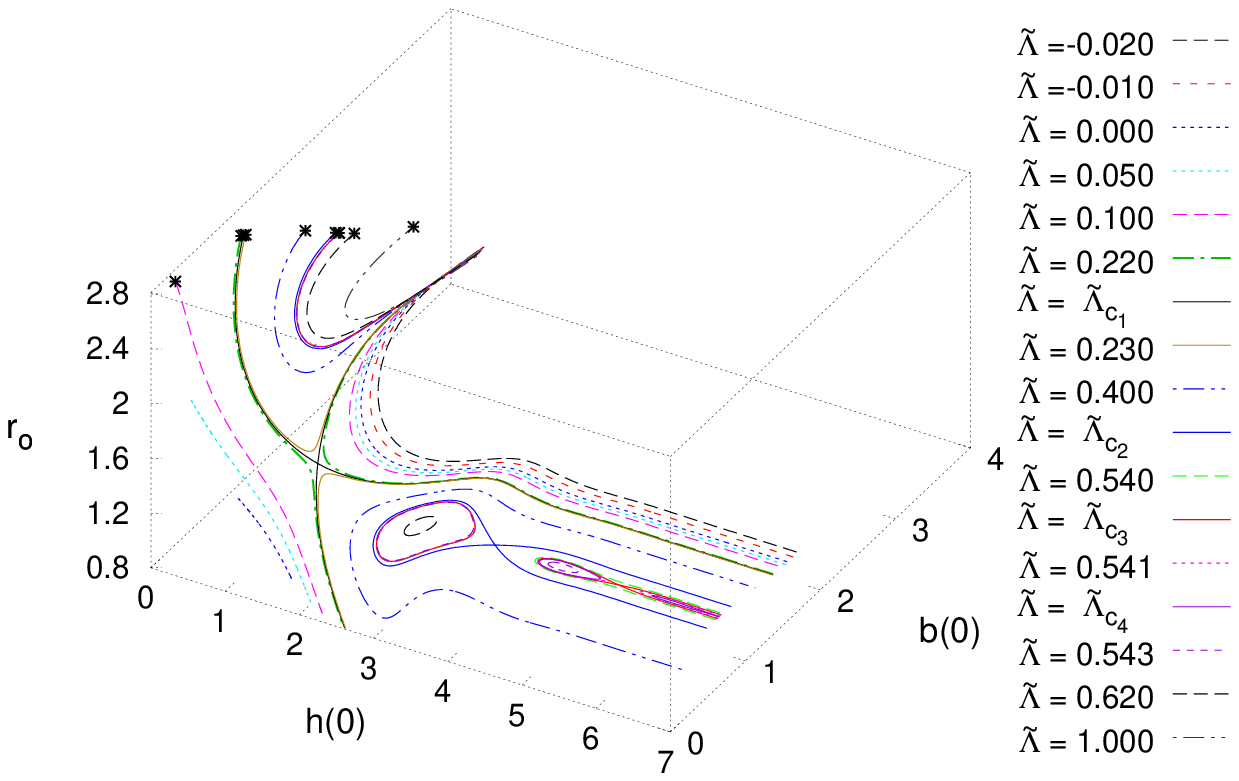}\label{fig:fe2}}\hspace{1cm}
        \subfigure[][]{\includegraphics[scale=0.64]{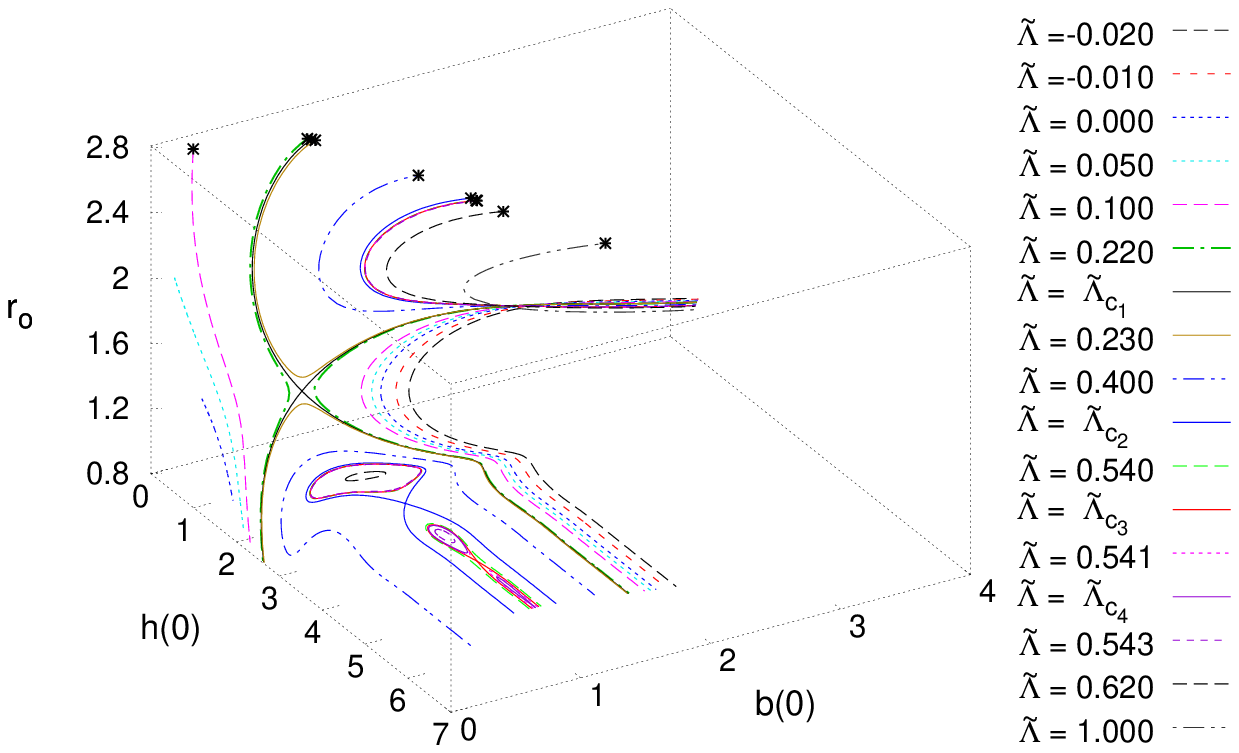}\label{fig:ff2}}}
    \end{center}
      \caption{Properties of gravitating boson stars:  Figs. (a), (b), (c), (d), (e) and (f) show the 3D plot (with six different viewing angles) of the scalar field $h$ and the U(1) guage field $b$ at the centre of the boson star versus the radius of the star  $\hat r_o$ for several values of the cosmological constant $\tilde\Lambda$ ). For the viewing angles denoted by $(\theta, \phi)$ we use the convention such that $\theta$ denotes the angle of rotation about the $ox$-axis in the anticlockwise direction and it can take values between 0 to $\pi$ and $\phi$ is the angle of rotation along the $oz^\prime$-axis also in the anticlockwise direction and it can take values between 0 to $2\pi$. Figs. (a)-(f)  correspond respectively to the values: $(\theta,\phi)\equiv(60,100),\ (112,17),\ (60,200),\ (35,245),\ (140,60)\mbox{ and } (130,30)$.\label{fig:2f}}
\end{figure*}

The conserved Noether current is given by
\begin{eqnarray}
j^\mu=-i \,e\,\left\{ \Phi(D^\mu \Phi)^*-\Phi^* (D^\mu \Phi) \right\}\,, \ \ \ \  
D_\mu\,j^{\mu} = 0\,.\ \ \ \ 
\end{eqnarray}
The charge $Q_{sh}$ of a boson shell is given by
\begin{eqnarray}
Q_{sh}=-\frac{1}{4\pi}\int_{\hat{r}_i} ^{\hat{r}_0} j^t \sqrt{-g} \,d\hat{r}\,d\theta\,d\phi  \,,\ \ \ \
j^t=-\frac{h^2\, b}{A^2\, N}\,.\nonumber
\end{eqnarray}
For a boson star the integration interval starts from zero.

For all the gravitating solutions we obtain the mass parameter M (in the units employed):
 \begin{equation}
M= \biggl(1-N(\hat{r}_o)+\frac{\alpha Q^2}{\hat{r}_o^2} -\frac{\Lambda}{3} \hat{r}_o^2\biggr)\frac{\hat{r}_o}{2}.
 \end{equation}

\begin{figure*}
\begin{center}
\mbox{\subfigure[][]{\includegraphics[scale=0.65]{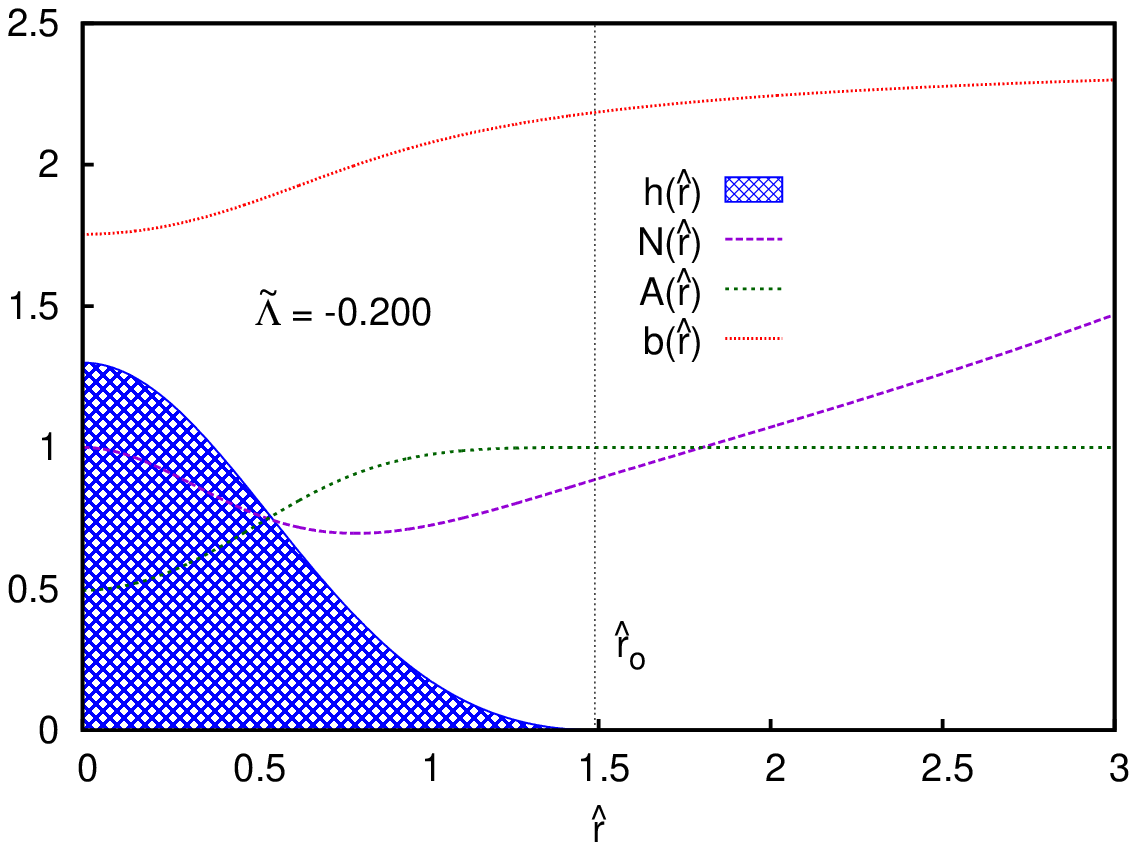}\label{fig:01}}\hspace{1cm}\subfigure[][]{\includegraphics[scale=0.65]{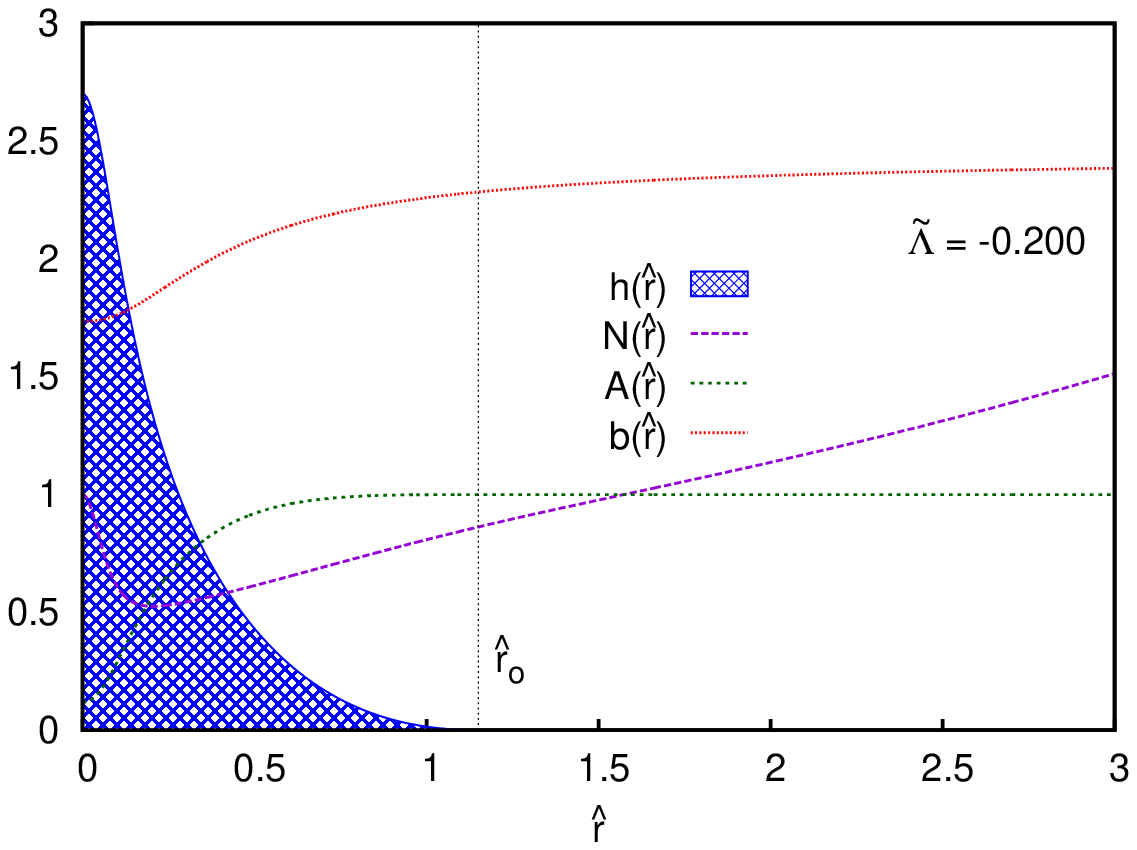}\label{fig:02}}}
\mbox{\subfigure[][]{\includegraphics[scale=0.65]{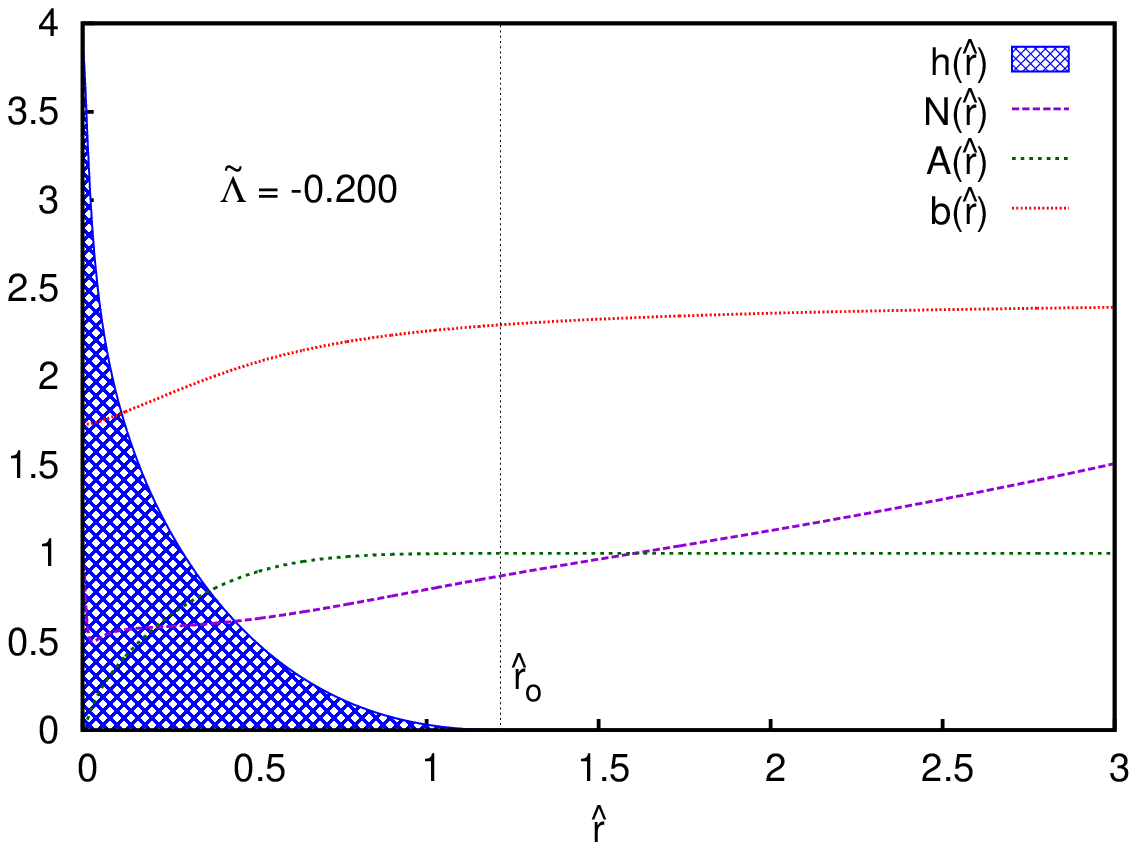}\label{fig:03}}\hspace{1cm}\subfigure[][]{\includegraphics[scale=0.65]{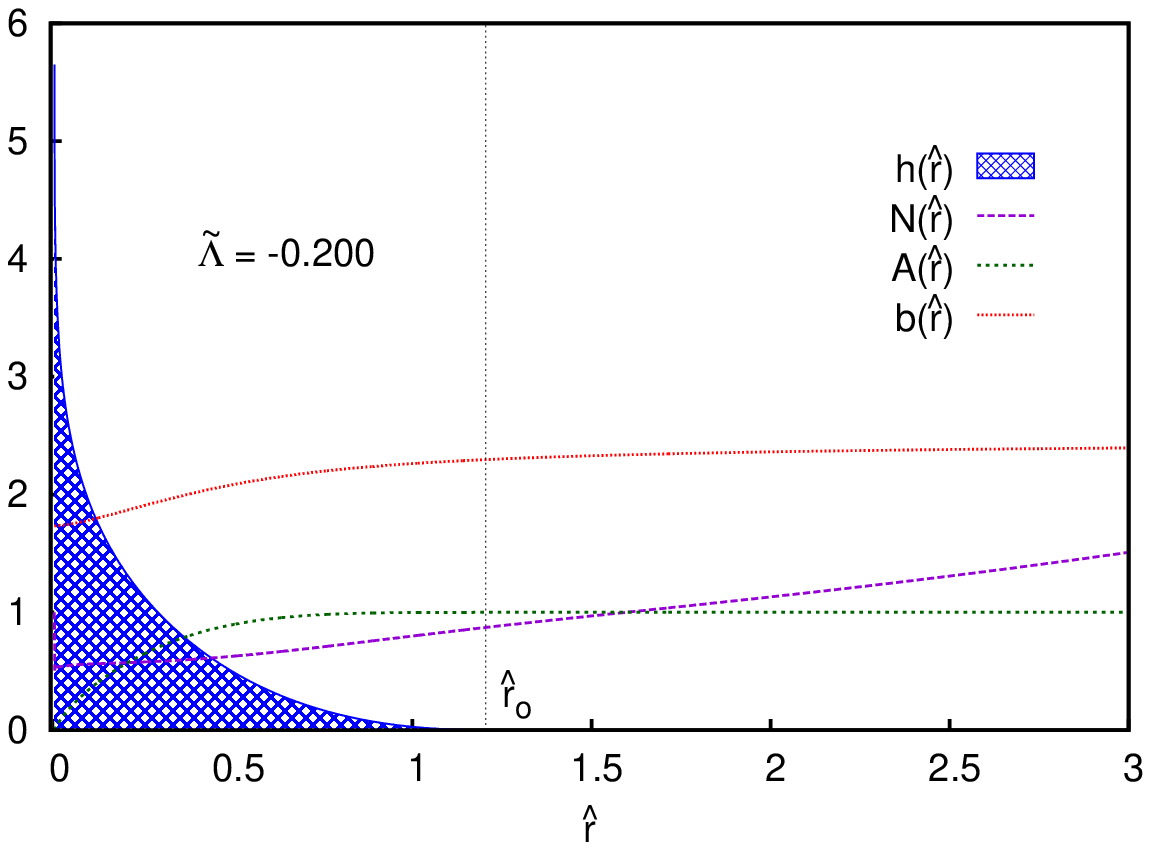}\label{fig:05}}}
\caption{Figs. (a)--(d) depict plots of $h(\hat{r})$, $N(\hat{r})$, $A(\hat{r})$ and $b(\hat{r})$ versus $\hat{r}$ for $\tilde{\Lambda}=-0.200$ (which corresponds to the AdS space).\label{fig:3f}}
\end{center}
\end{figure*}
\begin{figure*}
\begin{center}
\mbox{\subfigure[][]{\includegraphics[scale=0.65]{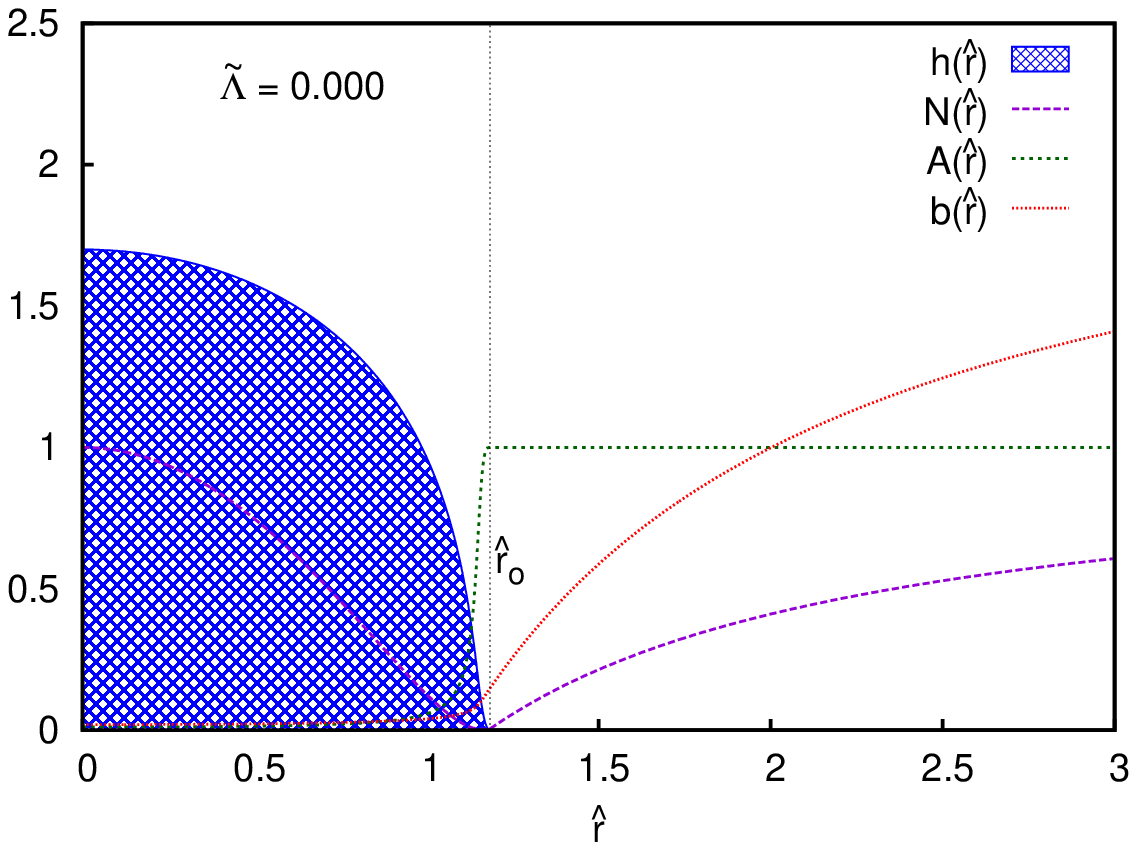}\label{fig:11}}\hspace{1cm}\subfigure[][]{\includegraphics[scale=0.65]{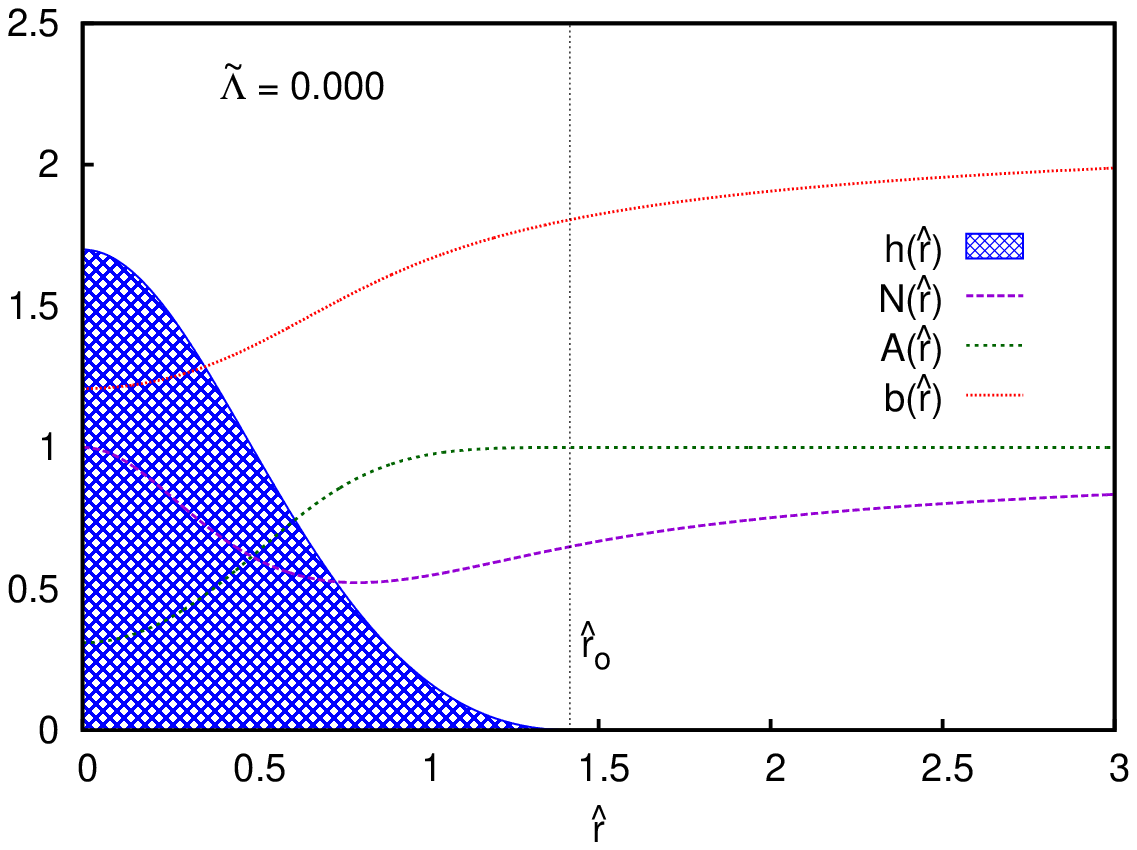}\label{fig:12}}}
\caption{Figs. (a) and (b) depict plots of $h(\hat{r})$, $N(\hat{r})$, $A(\hat{r})$ and $b(\hat{r})$ versus $\hat{r}$ for $\tilde{\Lambda}=0.000$.\label{fig:4f} }
\end{center}
\end{figure*}

\begin{figure*}
\begin{center}
\mbox{\subfigure[][]{\includegraphics[scale=0.65]{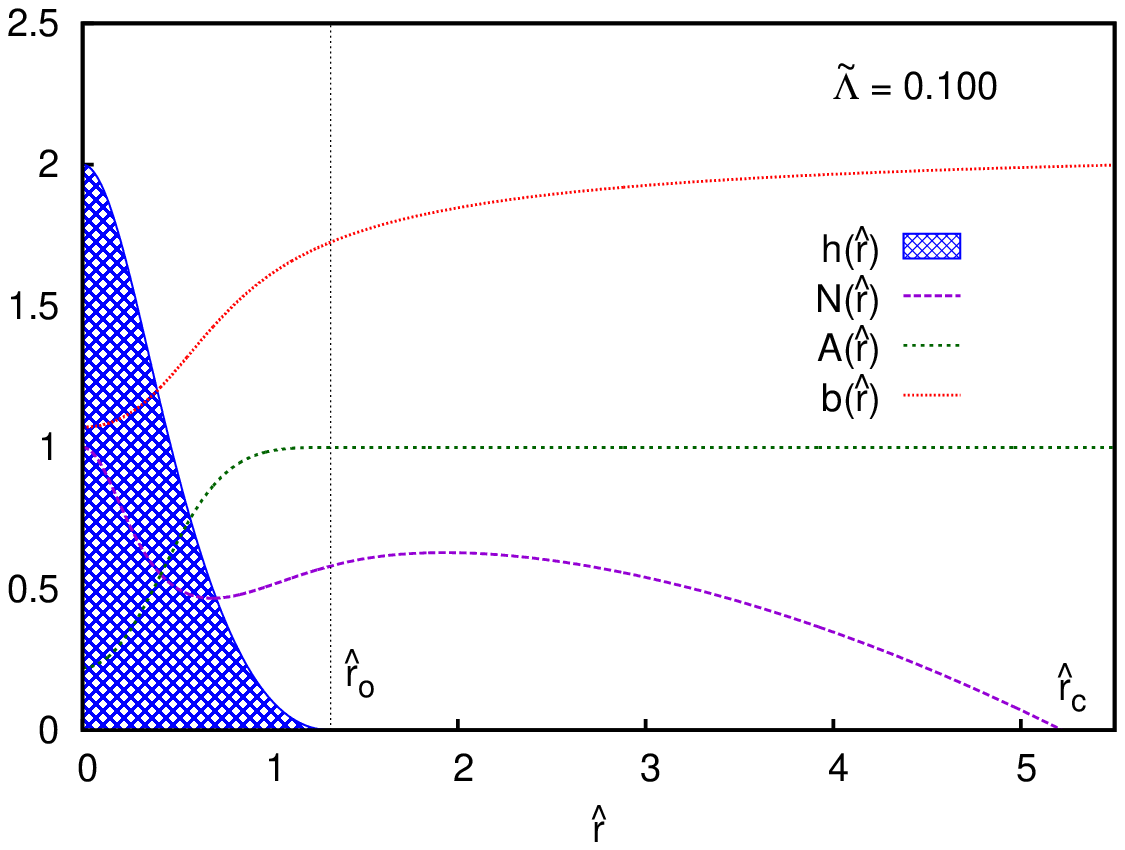}\label{fig:13}}\hspace{1cm}\subfigure[][]{\includegraphics[scale=0.65]{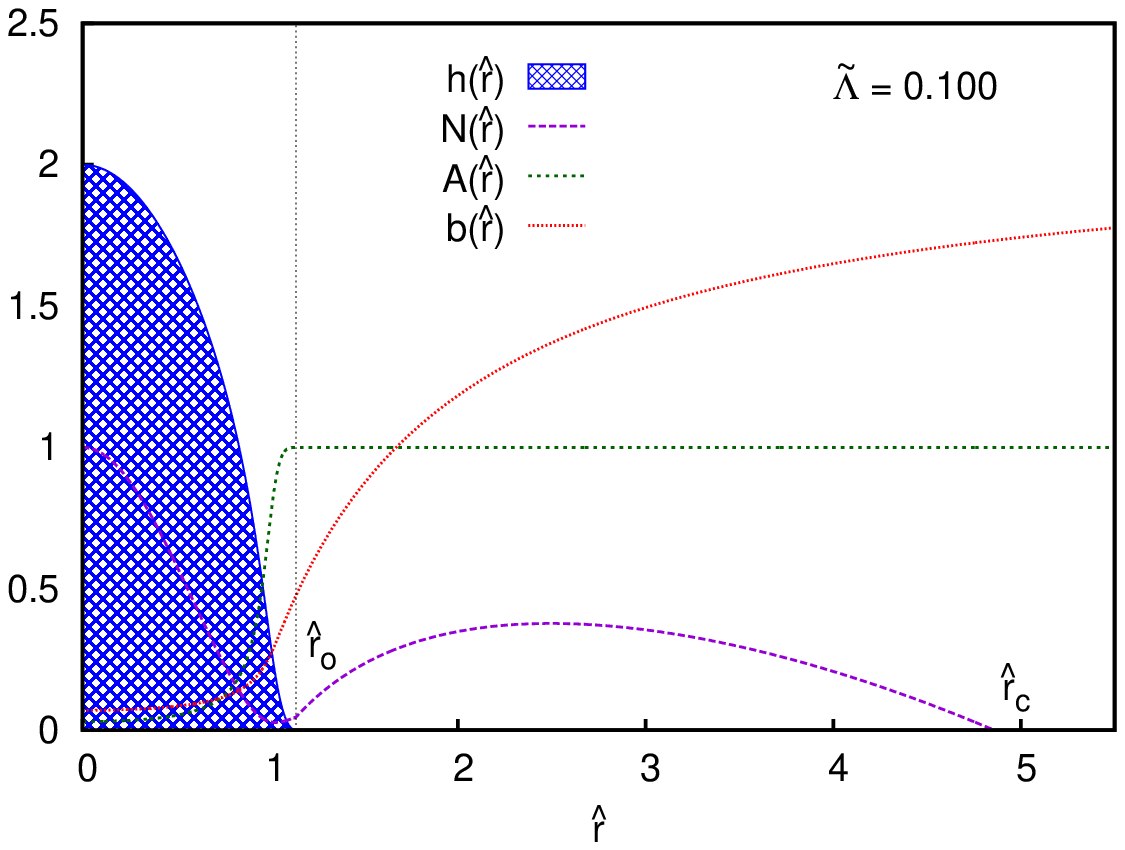}\label{fig:14}}}
\mbox{\subfigure[][]{\includegraphics[scale=0.65]{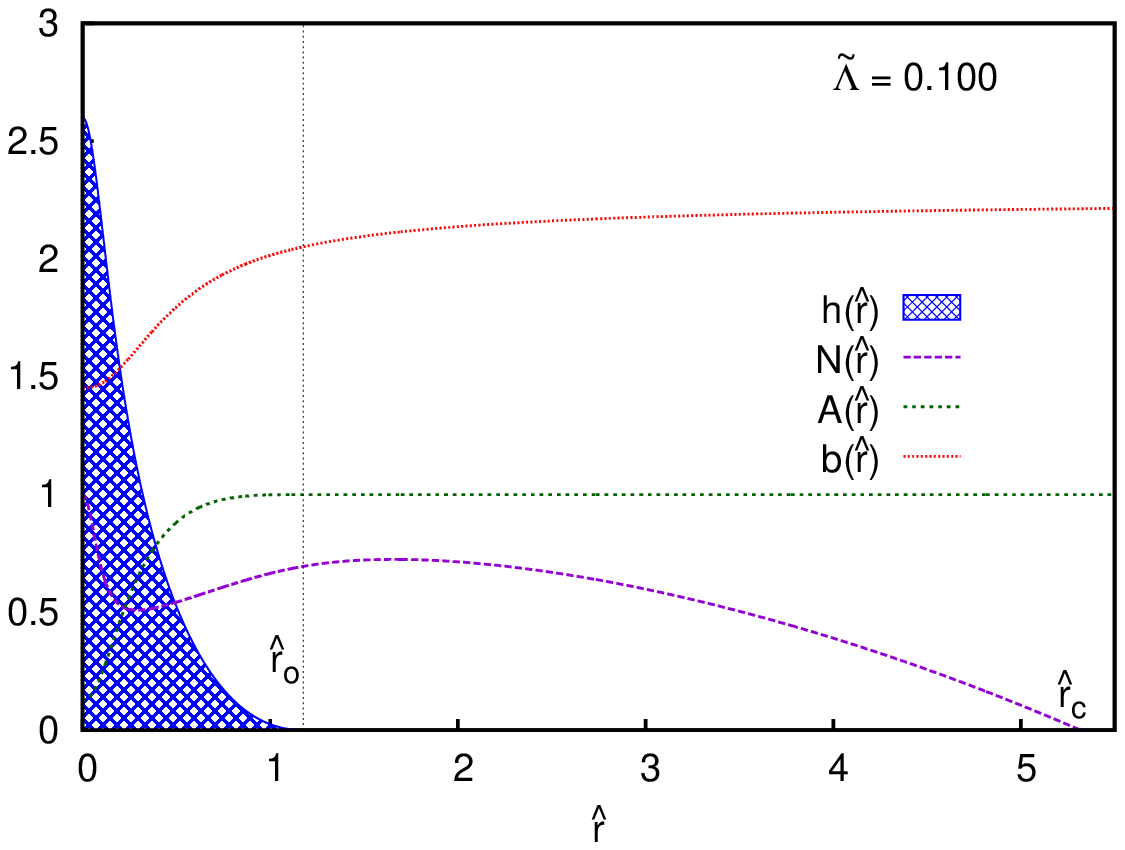}\label{fig:15}}\hspace{1cm}\subfigure[][]{\includegraphics[scale=0.65]{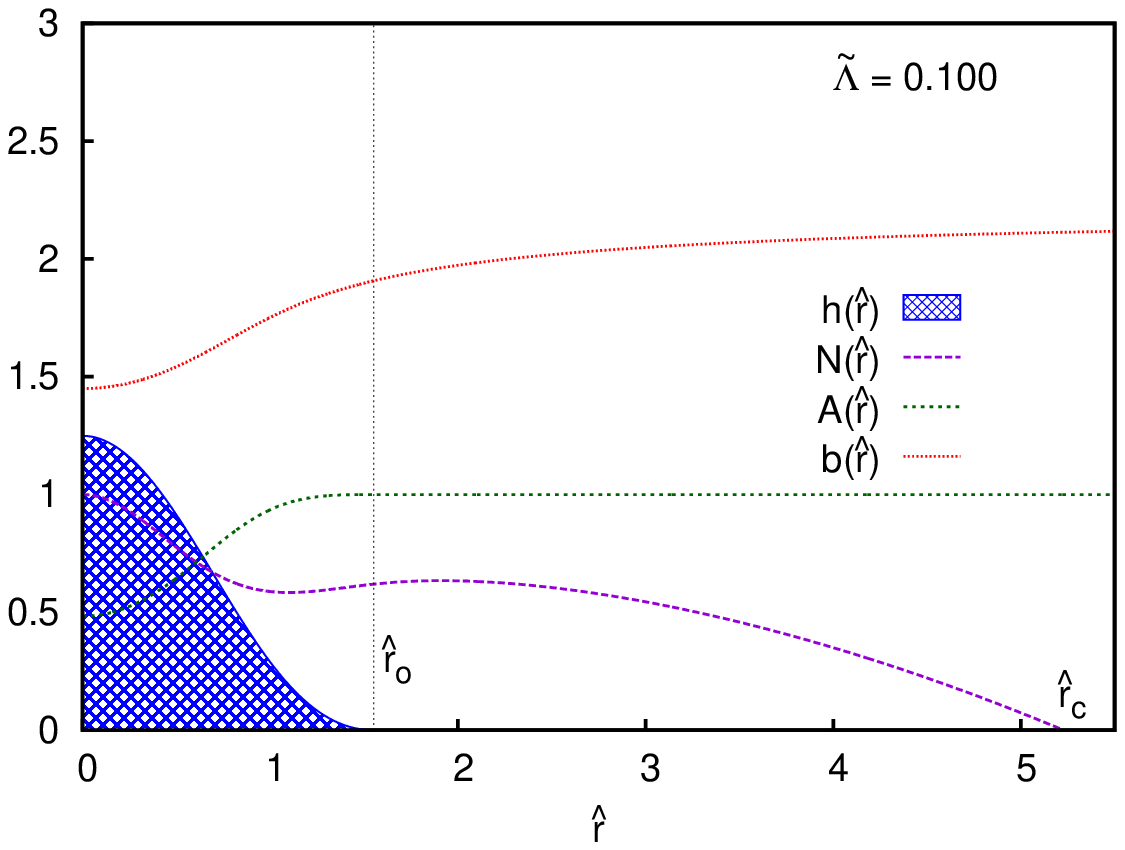}\label{fig:16}}}
\caption{Figs. (a) to (d) depict plots of $h(\hat{r})$, $N(\hat{r})$, $A(\hat{r})$ and $b(\hat{r})$ versus $\hat{r}$ for $\tilde{\Lambda}=0.100$ (which corresponds to the dS space).\label{fig:5f} }
\end{center}
\end{figure*}

\begin{figure*}
\begin{center}
\mbox{\subfigure[][]{\includegraphics[scale=0.65]{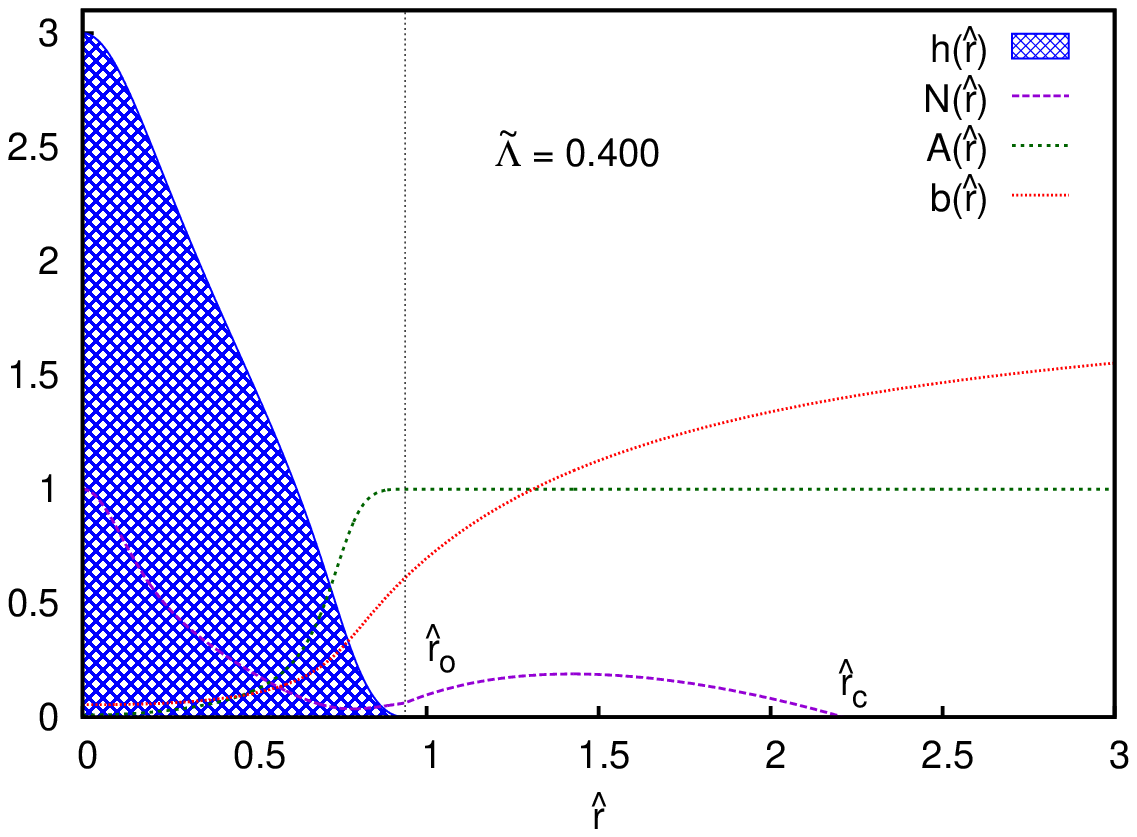}\label{fig:20}}\hspace{1cm}\subfigure[][]{\includegraphics[scale=0.65]{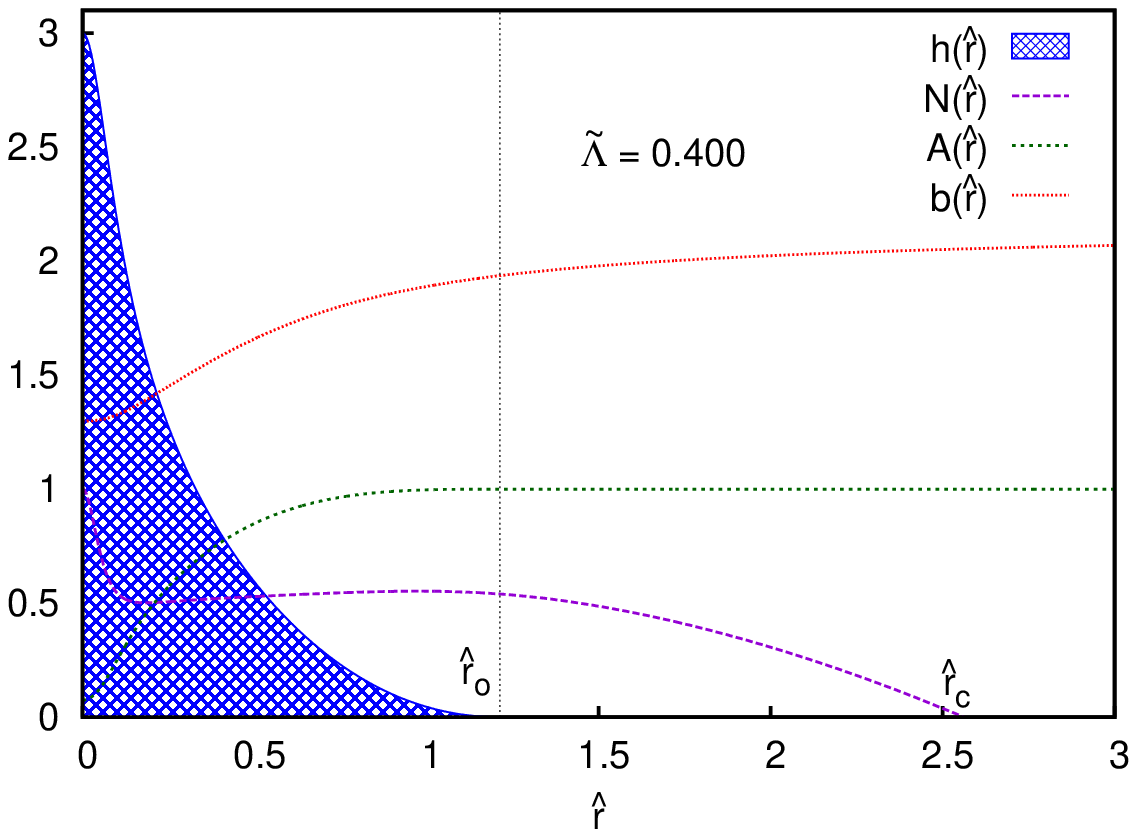}\label{fig:21}}}
\mbox{\subfigure[][]{\includegraphics[scale=0.65]{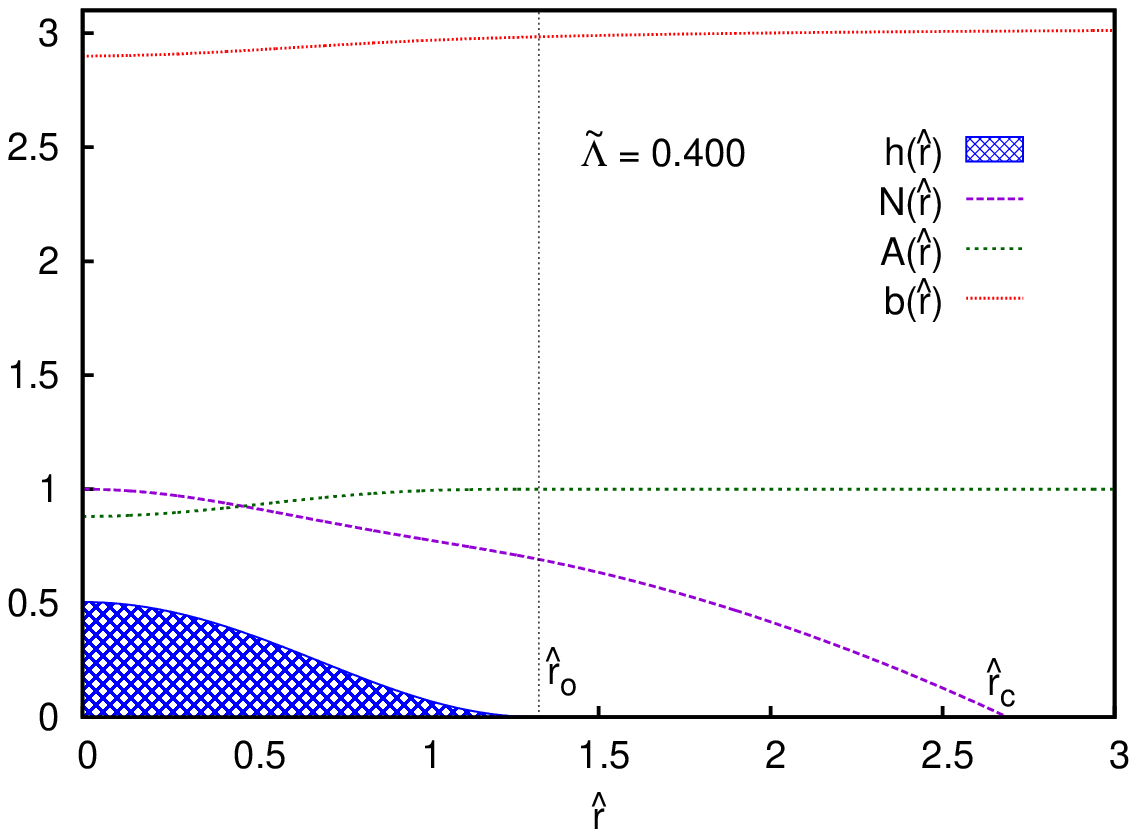}\label{fig:22}}\hspace{1cm}\subfigure[][]{\includegraphics[scale=0.65]{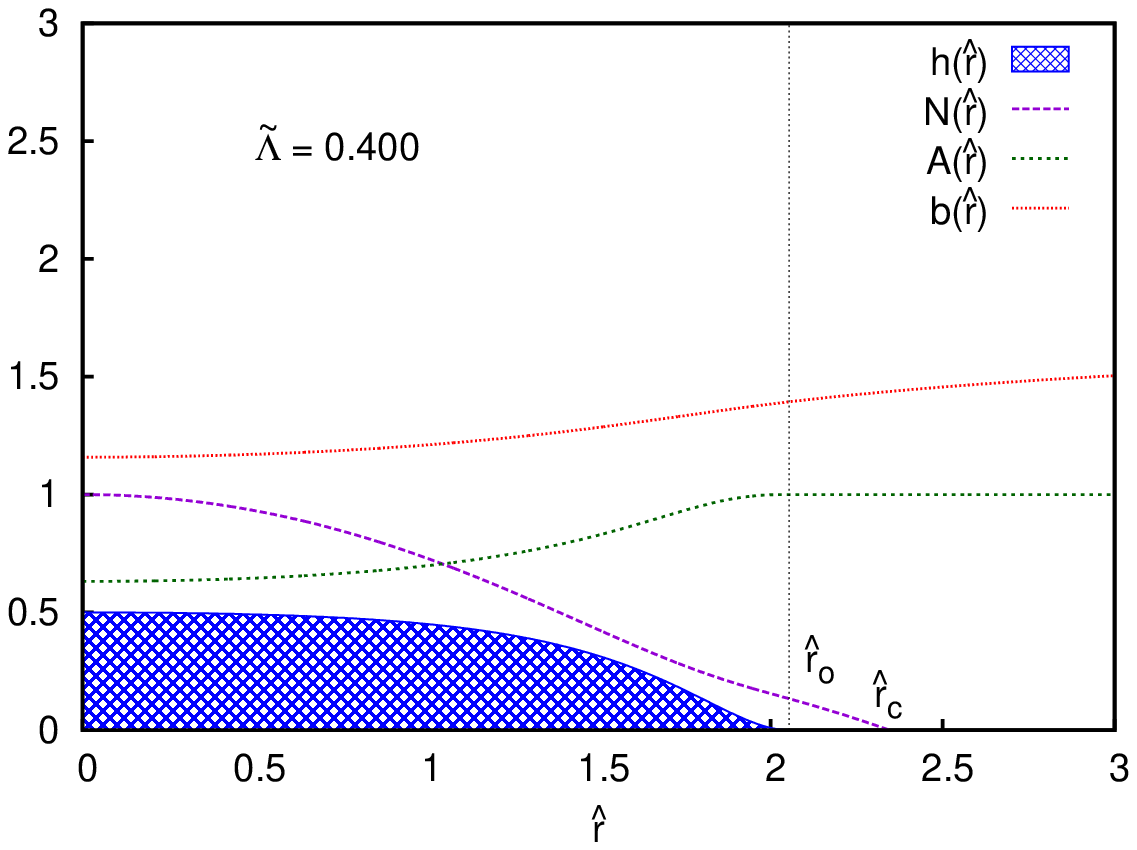}\label{fig:23}}}
\caption{Figs. (a) to (d) depict plots of $h(\hat{r})$, $N(\hat{r})$, $A(\hat{r})$ and $b(\hat{r})$ versus $\hat{r}$ for $\tilde{\Lambda}=0.400$ (which corresponds to the dS space).\label{fig:6f} }
\end{center}
\end{figure*}

\begin{figure*}
\begin{center}
\mbox{\subfigure[][]{\includegraphics[scale=0.68]{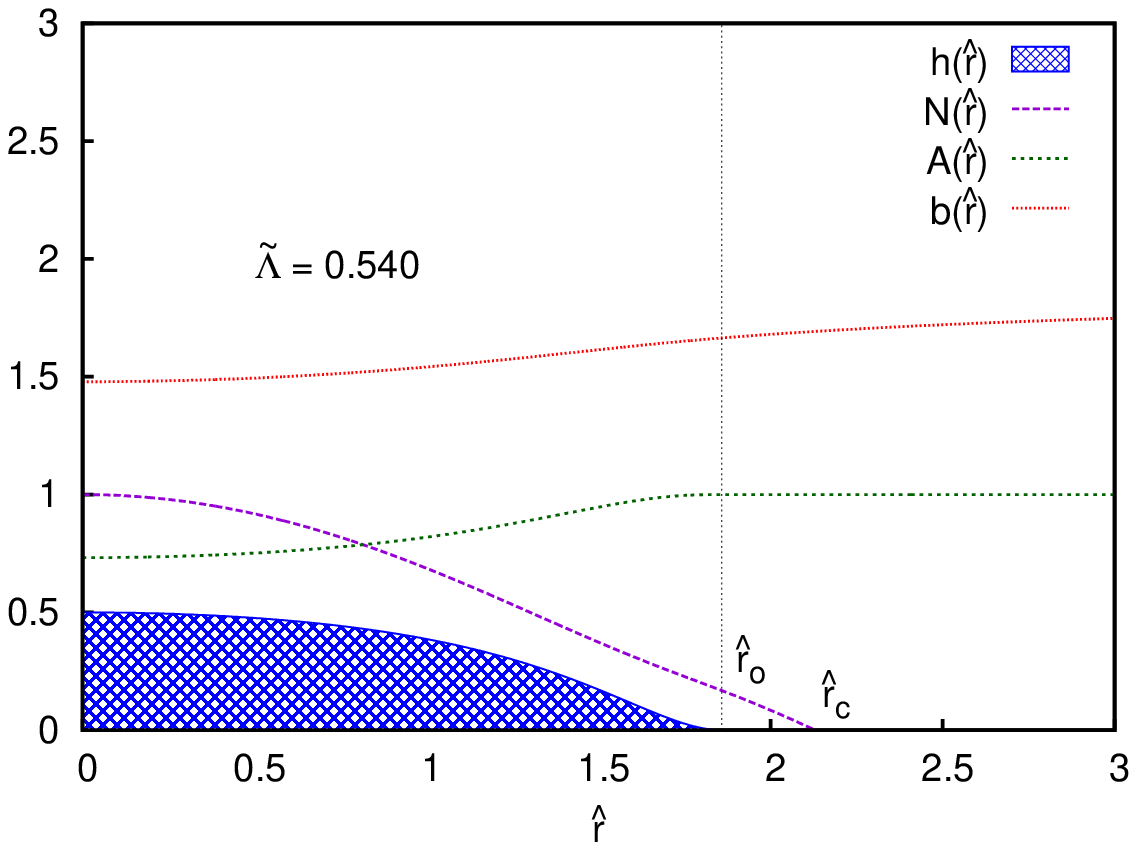}\label{fig:29}}\hspace{0.5cm}
\subfigure[][]{\includegraphics[scale=0.68]{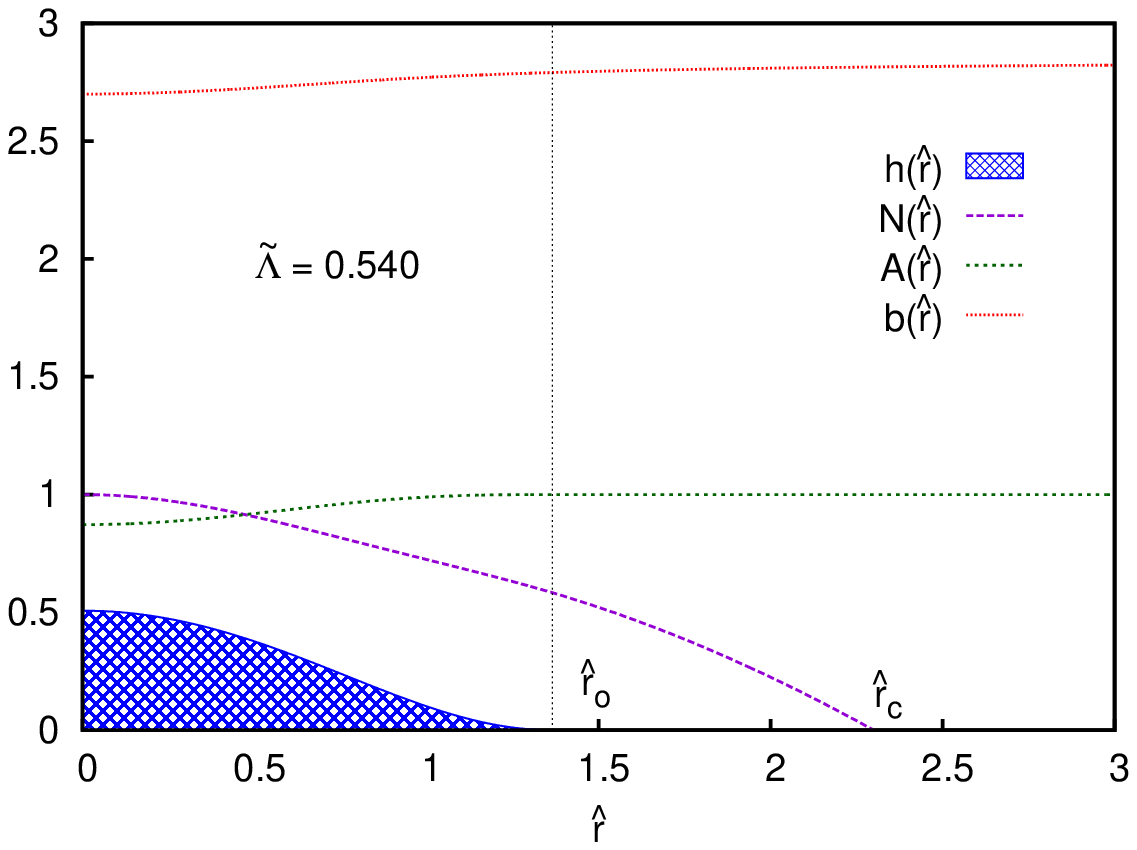}\label{fig:30}}}
\mbox{\subfigure[][]{\includegraphics[scale=0.68]{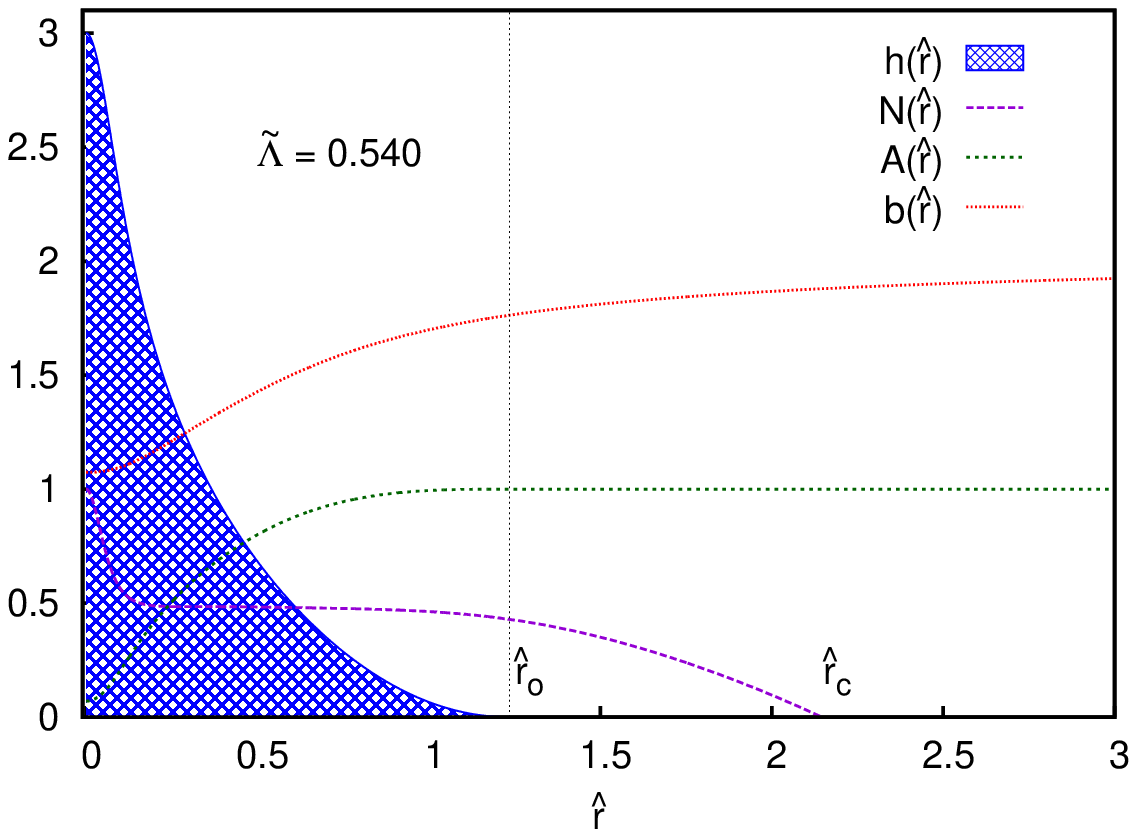}\label{fig:31}}\hspace{0.5cm}
\subfigure[][]{\includegraphics[scale=0.68]{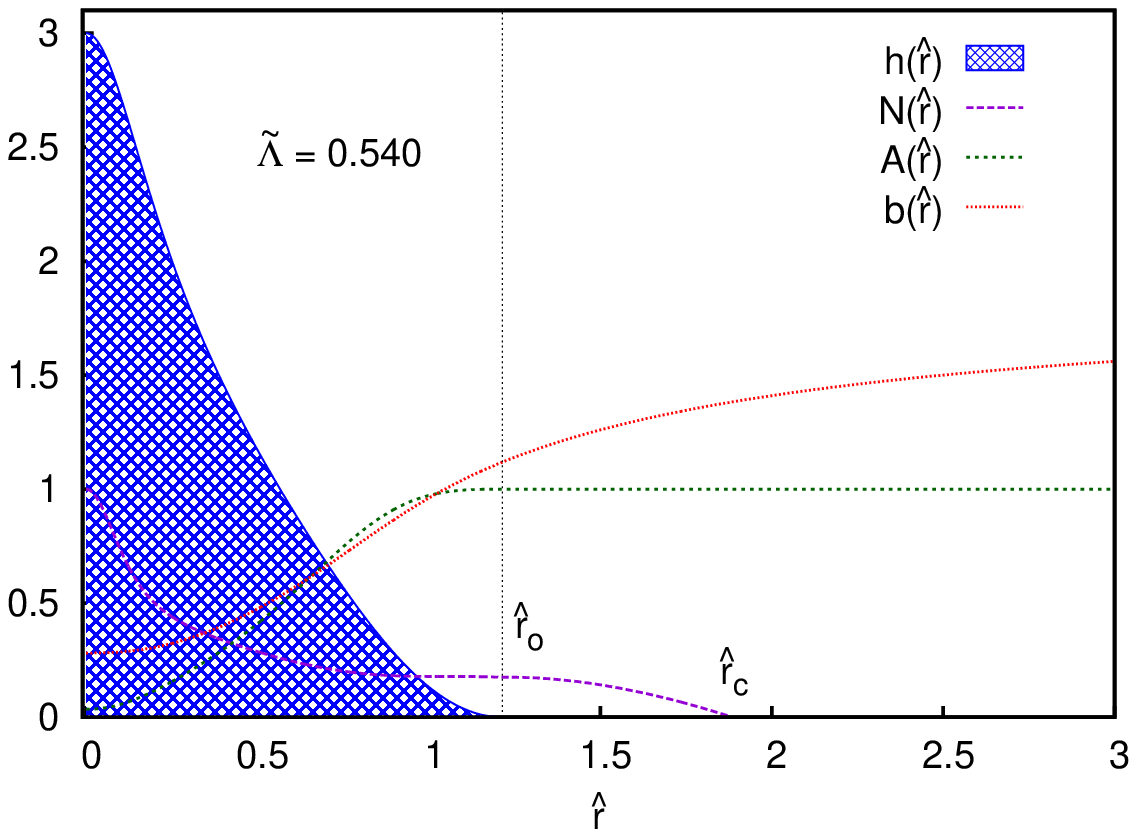}\label{fig:32}}}
\mbox{\subfigure[][]{\includegraphics[scale=0.66]{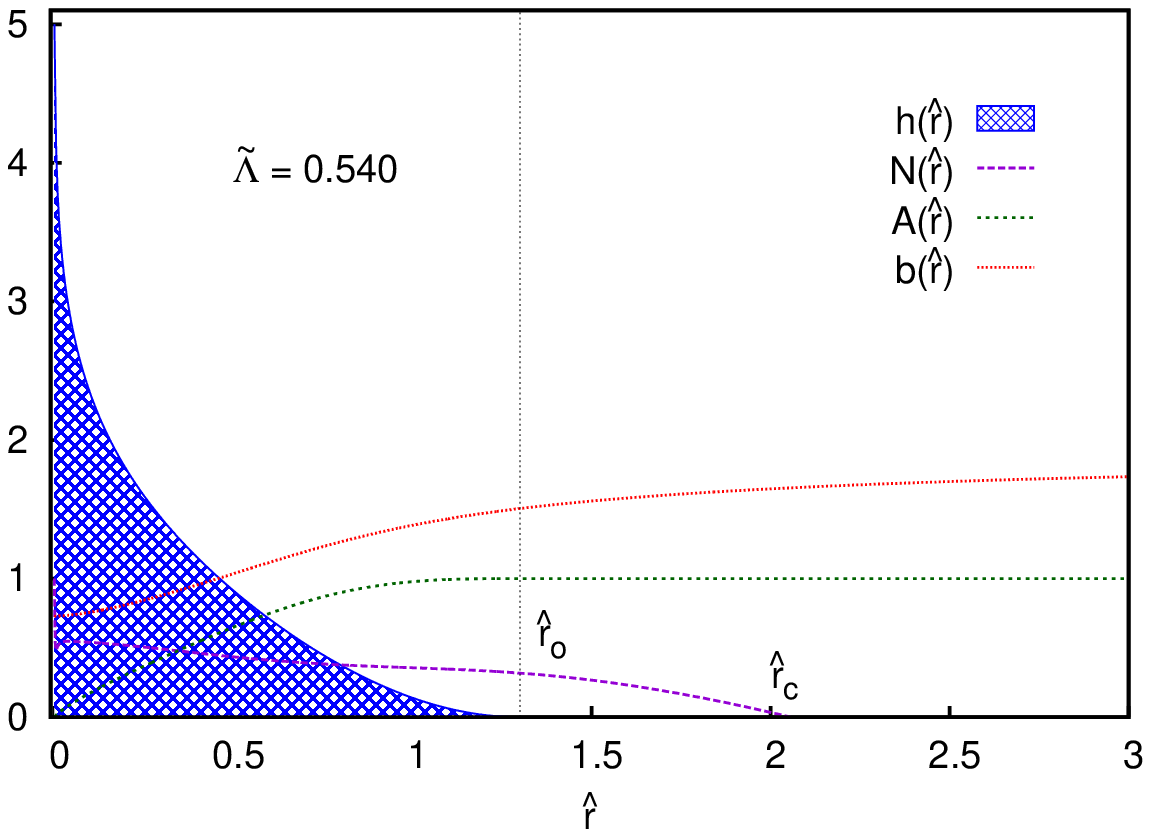}\label{fig:33}}\hspace{0.5cm}
\subfigure[][]{\includegraphics[scale=0.66]{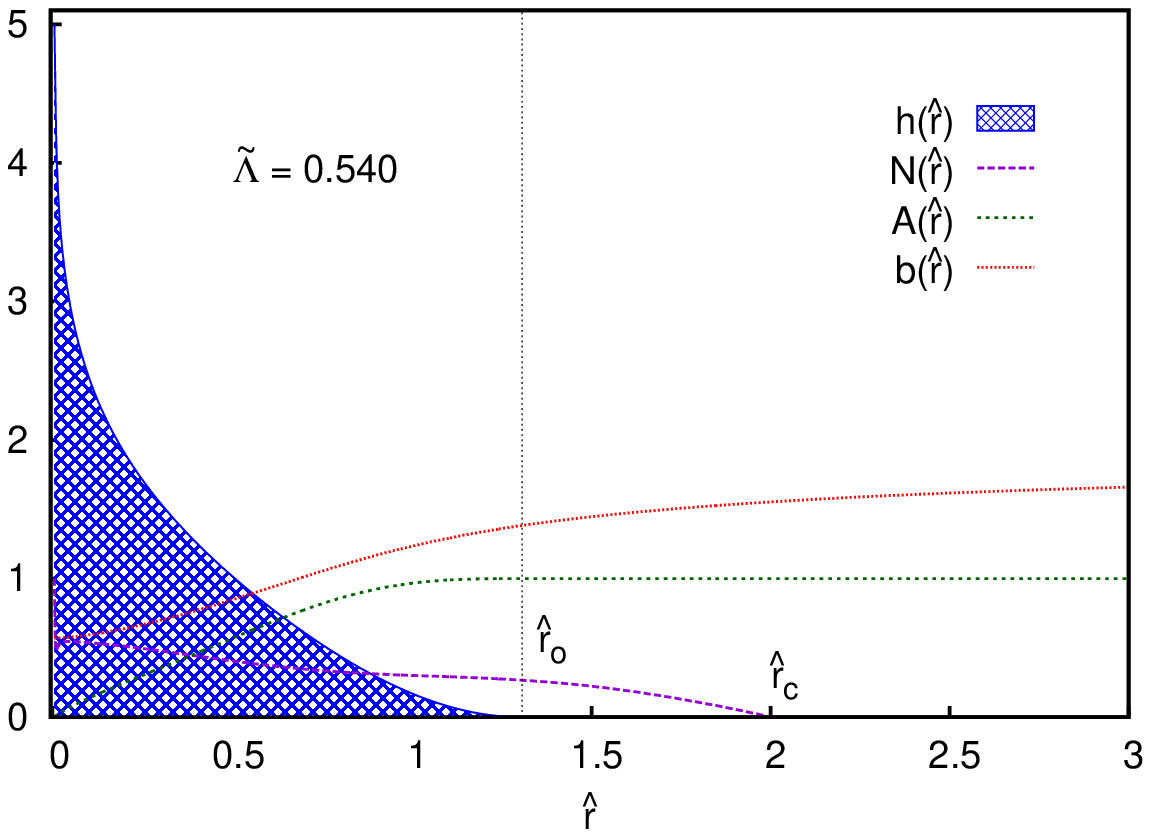}\label{fig:34}}}
\caption{Figs. (a) to (f) depict plots of $h(\hat{r})$, $N(\hat{r})$, $A(\hat{r})$ and $b(\hat{r})$ versus $\hat{r}$ for $\tilde{\Lambda}=0.540$ (which corresponds to the dS space).\label{fig:7f} }
\end{center}
\end{figure*}

\begin{figure*}
\begin{center}
\mbox{\subfigure[][]{\includegraphics[scale=0.68]{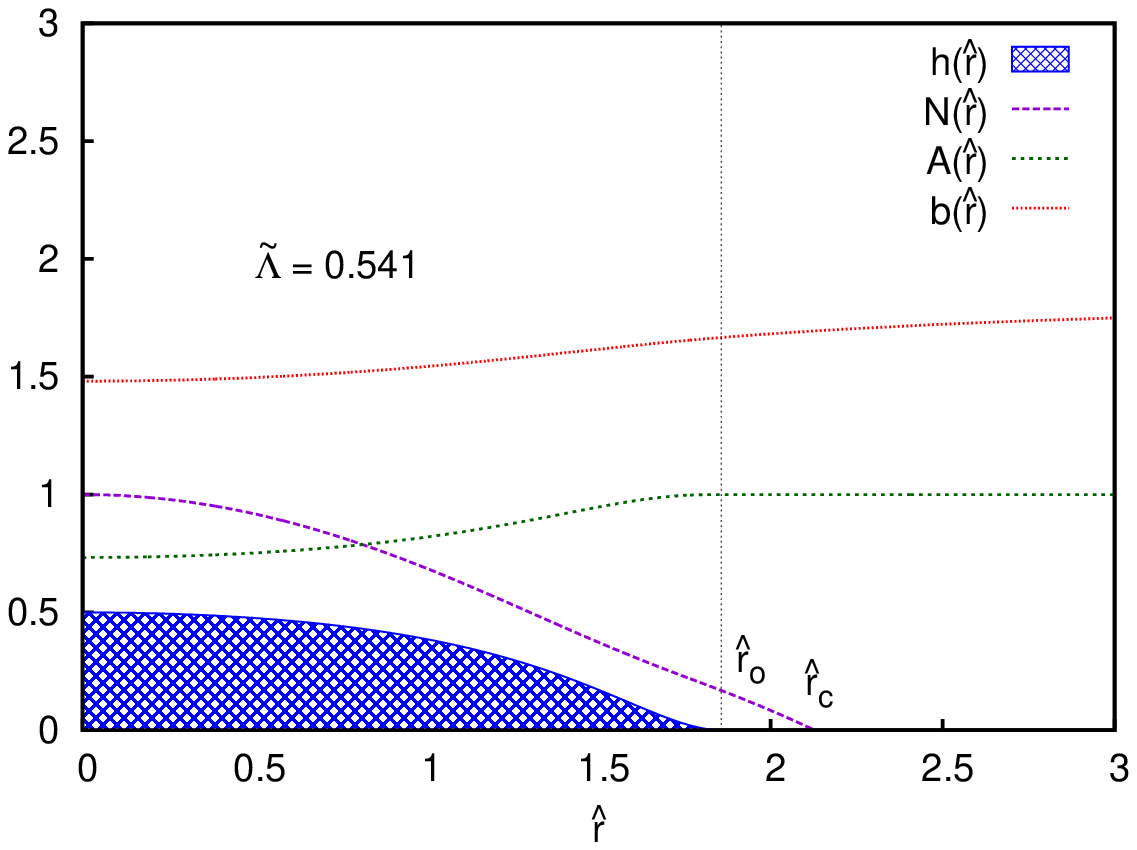}\label{fig:51}}\hspace{0.5cm}
\subfigure[][]{\includegraphics[scale=0.68]{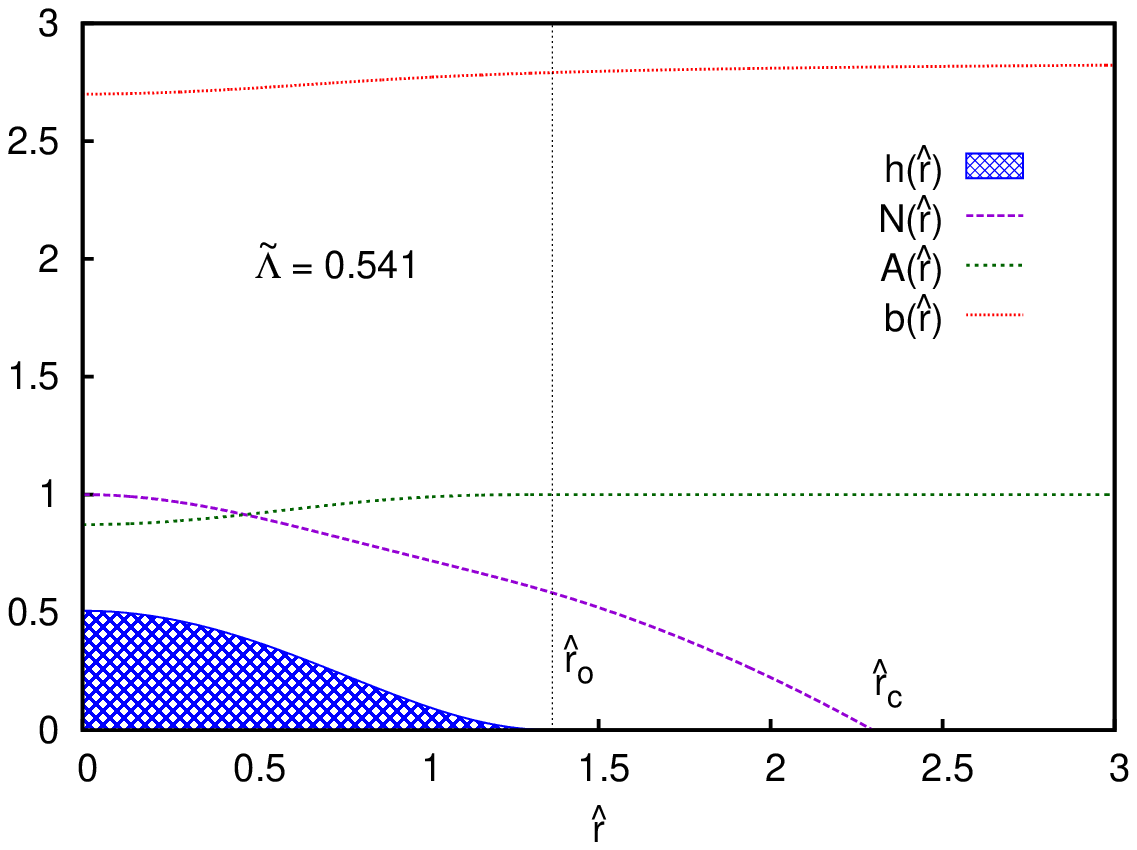}\label{fig:52}}}
\mbox{\subfigure[][]{\includegraphics[scale=0.68]{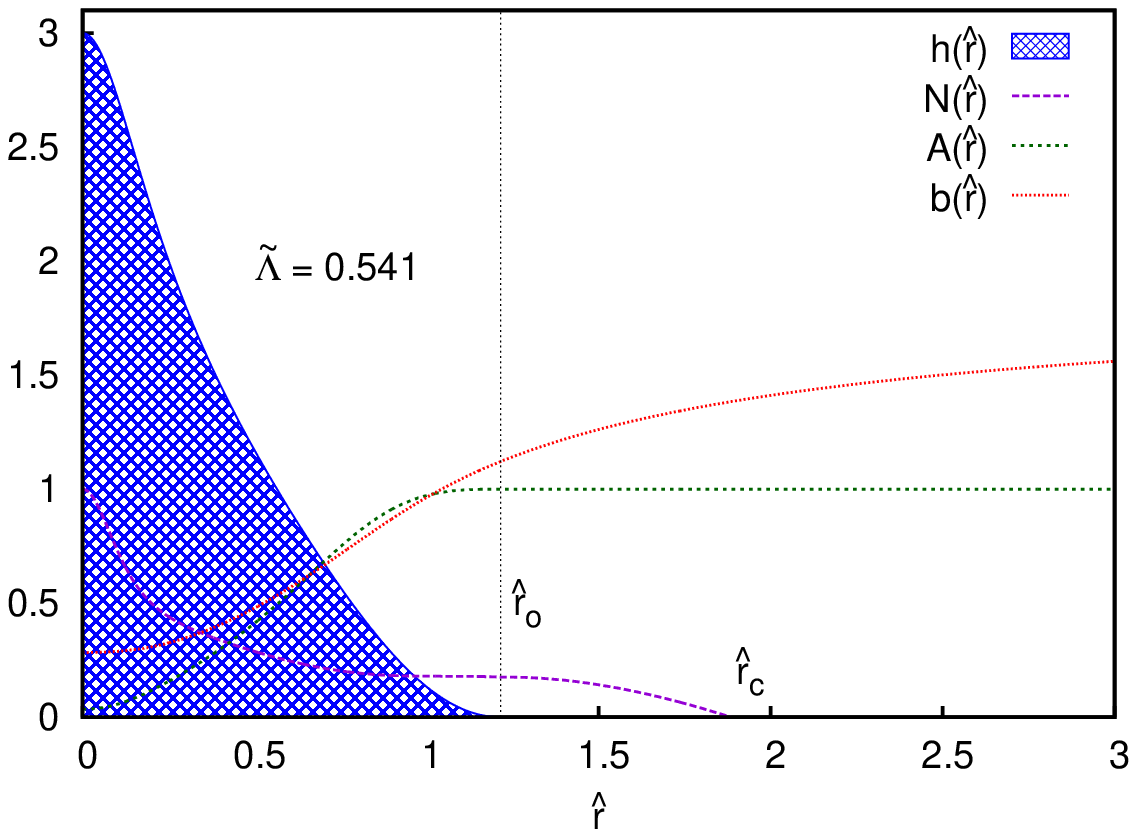}\label{fig:53}}\hspace{0.5cm}
\subfigure[][]{\includegraphics[scale=0.68]{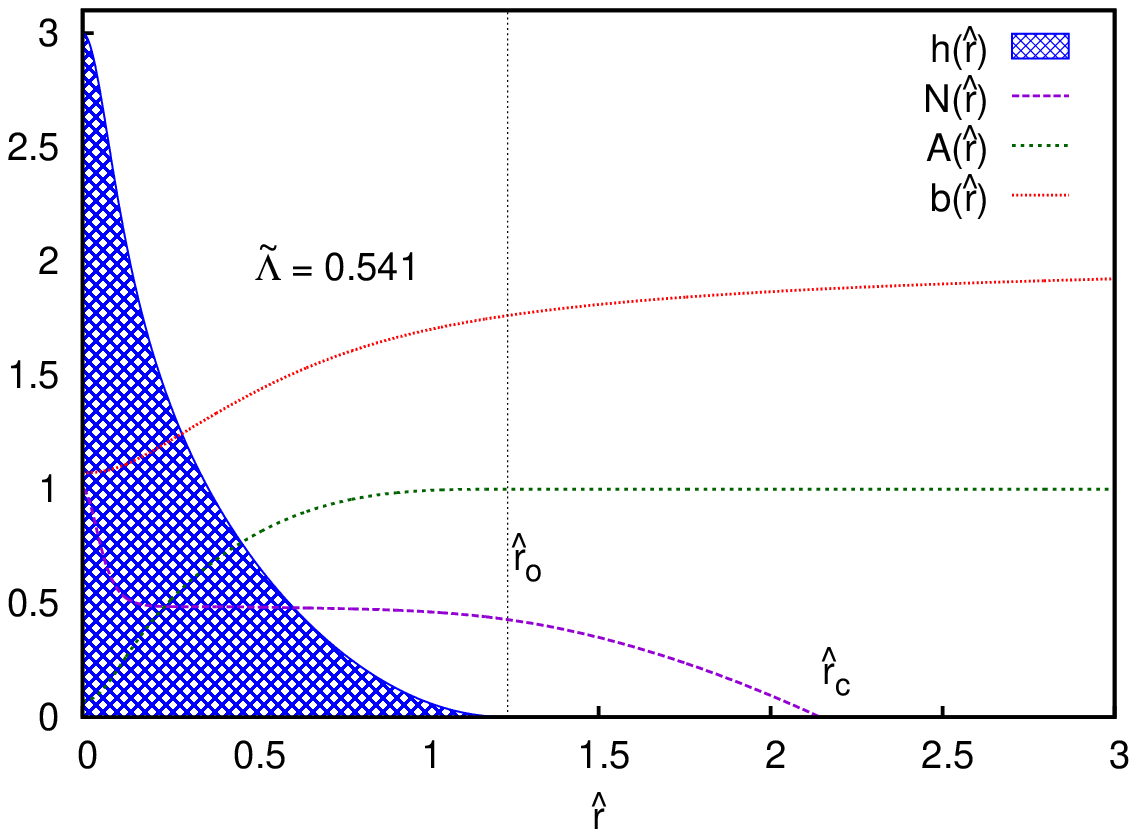}\label{fig:54}}}
\mbox{\subfigure[][]{\includegraphics[scale=0.65]{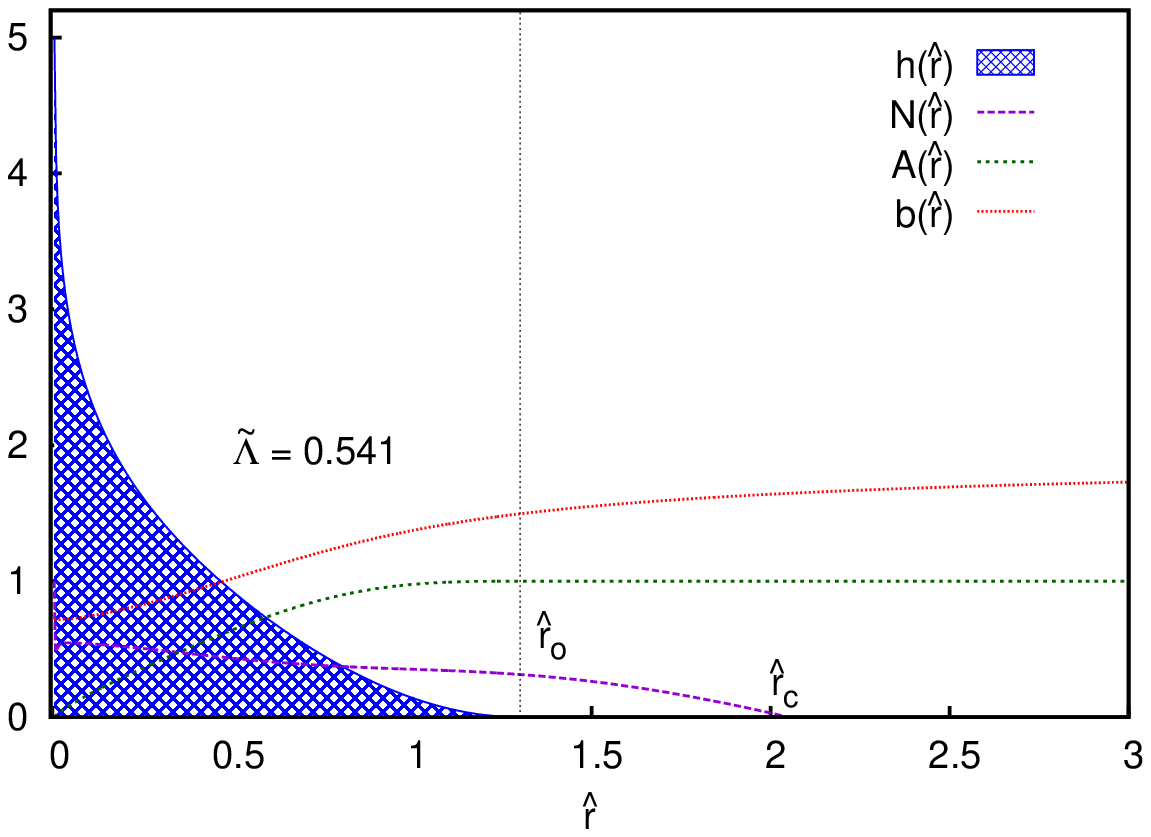}\label{fig:55}}\hspace{0.5cm}
\subfigure[][]{\includegraphics[scale=0.65]{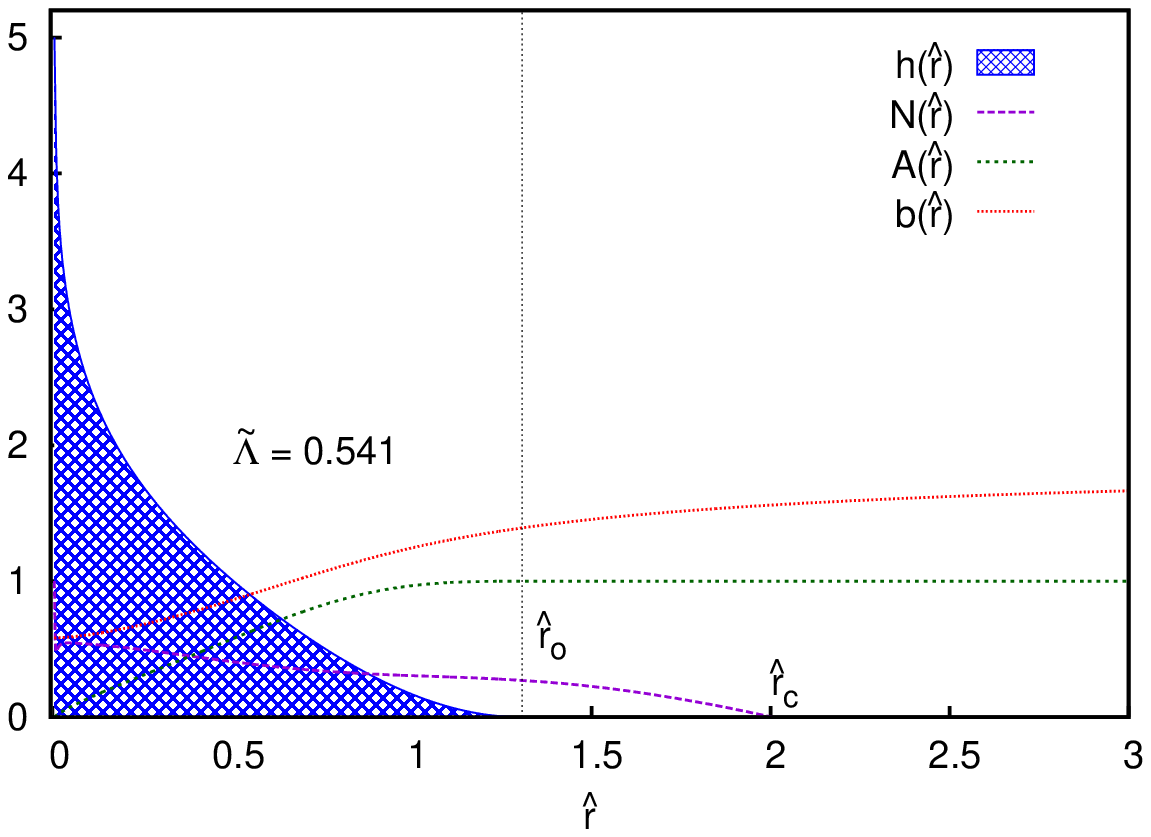}\label{fig:56}}}
\caption{ Figs. (a) to (f) depict plots of $h(\hat{r})$, $N(\hat{r})$, $A(\hat{r})$ and $b(\hat{r})$ versus $\hat{r}$ for $\tilde{\Lambda}=0.541$ (which corresponds to the dS space).\label{fig:8f}}
\end{center}
\end{figure*}

\begin{figure*}
\begin{center}
\mbox{\subfigure[][]{\includegraphics[scale=0.65]{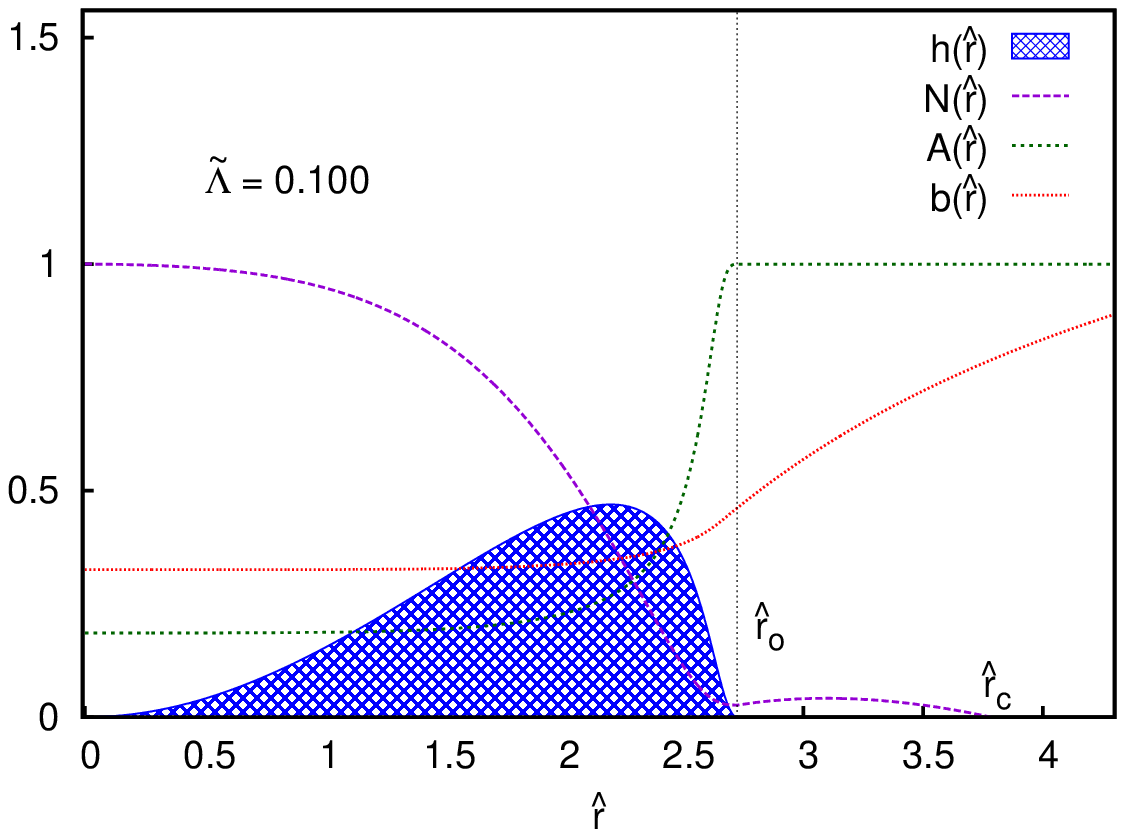}\label{fig:59}}\hspace{1cm}
\subfigure[][]{\includegraphics[scale=0.65]{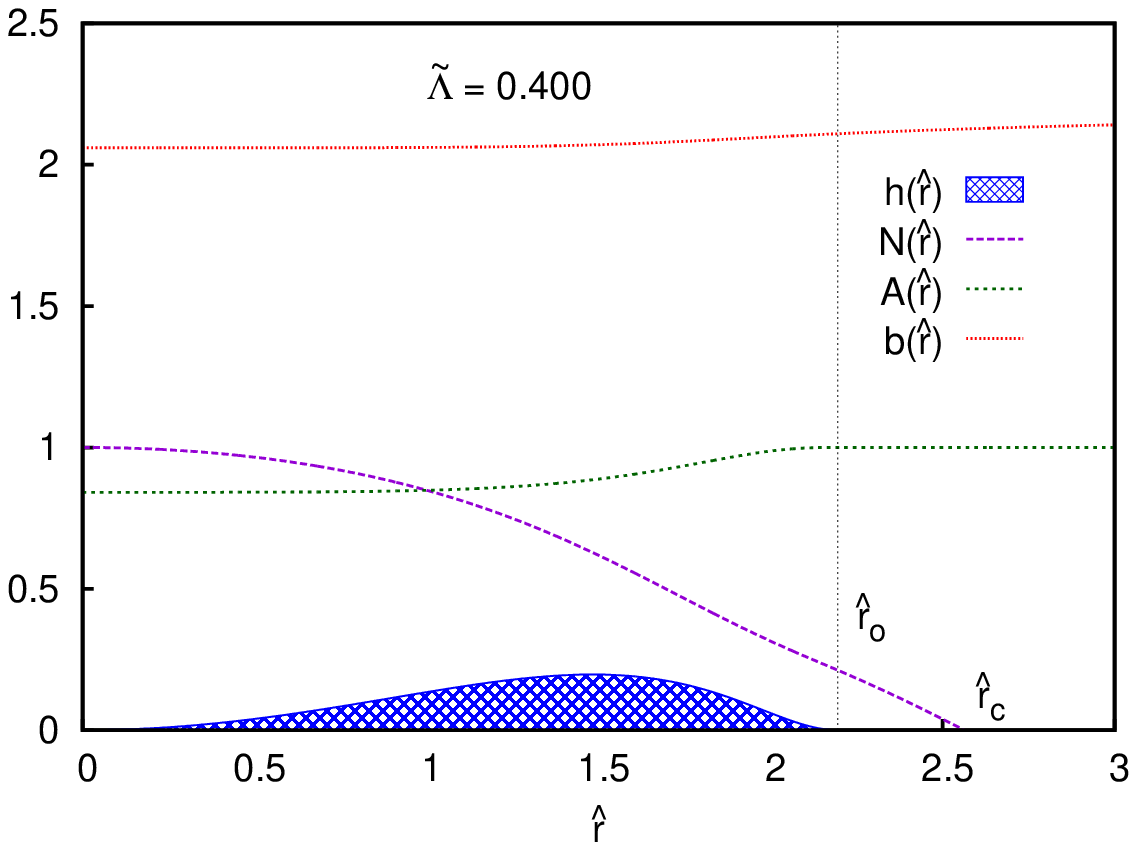}\label{fig:17}}}
\mbox{\subfigure[][]{\includegraphics[scale=0.65]{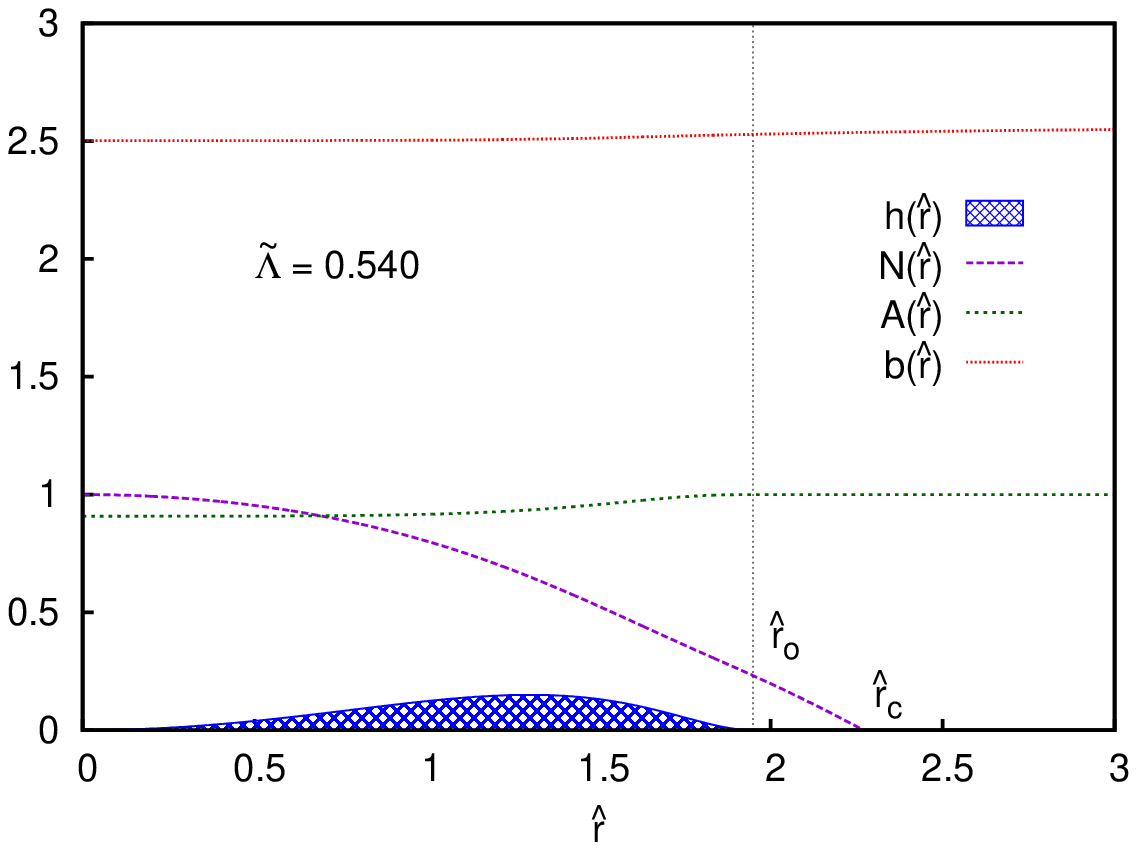}\label{fig:26}}\hspace{1cm}
\subfigure[][]{\includegraphics[scale=0.65]{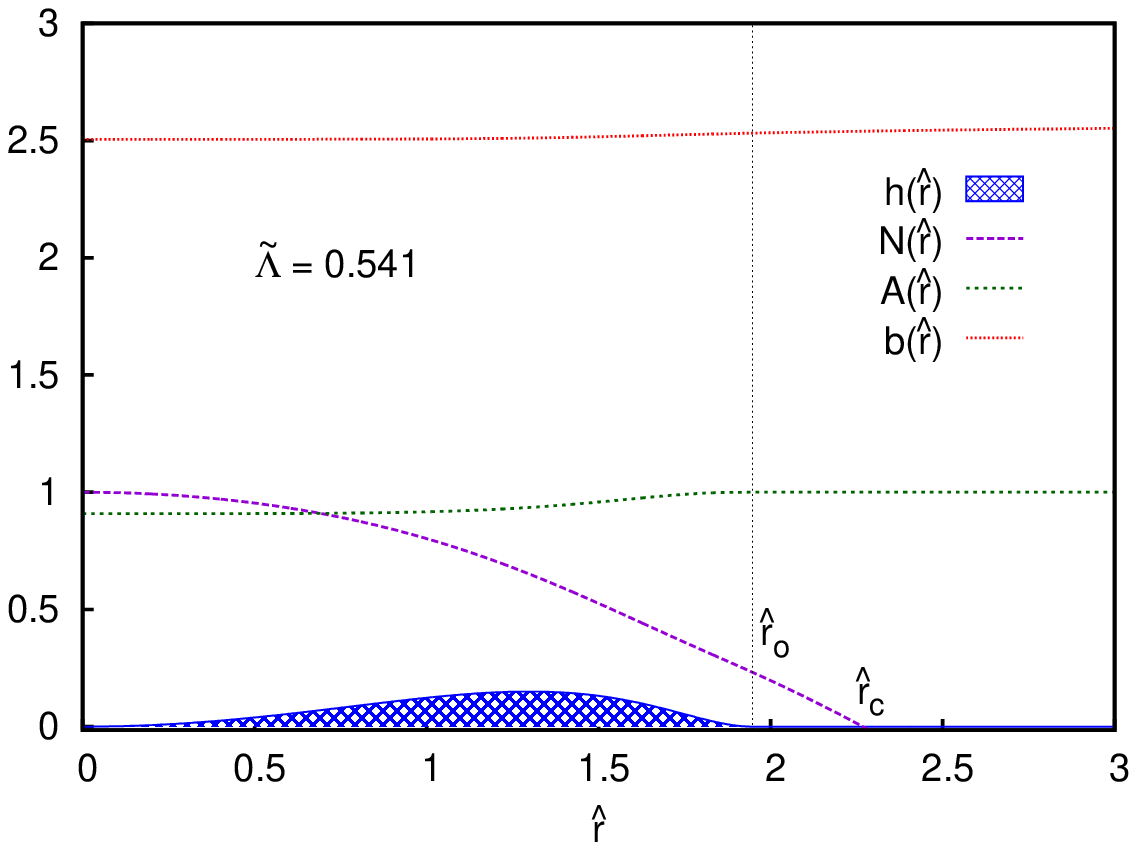}\label{fig:48}}}
\caption{Figs. (a) to (d) depict plots of $h(\hat{r})$, $N(\hat{r})$, $A(\hat{r})$ and $b(\hat{r})$ versus $\hat{r}$ for shown values of $\tilde{\Lambda}$ (all of which correspond to the dS space). Also, these figures correspond to the transition points of the theory from the bosons stars to the boson shells.\label{fig:9f} }
\end{center}
\end{figure*}

\begin{figure*}
\begin{center}
\mbox{\subfigure[][]{\includegraphics[scale=0.65]{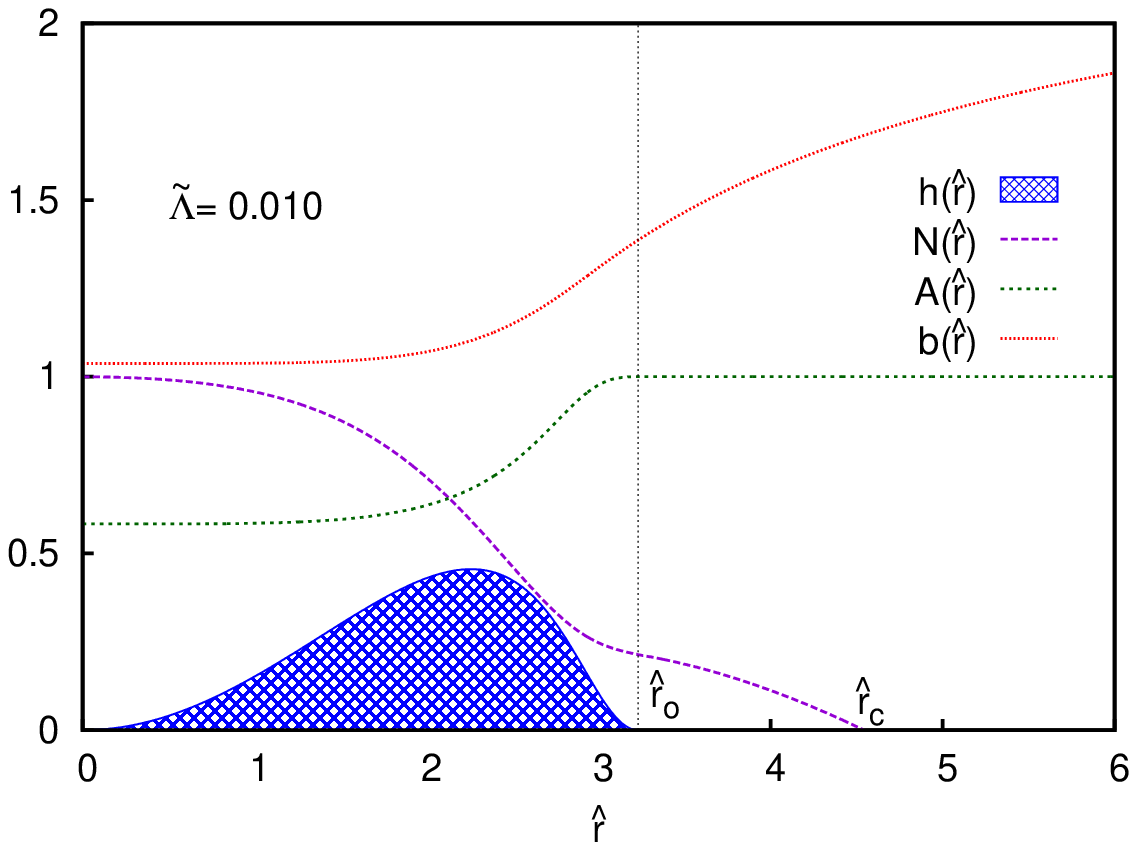}\label{fig:2-e}}\hspace{1cm}\subfigure[][]{\includegraphics[scale=0.65]{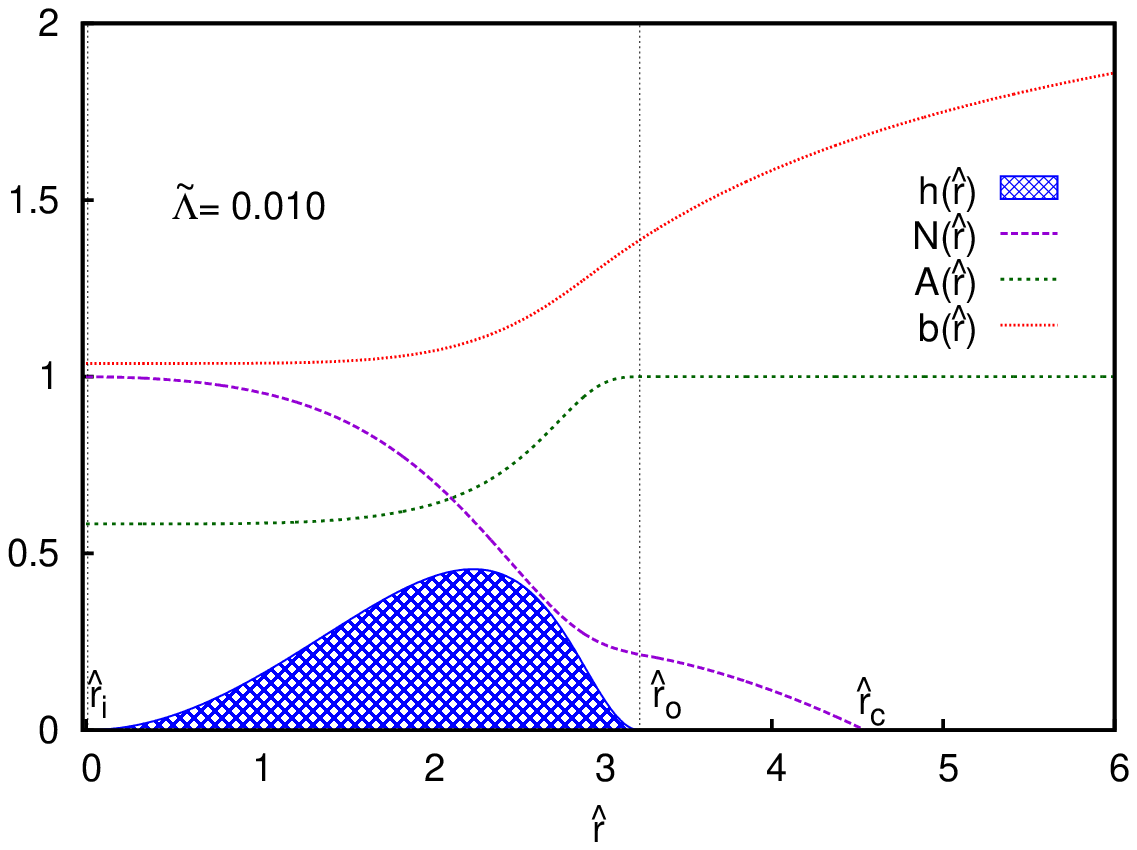}\label{fig:2-c}}}
\mbox{\subfigure[][]{\includegraphics[scale=0.65]{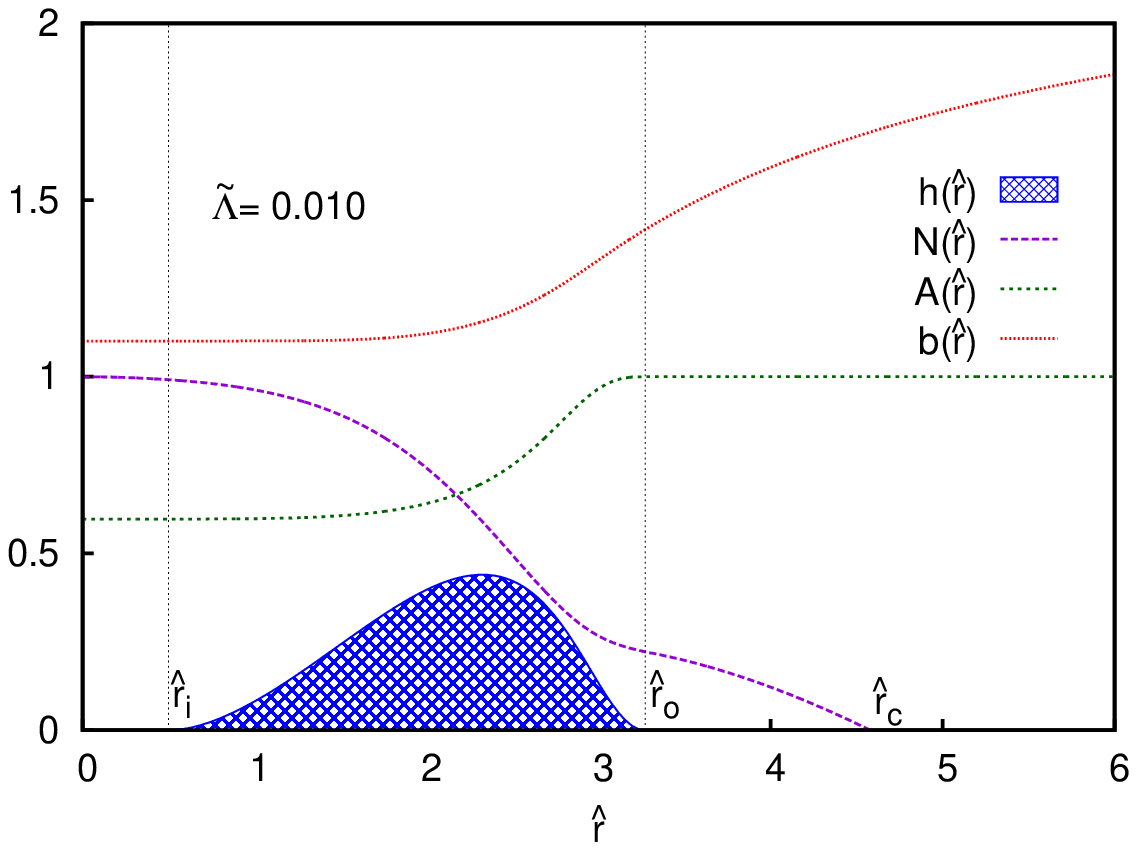}\label{fig:2-h}}\hspace{1cm}\subfigure[][]{\includegraphics[scale=0.65]{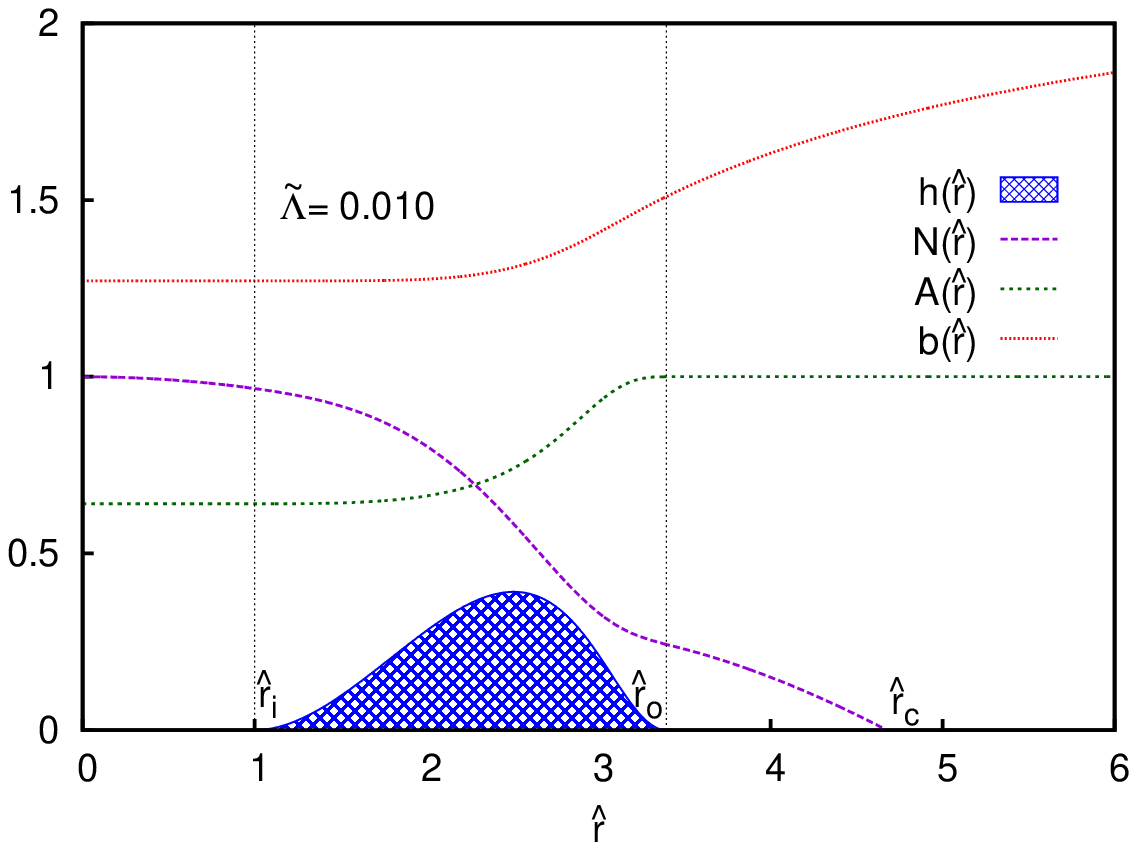}\label{fig:2-a}}}
\caption{ Figs. (a) to (d) depict plots of $h(\hat{r})$, $N(\hat{r})$, $A(\hat{r})$ and $b(\hat{r})$ versus $\hat{r}$ for  $\tilde{\Lambda}=0.010$ (which corresponds to the dS space).\label{fig:10f}}
\end{center}
\end{figure*}

\begin{figure*}
\begin{center}
\mbox{\subfigure[][]{\includegraphics[scale=0.65]{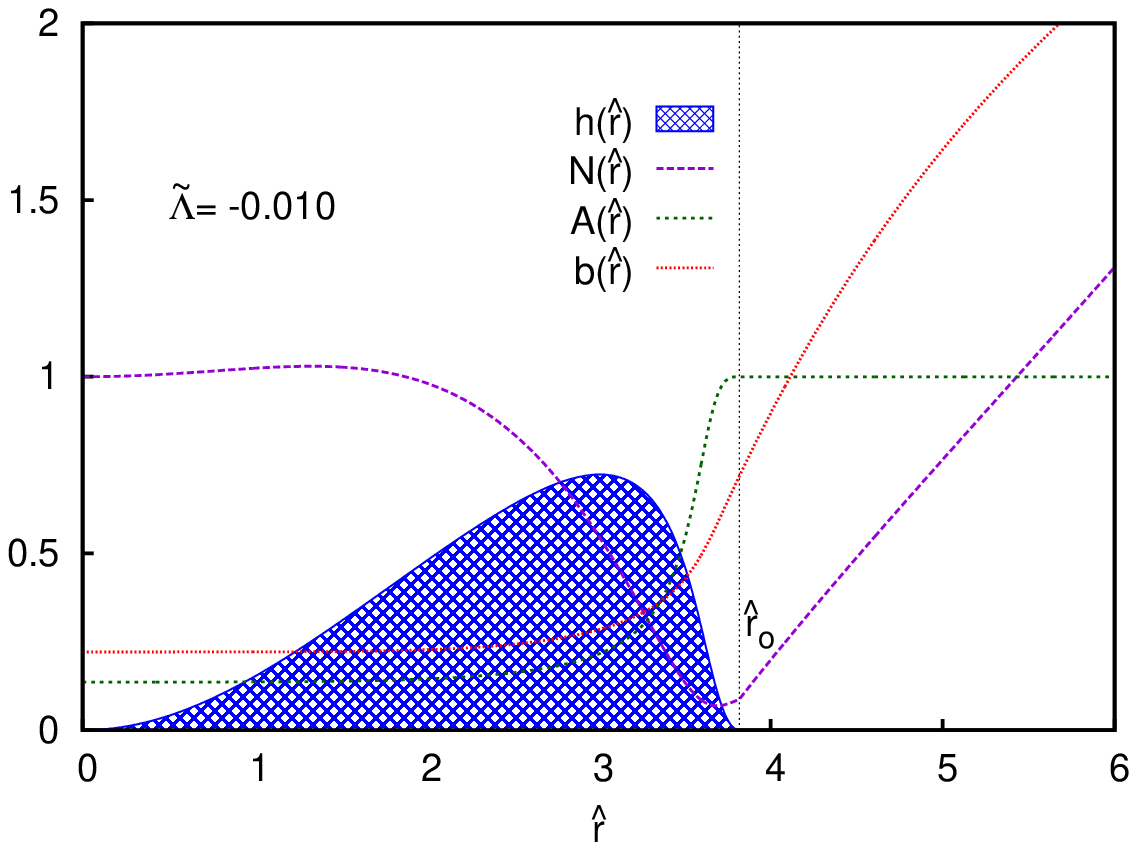}\label{fig:2-f}}\hspace{1cm}\subfigure[][]{\includegraphics[scale=0.65]{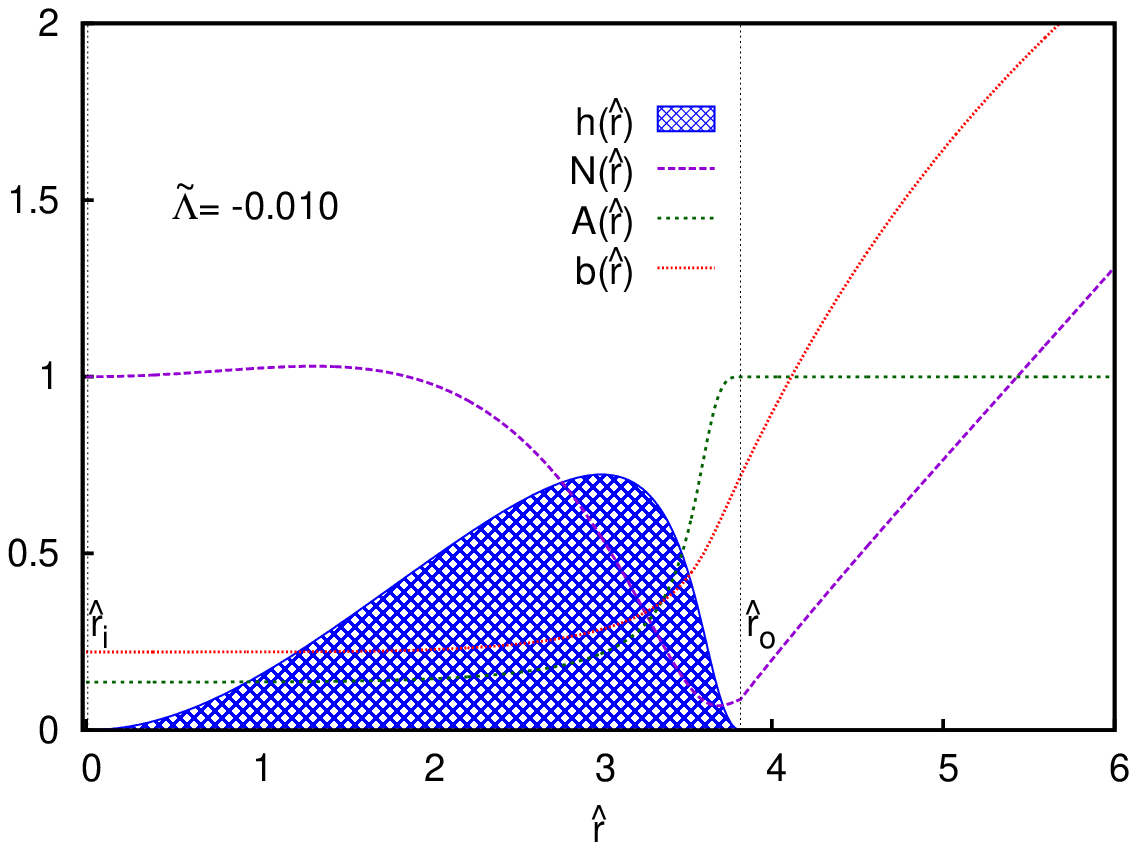}\label{fig:2-d}}}
\mbox{\subfigure[][]{\includegraphics[scale=0.65]{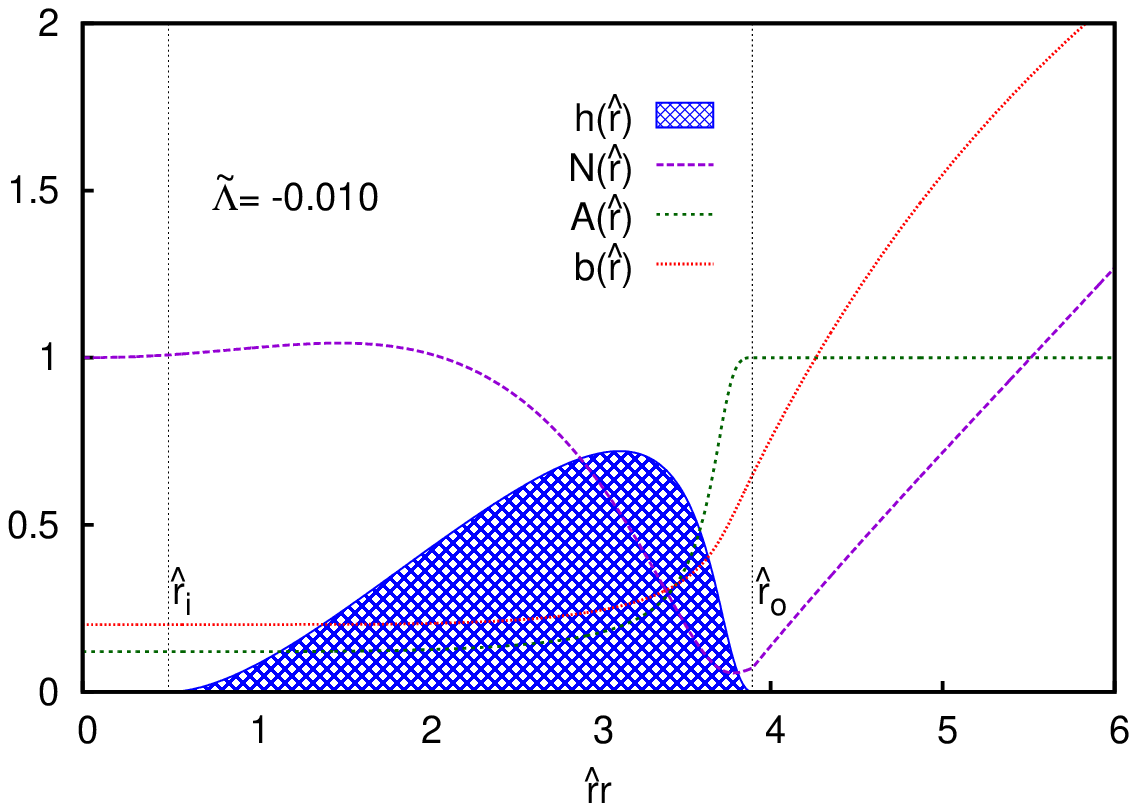}\label{fig:2-g}}\hspace{1cm}\subfigure[][]{\includegraphics[scale=0.65]{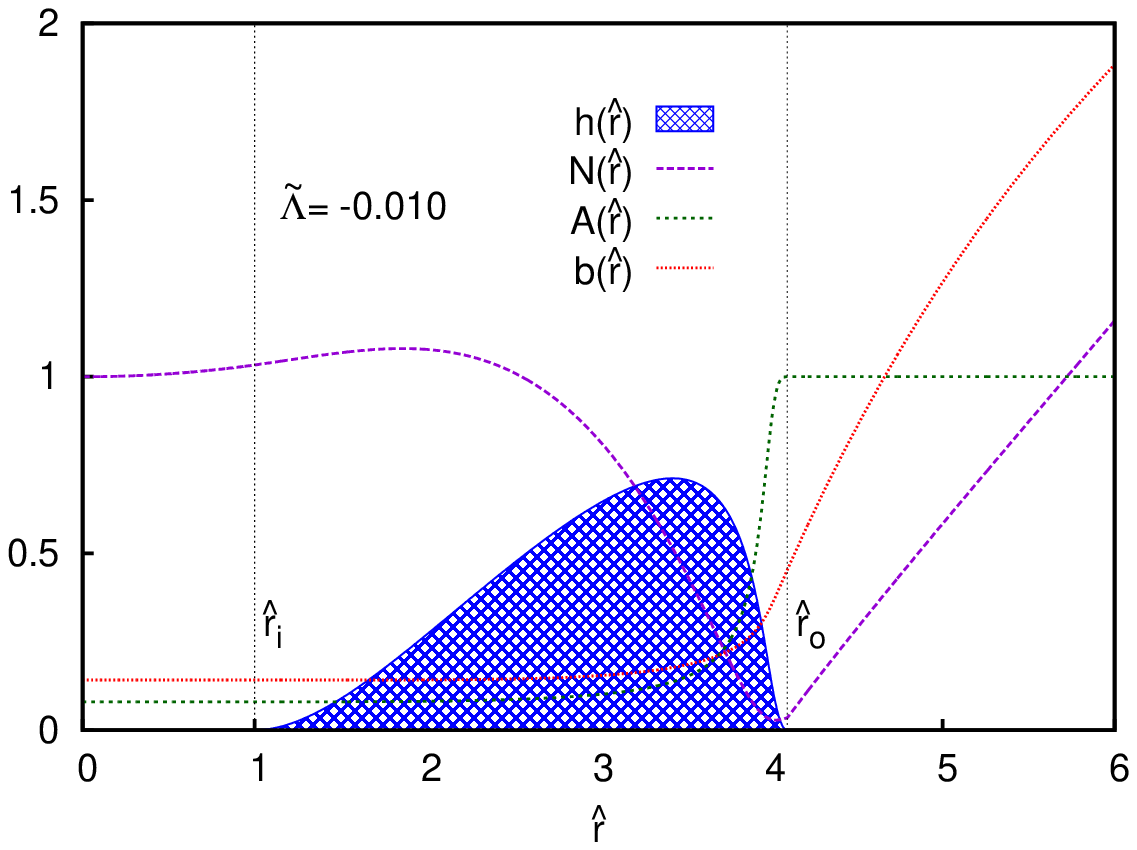}\label{fig:2-b}}}
\caption{Figs. (a)--(d) depict plots of $h(\hat{r})$, $N(\hat{r})$, $A(\hat{r})$ and $b(\hat{r})$ versus $\hat{r}$ for $\tilde{\Lambda}=-0.010$ (which corresponds to the AdS space).\label{fig:11f}}
\end{center}
\end{figure*}

\begin{figure*}
\begin{center}
	\mbox{\subfigure[][]{\includegraphics[scale=0.65]{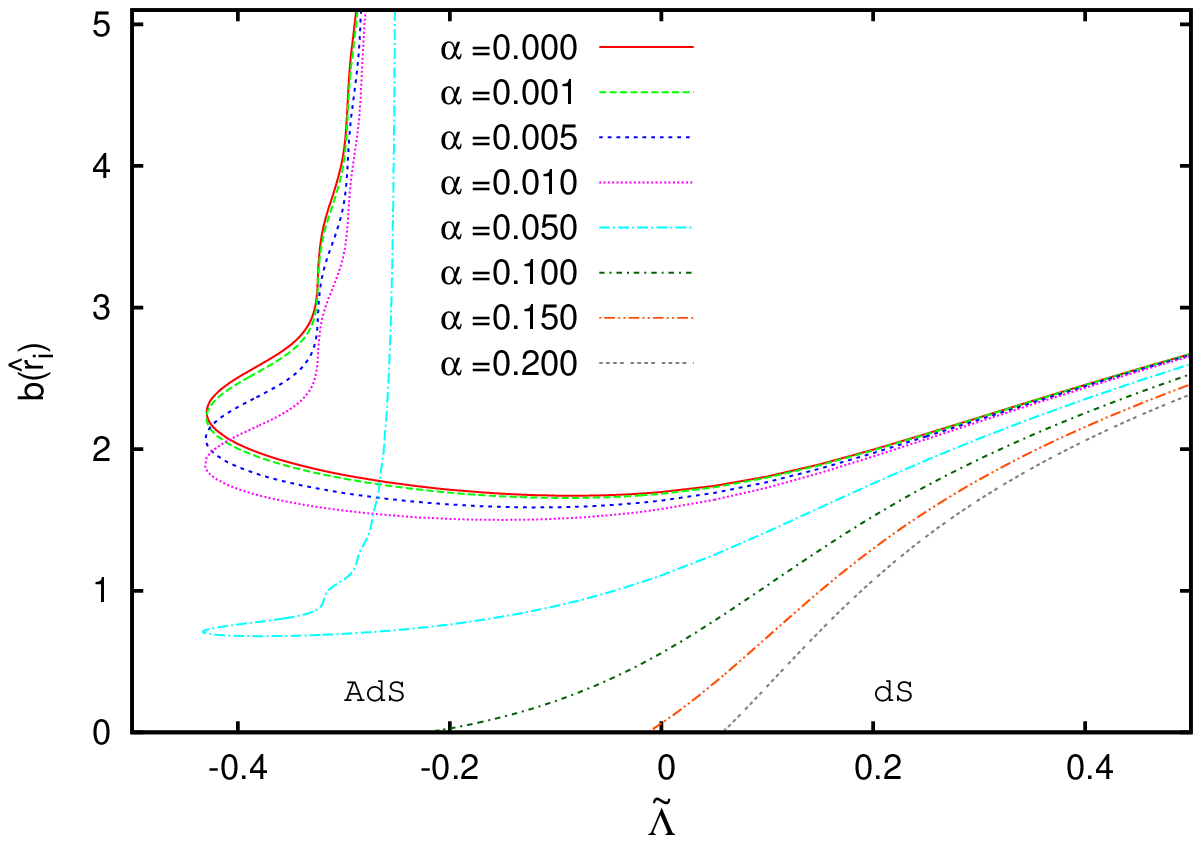}\label{fig:f2a}}\hspace{1cm}
\subfigure[][]{\includegraphics[scale=0.65]{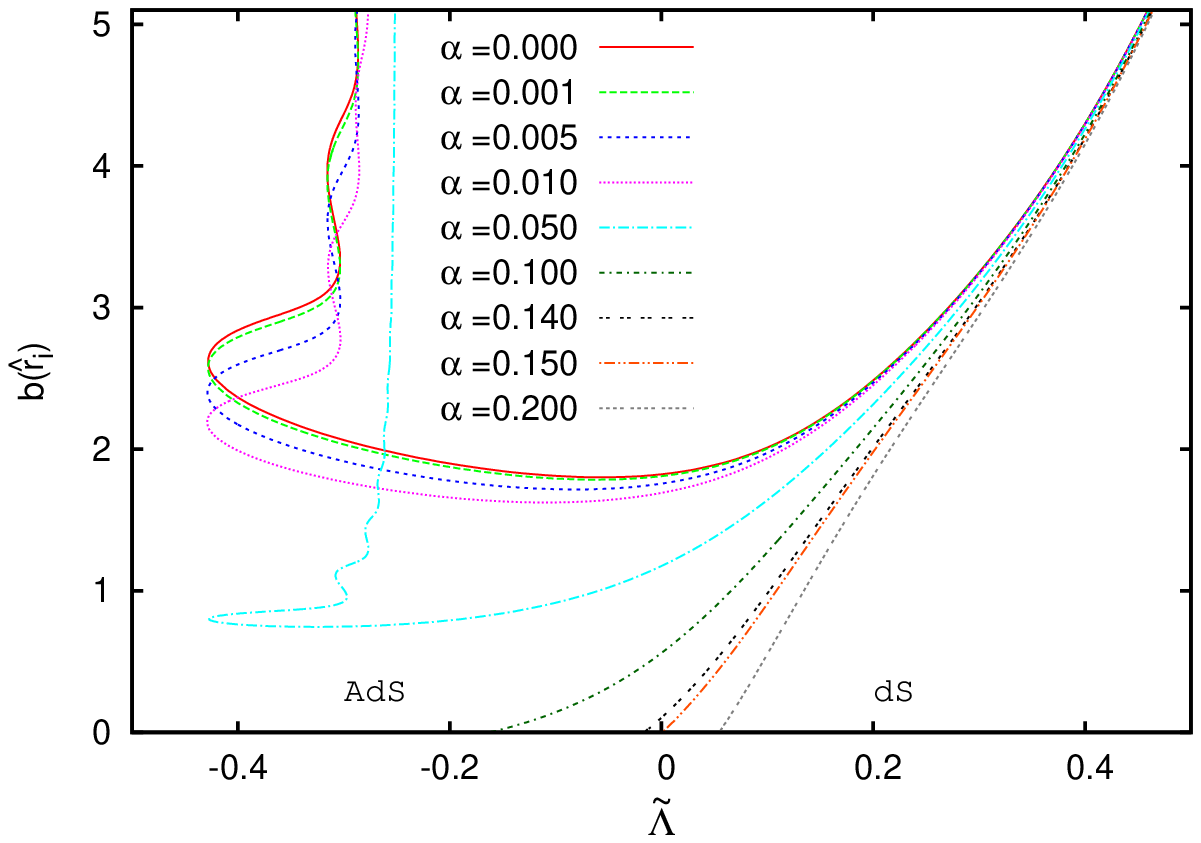}\label{fig:f2b}}}
		\mbox{\subfigure[][]{\includegraphics[scale=0.65]{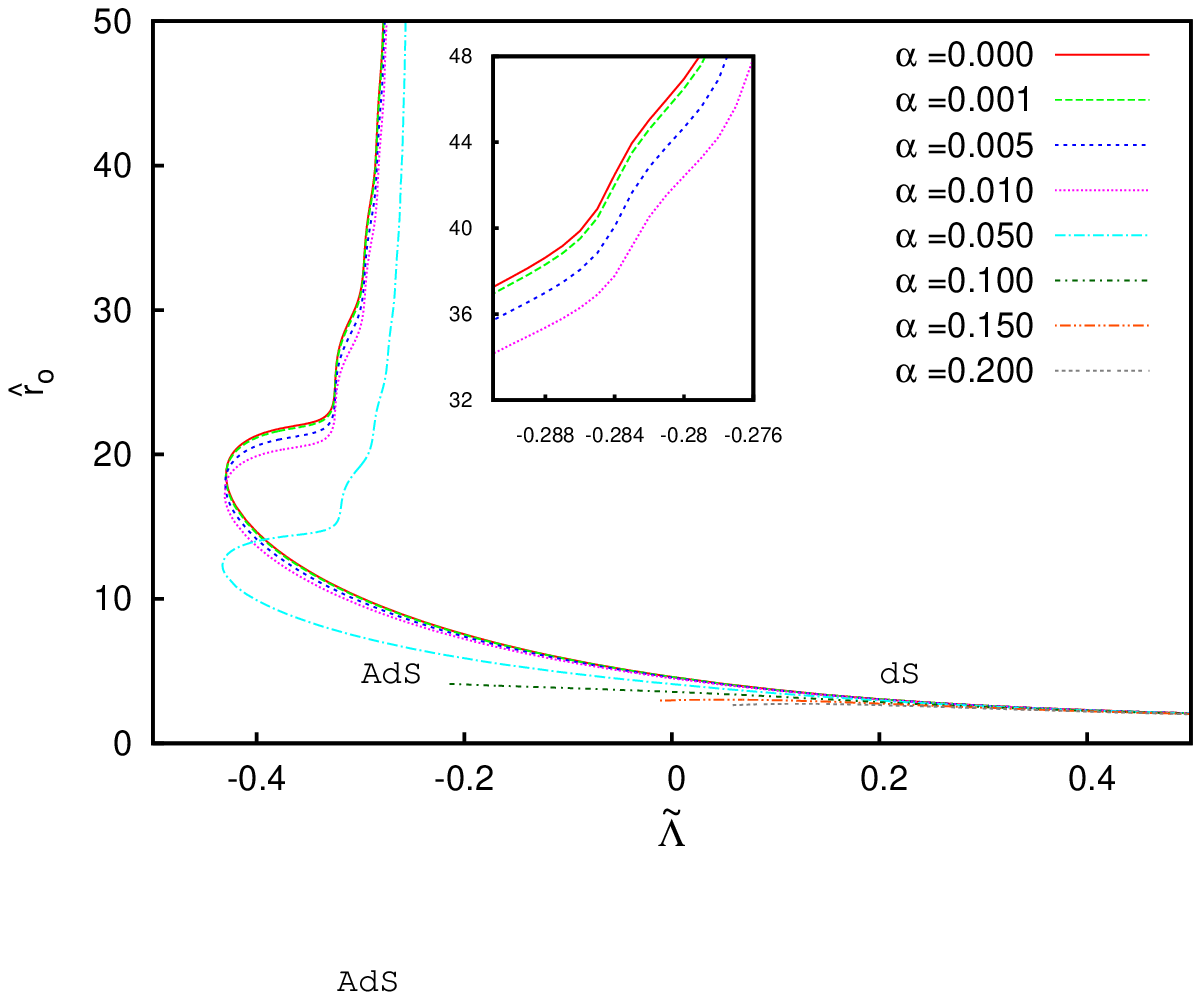}\label{fig:f3a}}\hspace{1cm}
		\subfigure[][]{\includegraphics[scale=0.65]{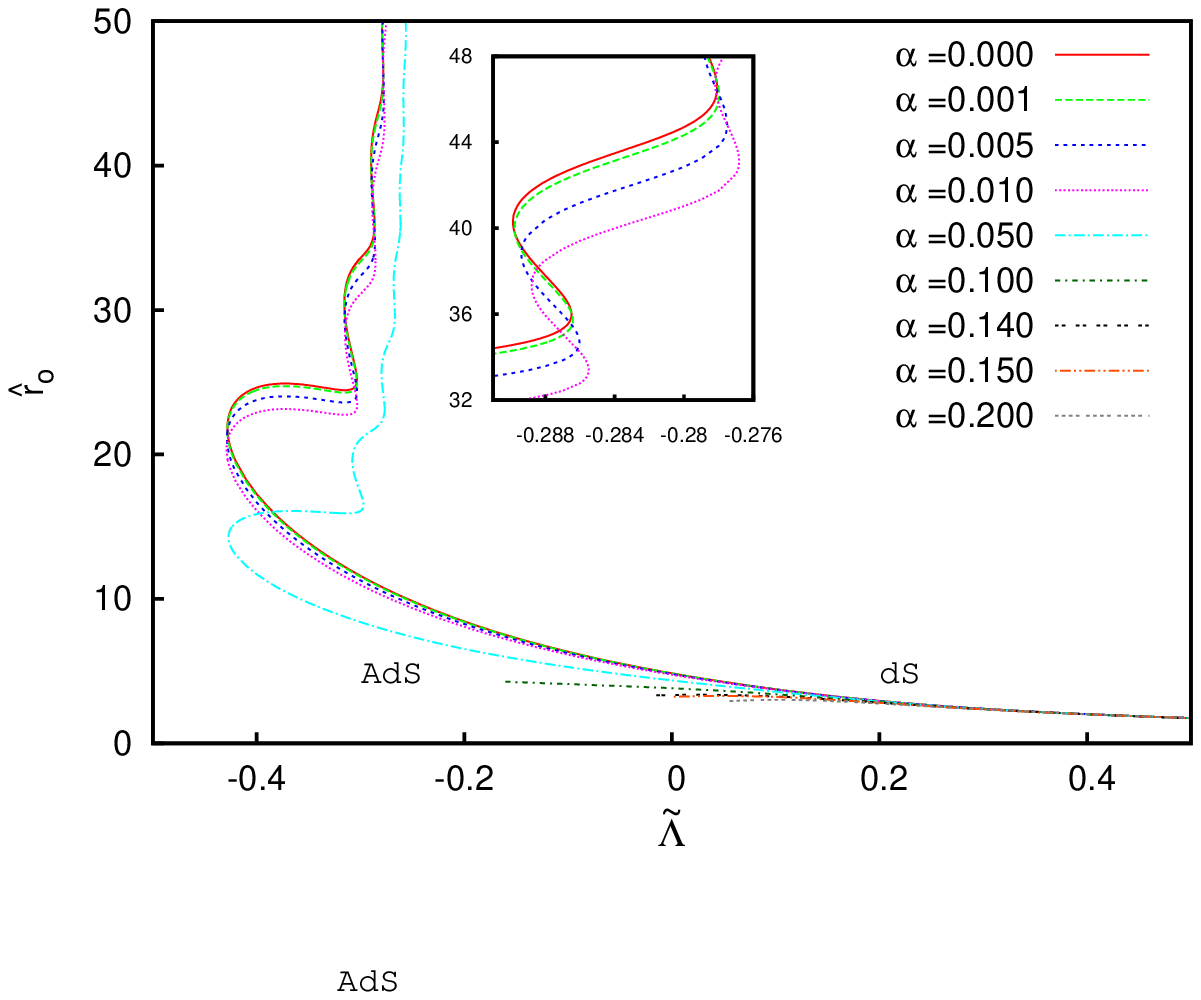}\label{fig:f3b}}}
\caption{Figs. (a) and (b) show respectively the properties of the gravitating boson stars and boson shells depicting the plot of $b(\hat{r}_i)$ versus $\tilde{\Lambda}$ for several values of $\alpha$ (where $\hat{r}_i=0$ for the boson stars). Figs. (c) and (d) show respectively the properties of the gravitating boson stars and boson shells depicting the plot of $\hat{r}_o$ versus $\tilde{\Lambda}$ for several values of $\alpha$. The figures in the inset show a part of the plots with a better precision for the depicted range of $\tilde{\Lambda}$.\label{fig:12f}}
\end{center}
\end{figure*}

\begin{figure*}
\begin{center}
	\mbox{\subfigure[][]{\includegraphics[scale=0.65]{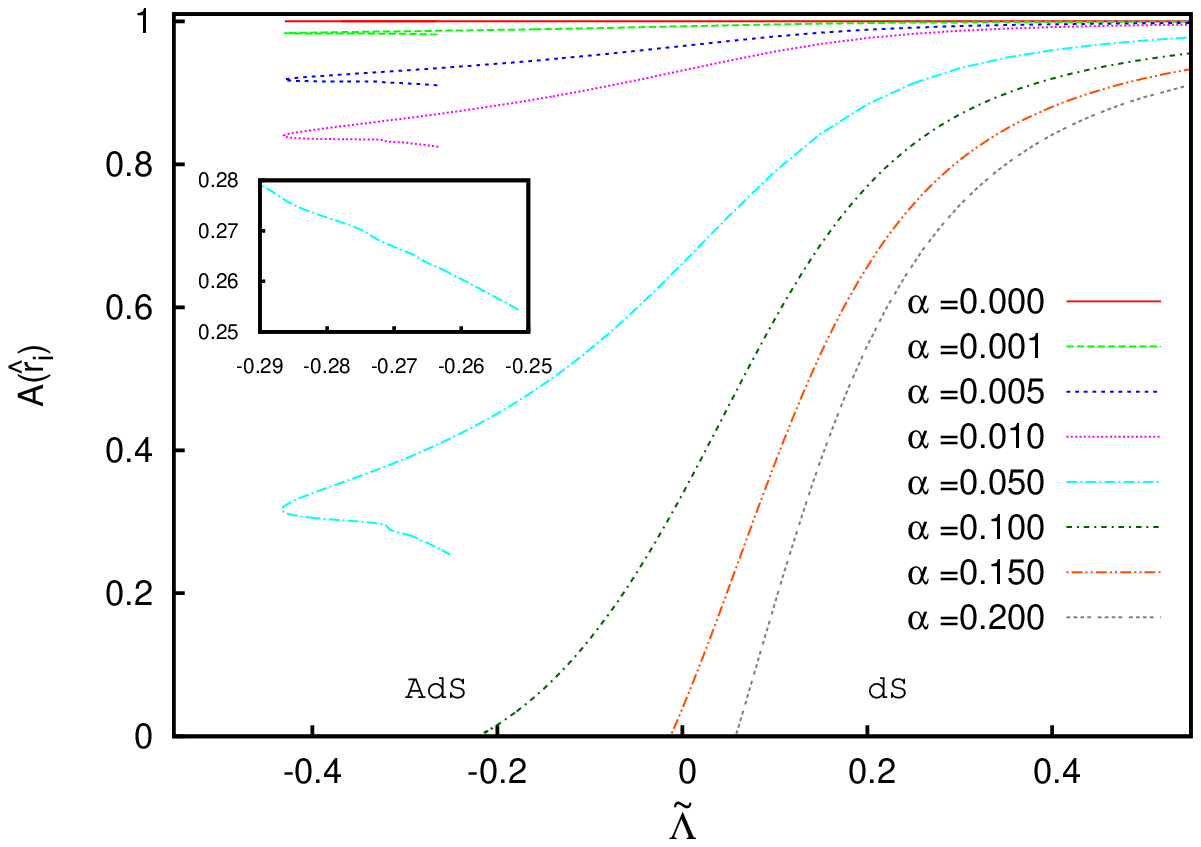}\label{fig:f4a}}\hspace{1cm}
	\subfigure[][]{\includegraphics[scale=0.65]{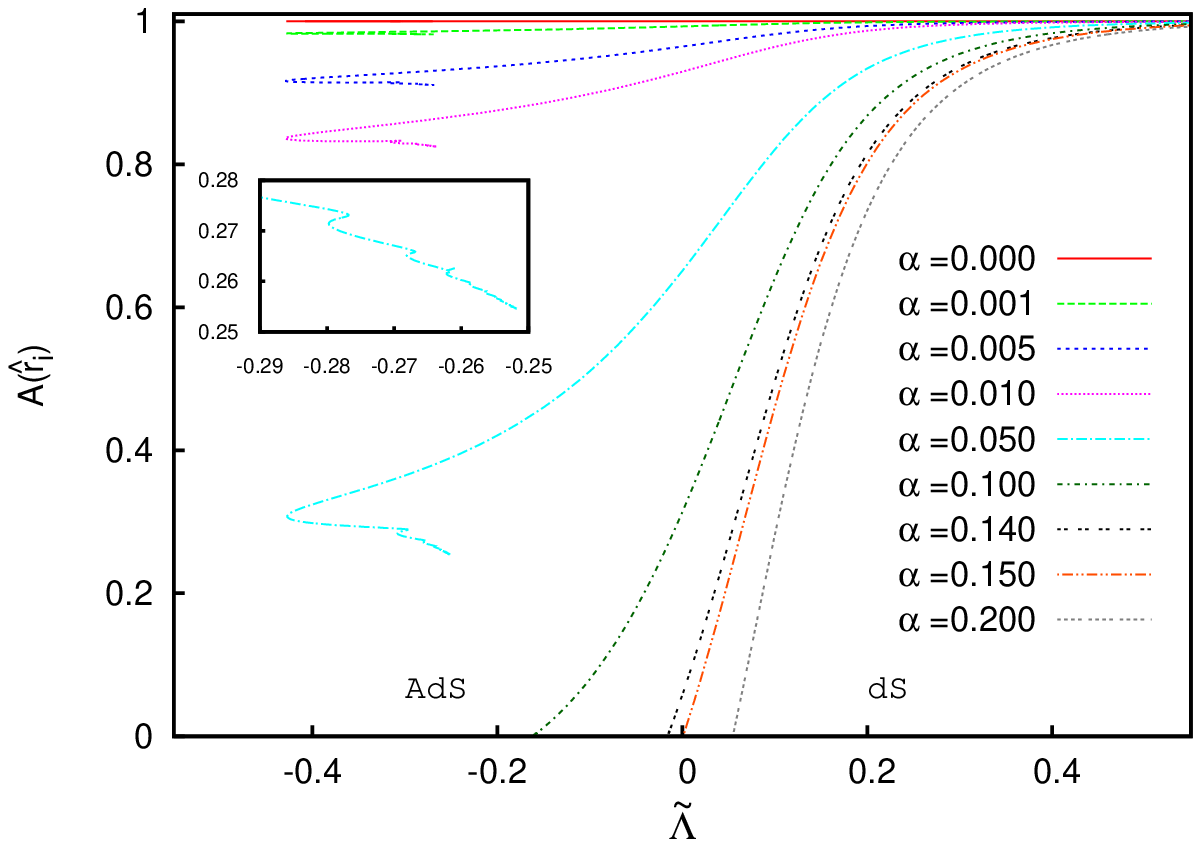}\label{fig:f4b}}}
		\mbox{\subfigure[][]{\includegraphics[scale=0.65]{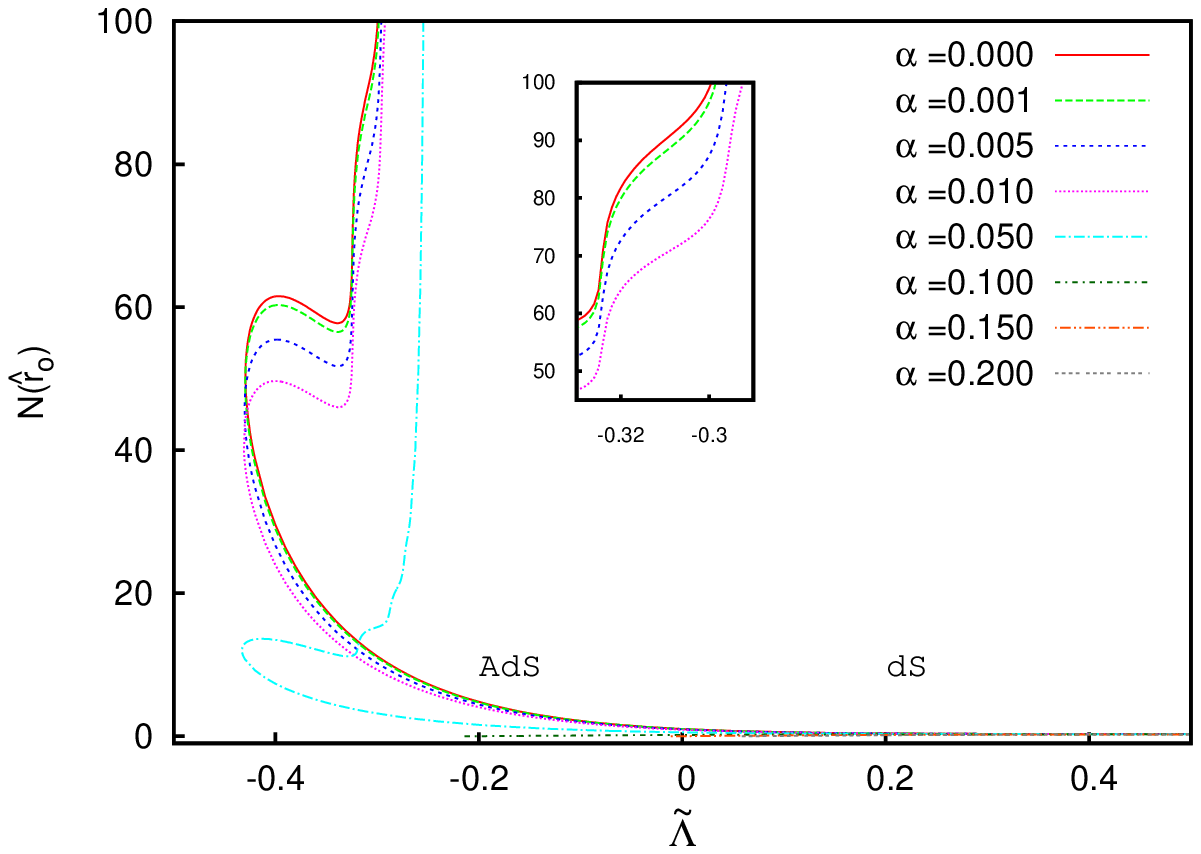}\label{fig:f5a}}\hspace{1cm}
		\subfigure[][]{\includegraphics[scale=0.65]{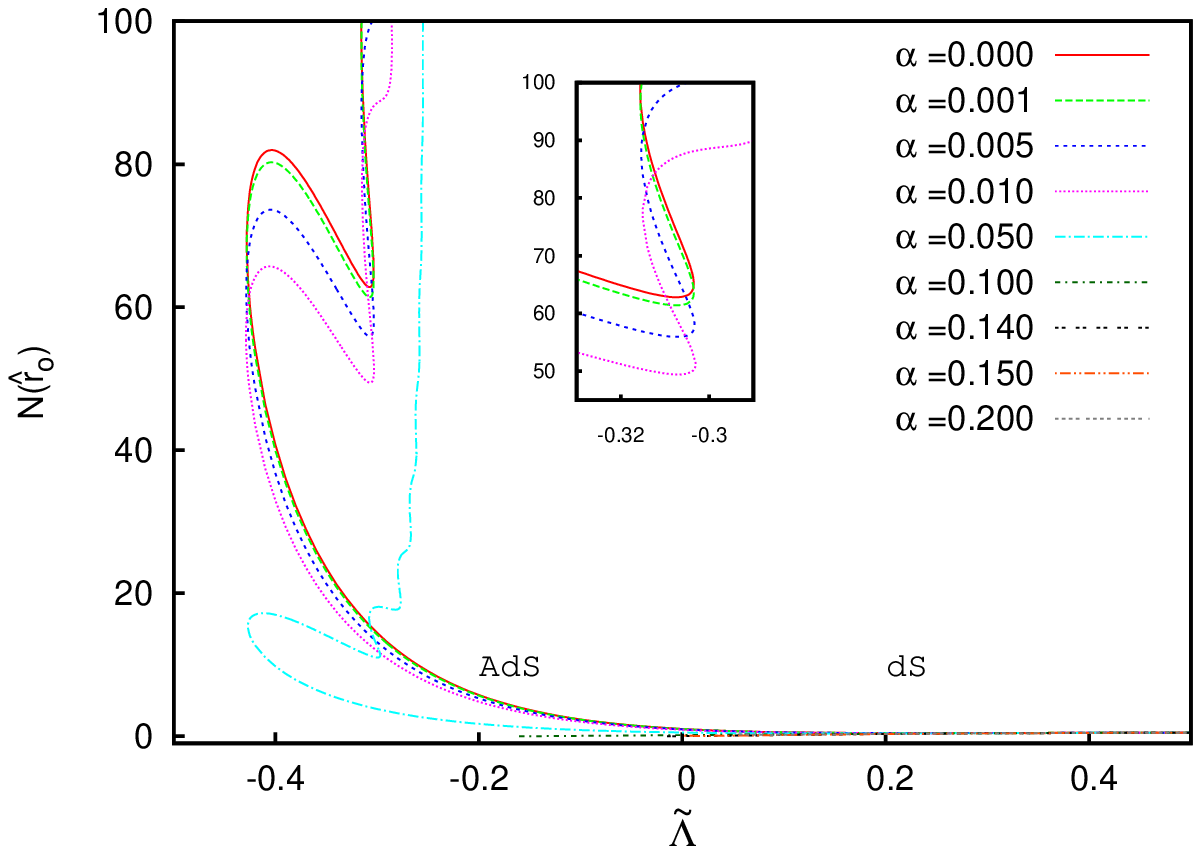}\label{fig:f5b}}}
\caption{Figs. (a) and (b) show respectively the properties of the gravitating boson stars and boson shells depicting the plot of $A(\hat{r}_i)$ versus $\tilde{\Lambda}$ for several values of $\alpha$ (where $\hat{r}_i=0$ for the boson stars). Figs. (c) and (d) show respectively the properties of the gravitating boson stars and boson shells depicting the plot of $N(\hat{r}_o)$ versus $\tilde{\Lambda}$ for several values of $\alpha$. The figures in the inset show a part of the plots with a better precision for the depicted range of $\tilde{\Lambda}$.\label{fig:13f} }
\end{center}
\end{figure*}

\section{Numerical solutions for boson stars and boson shells}
In this section we consider the numerical solutions for the boson stars and boson shells for the positive values of $\tilde{\Lambda}$ (corresponding to the dS space) as well as for the negative values of $\tilde{\Lambda}$ (corresponding to the AdS space).

For this we study the numerical solutions of eqs. (\ref{eq_H}), (\ref{eq_b}), (\ref{eq_N}), (\ref{eq_A}) (by introducing a coordinate $x$ defined by Eq. \ref{vcoord_x} under the boundary conditions given by eqs. \ref{aro} and \ref{bcstar} for boson stars and under \ref{aro} and \ref{bcshell} for boson shells, and we determine their domain of existance for some specific values of the parameters of the theory. We wish to mention that for our numerical investigations we have used the well-known Newton-Raphson scheme with the  adaptive stepsize Runge-Kutta method of order 4.

Our theory has three parameters $\alpha,\ \tilde \lambda$ and $\tilde \Lambda$ and we study the theory by keeping $\alpha$ and $\tilde\lambda$ fixed (namely $\alpha=0.2$ and $\tilde \lambda=1.0$) and we study the theory of boson stars and boson shells for different values of $\tilde \Lambda$ giving it positive as well as negative values, and we discuss the corresponding physics. In the first place we discuss in details our results for the case of boson stars. For this we study and discuss the phase diagrams of our theory for boson stars as shown in Figs \ref{fig:pa}, \ref{fig:pb} and \ref{fig:pd}. In Fig. \ref{fig:pa}, we study it for the fields $h(0)$ and $b(0)$, in Fig. \ref{fig:pb} we study it for the fields $A(0)$ and $b(0)$ and in Fig. \ref{fig:pd} we study it for the fields $A(0)$ and $h(0)$ (where $h(0)$, $b(0)$ and $A(0)$ denote the values of these fields at the centre of the star. Fig. \ref{fig:pc} depicts the plot of $\hat{r}_o $ versus $b(0)$ for different values of $\tilde \Lambda$ ranging from $\tilde \Lambda=-0.020$ to $\tilde \Lambda=+1.000$ (covering the AdS as well as dS spaces). Also the figures shown in the insets in Fig. \ref{fig:pa}--\ref{fig:pc} represent particular sections of these figures with better precision. The asterisks shown in  Figs. \ref{fig:pa}--\ref{fig:pc} corresponding to $h(0)=0$ represent the transition points from the boson stars to the boson shells. 

The phase diagram of the theory involving the fields $N(\hat{r}_o)$ and $b(0)$ is shown in Fig. \ref{fig:s1f1} (where $N(\hat{r}_o)$ denotes the value of the field at the outer radius of the star and $b(0)$ denotes the value of the gauge field at the centre of the boson star) for different values of $\tilde \Lambda$ ranging from $\tilde \Lambda =-0.020$ to $\tilde \Lambda =+1.000$ (covering the AdS as well as dS spaces).

Our present studies involve investigations corresponding to several properties of the boson stars and boson shells as explained in details in this section. In the case of boson stars we study not only the  phase diagrams of the theory for the scalar versus vector fields at the centre of the star (as studied in the work of Ref. \cite{Kumar:2016oop}) but also the phase diagrams of this theory for (i) the case of the metric field $A(\hat{r})$ at the centre of the boson star (i.e. $A(0)$) versus the vector field $b(\hat{r})$ at the centre of the star (i.e. $b(0)$) (depicted in Fig \ref{fig:pb}) and for (ii) the case of the metric field $A(\hat{r})$ at the centre of the boson star (i.e. $A(0)$) versus the scalar field $h(\hat{r})$ at the centre of the star (i.e. $h(0)$) (depicted in Fig \ref{fig:pd}) and for (iii) the case of the phase diagram of the theory involving the metric field $N(\hat r)$ at the outer radius of the boson star (i.e. $N(\hat{r}_o)$) versus $b(0)$ (depicted in Fig \ref{fig:s1f1}). Also all these studies are done for different values of $\tilde\Lambda$ ranging from $\tilde \Lambda=-0.020$ to $\tilde \Lambda=+1.000$ covering the AdS and dS spaces.


As discussed in details in Ref. \cite{Kumar:2016oop}  some interesting phenomenon occurs hear near some specific values of $\tilde \Lambda$ when the system is seen to have four bifurcation points $B_1\,,\,B_2\,,\,B_3$ and $B_4$ which correspond to four different values of the cosmological constant $\tilde{\Lambda}$: $\tilde{\Lambda}_{c_1}\simeq0.22521, ~\tilde{\Lambda}_{c_2}\simeq0.52605$ ,  $\tilde{\Lambda}_{c_3}\simeq0.54076 $ and $\tilde{\Lambda}_{c_4}\simeq0.541250 $ respectively. The theory is seen to have rich physics in the domain $\tilde{\Lambda} = +0.500$ to $\tilde{\Lambda}\simeq + 0.62$.

For a proper understanding of the richness of the physics observed in our phase diagrams obtained in our present investigations it is important to recapitulate some of the related important points from our earlier investigation (cf. work of Ref. \cite{Kumar:2016oop}). In fact, an understanding of the physics contents of the phase diagrams depicted in Fig. \ref{fig:pb} and Fig. \ref{fig:pd} necessitates a proper recount of the physics contents of the phase diagram depicted in Fig. \ref{fig:pa}.

In Fig. \ref{fig:pa}, we have divided our phase diagram into four regions denoted by IA, IB, IIA and IIB in the vicinity of $B_1$. The asterisks seen in Fig. \ref{fig:pa} coinciding with the axis $b(0)$ (i.e. corresponding to $h(0)=0$), represent the transition points from the boson stars to boson shells. 

The regions IA, IB and IIA  do not have any bifurcation points, however, the region IIB is seen to contain rich physics evidenced by the occurrence of more bifurcation points in this region. For better details, the region IIB is also plotted in Fig. \ref{fig:pb}. This region IIB is further divided into the regions IIB1, IIB2 and IIB3 in the vicnity of $B_2$ as seen in Fig. \ref{fig:pa}.

The region IIB3 is seen to have further bifurcation point $B_3$. In the vicinity of  $B_3$ we further subdivide the phase diagram into the regions IIB3a, IIB3b and  IIB3c. The region IIB3b is seen to have closed loops and the behavior of the phase diagram in this region is akin to that of the region IIB2. Also the figures shown in the inset represent part of the diagrams with better precision.

The region IIB3c is again seen to have further bifurcation point $B_4$, and in the vicinity of $B_4$, we again subdivide the phase diagram in to the regions IIB3c1, IIB3c2 and IIB3c3. The region IIB3c2 is again seen to have closed loops and the behavior of the phase diagram in this region is akin to that of the regions IIB2 and IIB3b.

The regions IA and IB could be divided into two sub-regions corresponding to positive and negative values of $\tilde{\Lambda}$, implying  the dS and AdS regions corresponding to positive and negative values of $\tilde{\Lambda}$. In the region IA, as we change the value of $\tilde{\Lambda}$ in the AdS region from $\tilde{\Lambda}=0.000 $ to $\tilde{\Lambda}=-0.020$ , we observe a continuous deformation of the curves in the phase diagram. In the region IB, as we change the value of $\tilde{\Lambda}$ in the domain $\tilde{\Lambda}=0.000$ to $\tilde{\Lambda}\simeq-0.02$ the theory is seen to have solutions for the boson stars only, without having transition points from boson stars to boson shells and the curves corresponding to the solutions disappear in the phase diagram of the theory for the values $\tilde{\Lambda}\lesssim-0.02$~. 

As we change the value of $\tilde{\Lambda}$ in the dS region from   $\tilde{\Lambda}=0.000 $ to $\tilde{\Lambda}=1.000$ , we observe a lot of new rich physics. While going from  $\tilde{\Lambda}=0.000 $ to some critical value $\tilde{\Lambda}=\tilde{\Lambda}_{c_1}$ , we observe that the solutions exist in two separate domains IA and IB ( as seen in Fig. \ref{fig:1f}). However, as we increase $\tilde{\Lambda}$ beyond $\tilde{\Lambda} =\tilde{\Lambda}_{c_1}$ the solutions of the theory are seen to exist in the regions IIA and IIB (instead of the regions IA and IB).

As we increase the value of $\tilde{\Lambda}$ from one critical value $\tilde{\Lambda}=\tilde{\Lambda}_{c_1}$ to another critical value $\tilde{\Lambda}=\tilde{\Lambda}_{c_2}$, we notice that the region IIA in the phase diagram shows a continuous deformation of the curves and the region IIB is seen to have its own rich physics as explained above. 

As we increase $\tilde{\Lambda}$ beyond $\tilde{\Lambda}_{c_2}$, we observe that in the region IIA there is again a continuous deformation of the curves all the way up to $\tilde{\Lambda}=1.000$. The occurence of more bifurcation points in the region IIB and the associated nature of the relevant phase trajectories in the sub-regions of the domain IIB of the phase diagram of the theory (cf. Figs. \ref{fig:pa} and \ref{fig:pb}) has been discussed in details in the foregoing (for many further details of our earlier work, we refer to Ref. \cite{Kumar:2016oop}). 

{\it We wish to emphasize here that a proper understanding of the richness of the physics contents of the theory like the occurence of multiple bifurcation points in the phase diagram shown in \ref{fig:pa} is very crucial for several reasons. This leads not only to a proper understanding of the physics contents of the 2D phase diagrams shown in Figs. \ref{fig:pb} and \ref{fig:pd}  on the lines parallel to those of Fig. \ref{fig:pa} but it also leads to a proper understanding of the physics contents of the 3D phase diagrams shown in Figs. \ref{fig:1df}.  This in fact, has necessitated an appropriate mentioning here of some of the important discussion given in our earlier work (cf. Ref. \cite{Kumar:2016oop}).}

A plot of the vector field at the center of the star $b(0)$ versus the radius $\hat{r}_o$ of the boson star is depicted in Fig. \ref{fig:pc}. As before, the point $B_1$ corresponds to the bifurcation point and the entire region depicted in Fig. \ref{fig:pc} is divided into four regions IA, IB and IIA, IIB in the vicinity of the bifurcation point $B_1$. The asterisks shown in Fig. \ref{fig:pc} represent the transition points from the boson stars to the boson shells. The spiral behavior of the solutions is visible in the regions IA and IIB. 

Figs. \ref{fig:spa} and \ref{fig:spb} represent the {\it simplified versions} of the Figs. \ref{fig:pa} and \ref{fig:pb} and depict the phase diagrams of the theory showing respectively the plots of $h(0)$ versus $b(0)$ and $A(0)$ versus $b(0)$ (where $h(0)$, $b(0)$ and $A(0)$  denote the values of these fields at the centre of the star). Fig. \ref{fig:s1f1} depicts the phase diagram of the theory involving $N(\hat{r}_o)$ and $b(0)$ (where $N(\hat{r}_o)$ denotes the value of the field at the outer radius of the star). Fig. \ref{fig:spc} again represents a {\it simplified version} of  Fig. \ref{fig:pc} and depicts a plot of $\hat{r}_o$ versus $b(0)$.  We wish to emphasize here that the Figs. \ref{fig:1f1} show the plots for 8 different values of $\tilde \Lambda$ ranging from $\tilde \Lambda =-0.020$ to $\tilde \Lambda =+1.000$ (covering the AdS as well as dS spaces). {\it It is to be noted here that Figs. \ref{fig:1f1} are being plotted only for  8 different values of $\tilde \Lambda$ whereas the Figs. \ref{fig:1f} are being plotted for 17 different value of $\tilde \Lambda$. In this sense Figs. \ref{fig:spa}, \ref{fig:spb} and \ref{fig:spc} represent the simplified versions of the Figs. \ref{fig:pa}, \ref{fig:pb} and \ref{fig:pc}}. The figures shown in the insets in Figs. \ref{fig:spa}--\ref{fig:spc} represent particular sections of these figures with better precision.

We have studied the properties of gravitating boson stars as depicted in  Figs. \ref{fig:fa}--\ref{fig:ff}  showing the 3D plots involving the scalar field $h$ and the U(1) guage field $b$ and the  metric field $A$ at the centre of the boson star for several values of the cosmological constant $\tilde\Lambda$ with six different viewing angles.  For the viewing angles denoted by $(\theta, \phi)$ we use the convention such that $\theta$ denotes the angle of rotation about the $ox$-axis in the anticlockwise direction and it can take values between 0 to $\pi$ and $\phi$ is the angle of rotation along the $oz^\prime$-axis also in the anticlockwise direction and it can take values between 0 to $2\pi$. Figs. \ref{fig:fa}--\ref{fig:ff} correspond respectively to the values: $(\theta,\phi)\equiv(60,60),\ (60,170),$ $(60,240),\ (130,30),$ $(130,70) \mbox{ and } (130,350)$.
 
We have also studied the properties of gravitating boson stars as depicted in Figs. \ref{fig:fa2}--\ref{fig:ff2} showing the 3D plot of the scalar field $h$ and the U(1) guage field $b$ at the centre of the boson star and the radius of the star  $\hat r_o$ for several values of the cosmological constant $\tilde\Lambda$ ) (with six different viewing angles).  Figs. \ref{fig:fa2}--\ref{fig:ff2} correspond respectively to the values: $(\theta,\phi)\equiv(60,100),\ (112,17),$ $(60,200),\ (35,245),$ $(140,60)\mbox{ and } (130,30)$.

In the following, we wish to explain that our 3D plots as depicted in Figs. \ref{fig:1df} and \ref{fig:2f} with some arbitrary values for $(\theta,\phi)$ could be reduced to the 2D plots as depicted in Fig. \ref{fig:1f} by choosing some appropriate values for $(\theta,\phi)$. For example, a choice of $(\theta,\phi)\equiv(0,0)$ for the Figs. \ref{fig:1df} and \ref{fig:2f} would reduce these 3D plots to the 2D plot shown in Fig. \ref{fig:pa} with $b(0)$ being plotted on the x-axis and $h(0)$ being plotted on the y-axis (and $A(0)$ and $\hat r_o$ being along the z-axis). Similarly, a choice of $(\theta,\phi)\equiv(90,0)$ for the Figs. \ref{fig:1df} and \ref{fig:2f} would reduce these 3D plots to the 2D plots shown respectively in the Fig. \ref{fig:pb}  (with $b(0)$ being plotted on the x-axis and $A(0)$ being plotted on the y-axis) and the Fig. \ref{fig:pc} (with $b(0)$ being plotted on the x-axis and $\hat r_o$ being plotted on the y-axis) ( and $h(0)$ being along the z-axis). Also, a choice of $(\theta,\phi)\equiv(90,90)$ for the Fig. \ref{fig:1df}  would reduce the 3D plots of Fig. \ref{fig:1df} to the 2D plot shown in Fig. \ref{fig:pd} with $h(0)$ being plotted on the x-axis and $A(0)$ being plotted on the y-axis (and $b(0)$ being along the z-axis). Also, the asterisks shown in  Figs. \ref{fig:1df}--\ref{fig:2f} corresponding to $h(0)=0$ represent the transition points from the boson stars to the boson shells.      

Further, we have invistigated in details the variations of all the four fields involved in our theory, namely $h(\hat r)$, $N(\hat r)$, $A(\hat r)$ and $b(\hat r)$ with $\hat r$ for several values of the cosmological constant $\tilde \Lambda$ covering the AdS as well as the dS spaces. The variations of our above investigations are plotted  in Fig. \ref{fig:3f} - Fig. \ref{fig:11f}.

Here, Figs. \ref{fig:01} to \ref{fig:05} depict plots of $h(\hat{r})$, $N(\hat{r})$, $A(\hat{r})$ and $b(\hat{r})$ versus $\hat{r}$ for $\tilde{\Lambda}=-0.200$ (which corresponds to the AdS space). In order to understand the Figs. \ref{fig:01} to \ref{fig:05} more properly we consider a hypothetical 5D parameter space spanned by $(\tilde \Lambda,\, N(0),\, b(0),\, h(0),\,A(0))$. We also remind ourselves that for a study of boson stars we have fixed $N(0)=1$ for all the cases under study. Next we pick up cases corresponding to different values of $\tilde \Lambda$. In particular, for the case under study represented by Figs. \ref{fig:01}-\ref{fig:05}, we choose the case corresponding to $\tilde \Lambda=-0.200$ ( which corresponds to AdS space). It is to be noted here that this particular curve lies entirely in the region IA shown in Figs. \ref{fig:pa} and \ref{fig:pb}. We then choose a particular value of $b(0)$ (namely $b(0)=1.735$) and from the phase diagram shown in Fig. \ref{fig:pa} we pick up four different points (on the curve corresponding to $\tilde \Lambda=-0.200$) corresponding to $h(0)=1.30 $, $h(0)=2.70 $, $h(0)=3.85 $ and $h(0)=5.65$ for the Figs. \ref{fig:01}-\ref{fig:05} respectively. However, we remind ourselves here that corresponding to these four values of $h(0)$, the corresponding value of $A(0)$ gets automatically fixed by virtue of the solutions of the nonlinear differential equations given by Eqs. \ref{eq_H}-\ref{eq_A}. We then investigate the variations of the four fields involved in the theory namely, $h(\hat r),\, N(\hat r),\, A(\hat r)$ and $b(\hat r)$ with respect to $\hat r$ for the ranges shown in Fig. \ref{fig:3f}.

This in a way gives us an idea not only about the values of the fields inside as well as outside the boson star but also the variations of these fields in the ranges shown in Fig. \ref{fig:3f}. In turn, it also explains our motivations behind these studies.

Figs. \ref{fig:11} and \ref{fig:12}  depict plots of $h(\hat{r})$, $N(\hat{r})$, $A(\hat{r})$ and $b(\hat{r})$ versus $\hat{r}$ for the case $\tilde{\Lambda}=0.000$ (which corresponds to the case of the absence of cosmological constant $\tilde \Lambda$ in the theory). For the Figs. \ref{fig:11} and \ref{fig:12} we choose two points: ($h(0)=1.70$, $b(0)=0.019532$) and ($h(0)=1.70$, $b(0)=1.20812$) which lie in the regions IB and IA respectively in the phase diagram of the theory (cf. Fig. \ref{fig:pa}) on the curves corresponding to $\tilde \Lambda=0.0$. It is to be noted here that the values of $A(0)$ for these two points get automatically fixed by virtue of the solutions of Eqs. \ref{eq_H}-\ref{eq_A}. The variations of the four fields involved in the theory namely, $h(\hat r),\, N(\hat r),\, A(\hat r)$ and $b(\hat r)$ are then investigated with respect to $\hat r$ for the ranges shown in Figs. \ref{fig:11}--\ref{fig:12}.

Figs. \ref{fig:13}--\ref{fig:16}  depict plots of $h(\hat{r})$, $N(\hat{r})$, $A(\hat{r})$ and $b(\hat{r})$ versus $\hat{r}$ for $\tilde{\Lambda}=0.100$ (which corresponds to the dS space). For the Figs. \ref{fig:13} and \ref{fig:14} we choose two points: ($h(0)=2.00,\, b(0)=1.071483$) and ($h(0)=2.00,\,b(0)=0.07008$) which lie in the regions IB and IA respectively. Also, For the Figs. \ref{fig:15} and \ref{fig:16} we choose another two points: ($h(0)=2.600,\;b(0)=1.450$) and ($h(0)=1.24925,\; b(0)=1.450$) both of which lie in the region IA of the phase diagram (Fig. \ref{fig:pa}). We then study the variations of the four fields involved in the theory namely, $h(\hat r),\, N(\hat r),\, A(\hat r)$ and $b(\hat r)$ versus $\hat r$ for the range shown in the Figs. \ref{fig:13}-\ref{fig:16}. It may be important to emphasize here that the cases corresponding to the dS space (having positive values of $\tilde \Lambda$) would have some definite values for the radius of the cosmological horizon $\hat r_c$ as depicted in Figs. \ref{fig:5f}--\ref{fig:10f}.
 
Figs. \ref{fig:20}--\ref{fig:23}  depict plots of $h(\hat{r})$, $N(\hat{r})$, $A(\hat{r})$ and $b(\hat{r})$ versus $\hat{r}$ for $\tilde{\Lambda}=0.400$ (which corresponds to the dS space). For the Figs. \ref{fig:20} and \ref{fig:21} we choose two points: ($h(0)=3.0,\; b(0)=0.05463$) and ($h(0)=3.0,\;b(0)=1.29837$). Both of these points lie in the region IIB1 in the phase diagram: Fig. \ref{fig:pa}. Similarly for the Figs. \ref{fig:22} and \ref{fig:23} we choose two points: ($h(0)=0.50,\;b(0)=2.900$ and $h(0)=0.50,\;b(0)=1.15874$. Both of these points lie in the region IIA. For these four cases, we then study the variations of the fields of the theory  $h(\hat r),\, N(\hat r),\, A(\hat r)$ and $b(\hat r)$ versus $\hat r$ for the ranges shown in the figures. 

Figs. \ref{fig:29}--\ref{fig:34} depict plots of $h(\hat{r})$, $N(\hat{r})$, $A(\hat{r})$ and $b(\hat{r})$ versus $\hat{r}$ for $\tilde{\Lambda}=0.540$ (which corresponds to the dS space). For the Figs. \ref{fig:29} and \ref{fig:30} we choose two points: ($h(0)=0.50,\; b(0)=1.4791 $) and ($h(0)=0.50,\;b(0)=2.700$), which lie in the region IIA in the phase diagram: Fig. \ref{fig:pa}. Similarly for the Figs. \ref{fig:31} and \ref{fig:32} we choose two points: ($h(0)=3.00,\;b(0)=1.0741$) and ($h(0)=3.00,\;b(0)=0.78123$), which lie in the region IIB2. Again for the Figs. \ref{fig:33} and \ref{fig:34} we choose two points: ($h(0)=5.00,\;b(0)=0.72877$) and ($h(0)=5.00,\;b(0)=0.57149$), which lie in the region IIB3a. Corresponding to these six points chosen we then study the varitions of the four fields of the theory  $h(\hat r),\, N(\hat r),\, A(\hat r)$ and $b(\hat r)$ versus $\hat r$ for the ranges shown in the figures.

Figs. \ref{fig:51} to \ref{fig:56}  depict plots of $h(\hat{r})$, $N(\hat{r})$, $A(\hat{r})$ and $b(\hat{r})$ versus $\hat{r}$ for $\tilde{\Lambda}=0.541$ (which corresponds to the dS space). For the Figs. \ref{fig:51} and \ref{fig:52} we choose two points: ($h(0)=0.50,\; b(0)=1.48133 $) and ($h(0)=0.50,\;b(0)=2.700$), which lie in the region IIA in the phase diagram: Fig. \ref{fig:pa}. Similarly for the Figs. \ref{fig:53} and \ref{fig:54} we choose two points: ($h(0)=3.00,\;b(0)=0.28354$) and ($h(0)=3.00,\;b(0)=1.07199$), which lie in the region IIB2. Again for the Figs. \ref{fig:55} and \ref{fig:56} we choose two points: ($h(0)=5.00,\;b(0)=0.71695$) and ($h(0)=5.00,\;b(0)=0.58354$), which lie in the region IIB3b. Corresponding to these six points chosen we then study the varitions of the four fields of the theory  $h(\hat r),\, N(\hat r),\, A(\hat r)$ and $b(\hat r)$ versus $\hat r$ for the ranges shown in the figures.
Figs. \ref{fig:59} to \ref{fig:48}  depict plots of $h(\hat{r})$, $N(\hat{r})$, $A(\hat{r})$ and $b(\hat{r})$ versus $\hat{r}$ for four values of $\tilde{\Lambda}$, namely $\tilde\Lambda= 0.100,\; 0.400,\; 0.540,$ and $0.541$ respectively (all of which correspond to the dS space).  Also, these figures correspond to the transition points of the theory from the bosons stars to the boson shells which implies $h(0)=0.00$ for all the four figures. Also, these figures actually correspond to $b(0)= 0.325587,\;2.06006,\;2.50273$ and $2.50550$ respectively. The line $h(0)=0.00$ in fact, marks one of the boundaries of the regions IB and IIA.  

Figs. \ref{fig:2-e}--\ref{fig:2-a}  depict plots of $h(\hat{r})$, $N(\hat{r})$, $A(\hat{r})$ and $b(\hat{r})$ versus $\hat{r}$  for the {\it boson shells} with different values of the inner radius namely, $\hat r_i=0.000,\;0.001,\;0.500,\;1.000$ respectively for  $\tilde{\Lambda}=0.010$ {\it (which corresponds to the dS space)}. 

Figs. \ref{fig:2-f}--\ref{fig:2-b} ) depict plots of $h(\hat{r})$, $N(\hat{r})$, $A(\hat{r})$ and $b(\hat{r})$ versus $\hat{r}$ for the {\it boson shells} with different values of the inner radius namely, $\hat r_i=0.000,\;0.001,\;0.500,\;1.000$ respectively for $\tilde{\Lambda}=-0.010$ {\it (which corresponds to the AdS space)}. 

We now explain the physics contents of Figs. \ref{fig:12f} and \ref{fig:13f} and highlight some important points (albeit salient features) of our investigations in the context of these figures.
Figs. \ref{fig:12f} and \ref{fig:13f} depict the physical properties of the compact boson star and boson shell solutions where the case $\alpha=0$ corresponds to the Q-shell solutions. We also determine the domains of existence of these physical properties for some specific values of the parameters of the theory. In our numerical calculations we have fixed the value of the parameter $\tilde{\lambda}$ to be equal to one for the boson stars as well as the boson shells solutions. The value of the parameter $\hat{r}_i$ has been fixed to one for the boson shells and $\hat{r}_i=0$ for the boson stars. 

 Figs. \ref{fig:f2a} and \ref{fig:f2b} show respectively the properties of the gravitating {\it boson stars} and {\it boson shells} depicting the plot of $b(\hat{r}_i)$ versus $\tilde{\Lambda}$ for several values of $\alpha$ (where $\hat{r}_i=0$ for the boson stars) and Figs. \ref{fig:f3a} and \ref{fig:f3b} show respectively the properties of the gravitating boson stars and boson shells depicting the plot of $\hat{r}_o$ versus $\tilde{\Lambda}$ for several values of $\alpha$. The figures in the insets show a part of these plots with a better precision for the depicted range of $\tilde{\Lambda}$.

 It is apparent from these figures that the minimum value of $b(\hat{r}_i)$ for a particular value of $\tilde{\Lambda}$ keeps decreasing with the increasing value of $\alpha$ until the value of $b(\hat{r}_i)$ reaches zero. In fact, the plot for a particular value of $\alpha$ continuously deforms itself as we increase the value of $\alpha$. This scenario is more transparent in the domain where the cosmological constant $\tilde{\Lambda}$ is negative (i.e. in the AdS space). Actually with the increase in the value of $\alpha$ the solutions cease to exit in the AdS region for $\alpha\ge 0.150$. For $\alpha\ge0.150$, the solutions do continue to exist however only in the dS region. Also within the AdS region the damped oscillations of the cosmological constant $\tilde{\Lambda}$ with increasing $b(\hat{r}_i)$ are seen in Figs. \ref{fig:f2a} and \ref{fig:f2b} for some smaller values of $\alpha$.

Figs. \ref{fig:f4a} and \ref{fig:f4b} show respectively the properties of the gravitating boson stars and boson shells depicting the plot of $A(\hat{r}_i)$ versus $\tilde{\Lambda}$ for several values of $\alpha$ (where $\hat{r}_i=0$ for the boson stars) and Figs. \ref{fig:f5a} and \ref{fig:f5b} show respectively the properties of the gravitating boson stars and boson shells depicting the plot of $N(\hat{r}_o)$ versus $\tilde{\Lambda}$ for several values of $\alpha$. Figures in the insets show a part of these plots with a better precision for the depicted range of $\tilde{\Lambda}$.

It is apperent from these figures that there is a continuous deformation of the curves as we increase or decrease the value of the cosmological constant $\tilde{\Lambda}$.

\section{Summary and Conclusions}
The boson stars and boson shells representing the localized self-gravitating solutions  have been studied in the literature in the presence of positive as well as negative values of the cosmological constant ${\Lambda}$. The theories in the dS space (corresponding to positive values of ${\Lambda}$) describe somewhat realistic  description of the compact stars and are relevant from observational point of view. They are also used in the dark energy models of the universe. However, the theories in the AdS space (with negative values of ${\Lambda}$) are important for the AdS/CFT theories.

In our earlier work (cf.  Ref. \cite{Kumar:2014kna}), we studied the boson stars and boson shells in a theory of complex scalar field coupled to  $U(1)$ gauge field $A_{\mu}$ and the gravity in the presence of a {\it positive} cosmological constant (i.e. in the dS space). However, in the present work we have studied this theory in the presence of {\it positive as well as negative values} of the cosmological constant treated as a free parameter. This allows us to study the theory in the dS as well as in the AdS space. Also in Ref. \cite{Kumar:2016oop}, we have studied the {\it boson stars} in a theory of complex scalar field coupled to the $U(1)$ gauge field $A_{\mu}$ and the gravity in the presence of a positive as well as negative cosmological constant. Whereas in the present work we have studied {\it not only the boson stars but also the boson shells} in this theory.

As in Ref. \cite{Kumar:2016oop}, in our present work also, we have studied the theory in the presence of a potential: $V(|\Phi|) := (m^2 |\Phi|^2 +\lambda |\Phi|) $ (with $m$ and $\lambda$ being constant parameters). We have investigated properties of the solutions of this theory and determined their domains of existence for some specific values of the parameters of the theory. 

It is important to emphasize here as also explained in the introduction that similar solutions have also been obtained by Kleihaus, Kunz, Laemmerzahl and List in a theory involving {\it massless} complex scalar fields coupled to the U(1) gauge field and gravity in a conical potential  in the {\it absence} of a cosmological constant ${\Lambda}$ \cite{Kleihaus:2009kr,Kleihaus:2010ep}. They have obtained explicitly the domain of existence of compact boson stars and boson shells. They have also considered the boson shells, which do  not have an empty inner region $r<r_i$, but instead they harbour a Schwarzschild black hole or a Reissner-Nordstr\"om black hole in the region $r<r_i$ \cite{Kleihaus:2009kr,Kleihaus:2010ep}. 

As mentioned in the foregoing, polynomial potentials have also been used in the literature for a study of the boson stars. For a study of the {\it compact} boson stars, presence of the {\it conical piece} of the potential is very crucial. However, the choice of the massless scalar fields (instead of the massive scalar fields as considered in the present work) in the theory reduces the number of free parameters of the theory by one and is expected to produce the results conceptually somewhat similar to our present results although with different numerical details (albeit, with different domains of existence). These studies are currently underway and would be reported later. The inclusion of the quartic and/or sextic terms in the potential is expected to bring-in some new additional features in the results. We propose to undertake such studies in the near future.  

In the present work, we have constructed the boson star and boson shell solutions of this theory numerically and we have studied their properties by assuming the interior of the shells to be empty space (dS-like or AdS-like). 

In this work, we have shown that these charged shell-like solutions persist in the presence of not only the {\it positive} cosmological constant (corresponding to dS space) but also for the {\it negative} cosmological constant (corresponding to AdS space).

The self-gravitating compact boson shells constructed in our present work possess an empty dS-like or AdS-like interior region: $\hat{r}<\hat{r}_i$, and a Reissner-Nordstr\"om-dS or Reissner-Nordstr\"om-AdS exterior region: $\hat{r}>\hat{r}_o$. 

In our future work, we also propose to consider the possibility of filling of the interior region of the boson shells with black holes, analogous to the study in Ref. \cite{Kleihaus:2010ep}. It would also be interesting to extend our present studies of the charged compact boson star and boson shell solutions to other dimensions. 

\section{Acknowledgment}
We thank Jutta Kunz, Burkhard Kleihaus and James Vary for several helpful educative discussions and encouragements. We also like to extend our sincere thanks to the esteemed referee for his/her highly constructive comments and suggestions.This work was supported in part by the US Department of Energy under Grant No. DE-FG02-87ER40371 and by the US National Science Foundation under Grant No. PHY-0904782.
\bibliography{prdr}
\end{document}